\pdfoutput=1

\documentclass[%
reprint,
 amsmath,amssymb,
 aps,
superscriptaddress
]{revtex4-2}

\usepackage{graphicx}%
\usepackage{dcolumn}%
\usepackage{bm}%
\usepackage{hyperref}%
\usepackage{xcolor}
\usepackage{url}

\usepackage{xr-hyper}
\usepackage{pdfpages}
\makeatletter
\AtBeginDocument{\let\LS@rot\@undefined}
\makeatother

\begin{document}

\preprint{APS/123-QED}

\title{Robust Wannierization including magnetization and spin-orbit coupling via projectability disentanglement}

\author{Yuhao Jiang}
\affiliation{PSI Center for Scientific Computing, Theory and Data, 5232 Villigen PSI, Switzerland}
\affiliation{
  Fert Beijing Institute,
  School of Integrated Circuit Science and Engineering,
  Beihang University, 100191 Beijing, China
}

\author{Junfeng Qiao}%
\affiliation{%
  Theory and Simulations of Materials (THEOS),
  and National Centre for Computational Design and Discovery of Novel Materials (MARVEL),
  École Polytechnique Fédérale de Lausanne, 1015 Lausanne, Switzerland
}%

\author{Nataliya Paulish}
\affiliation{PSI Center for Scientific Computing, Theory and Data, 5232 Villigen PSI, Switzerland}

\author{Weisheng Zhao}
\email{weisheng.zhao@buaa.edu.cn}
\affiliation{
  Fert Beijing Institute,
  School of Integrated Circuit Science and Engineering,
  Beihang University, 100191 Beijing, China
}
\affiliation{
  National Key Lab of Spintronics,
  Institute of International Innovation,
  Beihang University, 311115 Hangzhou, China
}

\author{Nicola Marzari}
\affiliation{PSI Center for Scientific Computing, Theory and Data, 5232 Villigen PSI, Switzerland}
\affiliation{%
  Theory and Simulations of Materials (THEOS),
  and National Centre for Computational Design and Discovery of Novel Materials (MARVEL),
  École Polytechnique Fédérale de Lausanne, 1015 Lausanne, Switzerland
}%

\author{Giovanni Pizzi}
\email{giovanni.pizzi@psi.ch}
\affiliation{PSI Center for Scientific Computing, Theory and Data, 5232 Villigen PSI, Switzerland}
\affiliation{%
  Theory and Simulations of Materials (THEOS),
  and National Centre for Computational Design and Discovery of Novel Materials (MARVEL),
  École Polytechnique Fédérale de Lausanne, 1015 Lausanne, Switzerland
}%

\begin{abstract}
Maximally-localized Wannier functions (MLWFs) are widely employed as an essential tool for calculating the physical properties of materials
due to their localized nature and computational efficiency.
Projectability-disentangled Wannier functions (PDWFs) have recently emerged as a reliable and efficient approach
for automatically constructing MLWFs that span both occupied and lowest unoccupied bands.
Here, we extend the applicability of PDWFs to magnetic systems and/or those including spin-orbit coupling, and implement such extensions in automated workflows.
Furthermore, we enhance the robustness and reliability of constructing PDWFs by defining an extended protocol that automatically expands the projectors manifold, when required,
by introducing additional appropriate hydrogenic atomic orbitals.
We benchmark our extended protocol on a set of 200 chemically diverse materials, as well as on the 40 systems with the largest band distance obtained with the standard PDWF approach,
showing that on our test set the present approach delivers a 100\% success rate in obtaining accurate Wannier-function interpolations, i.e., an average band distance below 15~meV between the DFT and Wannier-interpolated bands, up to 2 eV above the Fermi level.
\end{abstract}

\maketitle

\section{\label{sec:intro}Introduction}

In materials science calculations,
density-functional theory (DFT)~\cite{Hohenberg1964} is nowadays an established and highly versatile method,
widely used for calculating various physical properties of extended systems and molecules.
However, to accurately capture several properties of crystalline materials,
such as for instance the anomalous Hall effect~\cite{Nagaosa2010} or the spin Hall effect~\cite{Sinova2015, Derunova2019},
calculations on a very  dense $k$-point mesh in the Brillouin zone are required,
often involving even millions of $k$-points~\cite{Yao2004,Guo2008} to achieve convergence.
Solving the Kohn-Sham equations~\cite{Kohn1965} independently for such an extensive number of $k$-points
incurs significant computational cost.

An alternative approach is to construct a tight-binding model using a real-space basis of  Wannier functions (WFs)~\cite{Marzari2012},
which are efficiently obtained from Bloch functions via a unitary transformation between the Bloch wavefunctions at every $k$-point, and then Fourier-transformed to generate maximally localized Wannier functions (MLWFs)~\cite{Marzari1997}.
These functions allow for efficient interpolation of wavefunctions onto arbitrary $k$-point meshes at a low computational cost.
This approach has been widely adopted in particular for calculating physical quantities requiring extensive $k$-point integration~\cite{Marrazzo_2024},
such as the density of states (DOS)~\cite{Zhang2009}, Boltzmann transport~\cite{Pizzi_2014},
anomalous Hall effect~\cite{Wang2006},
orbital magnetic moments~\cite{Thonhauser2005},
and the spin Hall effect~\cite{Qiao2018,Ryoo2019}.

WFs $|w_{n\bf{R}}\rangle$ are obtained via a Fourier transformation of
the Bloch states $|\psi_{n\bf{k}}\rangle$ associated to the same band $n$,

\begin{equation}
  \label{eq:wan_from_bloch}
  |w_{n \bf{R}} \rangle =
  \frac{V}{(2\pi)^3}\int_{BZ} d {\bf k} e^{-i\bf{kR}} |\psi_{n \bf{k}}\rangle,
\end{equation}
where $V$ is the volume of the primitive cell, and $\bf k$ and $\bf R$ are the Bloch quasi-momentum in the BZ and 
a real-space lattice vector, respectively.
However, there is a gauge freedom of the Bloch functions where each Bloch state can be multiplied by a phase factor $e^{i\phi_{n}(\bf k)}$, dependent both on $n$ and $\bf k$, 
without changing the Hilbert space but changing the shape of the WFs (and in particular their localization in real space). MLWFs utilize such a gauge freedom to obtain the most localized WFs~\cite{Marzari1997} by minimizing a quadratic spread functional
\begin{equation}
  \label{eq:quadratic_spread}
  \Omega = \sum_{n=1}^{J}[
    \langle w_{n\bf 0} | {\bf r}^2 | w_{n\bf 0}\rangle - 
    |\langle w_{n\bf 0} | {\bf r} | w_{n\bf 0}\rangle|^2
  ],
\end{equation}
where $J$ is the number of target Wannier bands.
For multi-band systems, the gauge freedom is further generalized to allow mixing between different bands, rather than being limited to a simple exponential phase factor. This freedom can be
encoded in a set of unitary matrices $U_{mn\bf{k}}$, so that MLWFs can be expressed as
\begin{equation}
  \label{eq:wan_from_bloch_gauge}
  |w_{n \bf{R}} \rangle =
  \frac{V}{(2\pi)^3}\int_{BZ} d {\bf k} e^{-i\bf{kR}}
  \sum_{m=1}^{J_{\bf k}}|\psi_{m \bf{k}}\rangle U_{mn\bf{k}}.
\end{equation}
For an isolated set of bands, such as the valence bands of semiconductors or insulators, $J_{\bf{k}}$ is a constant
that equals to $J$, and the $U_{mn\bf{k}}$ are unitary square matrices.
For metallic systems, where the energy bands are entangled, one should select more bands and perform a disentanglement procedure~\cite{Souza2002}.
The number $J_{\bf k}$ of selected bands is $k$-dependent, and $U_{mn\bf k}$ are semi-unitary
rectangular matrices.

In practice, the algorithm to obtain MLWFs by minimizing Eq.~(\ref{eq:quadratic_spread}) is typically implemented via an iterative algorithm, for which an adequately localized
initial guess must be provided. This initial guess needs to be sufficiently close to the final Wannier functions to achieve convergence and, until very recently, its selection required physical intuition.
One common approach is to project Bloch functions onto hydrogenic wave functions $|g_n\rangle$~\cite{Marzari1997}, to obtain
\begin{equation}
  \label{eq:initial_guess}
  |\phi_{n\bf k} \rangle = \sum_{m=1}^{J_{\bf k}} |\psi_{m\bf k}\rangle\langle\psi_{m\bf k} | g_n\rangle.
\end{equation}
Since the $U_{mn\bf k}$ should be unitary matrices,
the projection matrices $A_{mn\bf k} = \langle \psi_{m\bf k} | g_n \rangle$
are further orthonormalized through the L\"{o}wdin orthonormalization algorithm~\cite{Lowdin1950} to obtain the starting $U_{mn\bf k}$ matrices for the minimization procedure.
Furthermore, for entangled bands the conventional approach [that we label as energy disentanglement (ED)]
is to set an energy window to select the disentanglement manifold~\cite{Souza2002}. An outer window is first defined, including all Bloch states that can be linearly combined to obtain a smaller disentangled manifold. A smaller inner window is then often also used to define frozen states, which are kept unchanged during the disentanglement process. 
However, this requires manual setting of input parameters, such as the number of bands
and target MLWFs, and the parameters determining the shape of hydrogenic projectors, 
which limits its integration into high-throughput (HT) calculations.
Recently, new algorithms have been proposed to address the challenges associated with
HT computations~\cite{Gresch2018,Vitale2020,Qiao2023,Qiao2023a}.
Among these, projectability disentanglement (PD)~\cite{Qiao2023} emerged an efficient and accurate algorithm that can be easily automated.
The PD method uses a criterion based on the value of the projectability of each state onto a set of localized pseudo-atomic orbitals
(PAOs)~\cite{Agapito2016} to determine the
selection of bands used to construct the initial guess. Typically, PAOs are extracted from the pseudopotentials used in DFT
calculations. In this context, the value of the projectability $p_{m\bf{k}}$ refers to the projection of the Bloch
wave function $|\psi_{m\bf{k}}\rangle$ onto the PAOs $|g_n\rangle$, as expressed by
\begin{equation}
  \label{eq:projectability}
  p_{m\bf{k}} = \sum_n \langle \psi_{m\bf{k}}| g_n \rangle \langle g_n|\psi_{m\bf{k}} \rangle.
\end{equation}
The main idea of the PD method can be summarized as follows: states with projectability {$p_{m\bf{k}} \approx 1$} are kept unchanged (in such a case, the set of projectors $|g_n\rangle$ is an almost complete set for the Bloch state $|\psi_{m\bf{k}}\rangle$);
states with projectability $p_{m\bf{k}} \approx 0$ can instead be neglected, as they are
essentially not included in the space spanned by the projectors. The remaining states are instead combined, as prescribed by the disentanglement procedure, to construct the disentangled manifold.

HT calculations based on PDWFs have demonstrated that this method can efficiently produce highly accurate
tight-binding (TB) models~\cite{Qiao2023}. 
Current automated PDWF implementations have only been performed on spin-unpolarized systems,
without considering 
spin degrees of freedom needed to describe, e.g., ferromagnetic, antiferromagnetic~\cite{Baltz2018,Xiong2022} and ferrimagnetic~\cite{Kim2022} structures.
Furthermore, when relativistic effects are also taken into account by introducing spin-orbit coupling (SOC),
higher-order magnetic interactions~\cite{Brinker2020,Ham2021}, topological structures~\cite{Armitage2018, Hasan2010},
intricate magnetic structures in real space~\cite{Zhou2024}, spin textures in momentums space~\cite{Manchon2015}
and other SOC-dominated physical phenomena can be described.
Therefore, in this paper we aim to extend the PDWF approach to magnetic systems and requiring SOC. In doing so, we also extend the PDWF algorithm by defining a protocol to introduce additional projections in the form of hydrogenic atomic orbitals. As a result, we enhance the overall robustness of the Wannierization process, achieving a remarkable 100\% success rate in obtaining accurate Wannier interpolations for all materials in our test set.

\section{\label{sec:Results}Results}

To quantitatively measure the quality of the band structures obtained from 
Wannier interpolation with respect to the DFT band structures,
we employ the definitions from Ref.~\cite{Prandini2018}
to compute the average band distance
\begin{equation}
  \label{eq:average_bandsdistance}
  \eta_\nu = \sqrt{
    \frac{
      \sum_{n\bf{k}}\tilde{f}_{n\bf{k}}
      \left(\epsilon_{n\bf{k}}^{\text{DFT}} - \epsilon_{n\bf{k}}^{\text{Wan}}\right)^2
    }
    {\sum_{n\bf{k}}\tilde{f}_{n\bf{k}}}
  }
\end{equation} and the maximum band distance
\begin{equation}
  \eta_\nu^{max} = \max\limits_{n\bf{k}}
  \left(
    \tilde{f}_{n\bf{k}}
    |\epsilon_{n\bf{k}}^{\text{DFT}} - \epsilon_{n\bf{k}}^{\text{Wan}}|
  \right),
\end{equation}
where 
\(\tilde{f}_{n\bf{k}} = \sqrt{
  f_{n\bf{k}}^{\text{DFT}}(E_F+\nu, \sigma)f_{n\bf{k}}^{\text{Wan}}(E_F+\nu, \sigma)
}\) is an effective Fermi-Dirac distribution, and $f(E_F + \nu, \sigma)$ is
the Fermi-Dirac distribution for DFT and Wannier interpolated states, with $E_F$ being the Fermi level of the system.
In the following, we choose $\nu=2$~eV and $\sigma=0.1$~eV (the same as in Ref.~\cite{Qiao2023}) in order to consider band differences only for those bands with energy (approximately) below $E_F + 2$~eV. This includes the valence bands and a few conduction bands near the Fermi level, which are typically the relevant ones to determine most physical properties.
Furthermore, in the calculations presented in this work, all computations based on PDWFs were configured with an additional frozen energy window,
which was set to include all states up to 2~eV above the Fermi level.
This approach (also named PD+ED in Ref.~\cite{Qiao2023}) is the recommended approach when using the PDWF method (see Ref.~\cite{Qiao2023}) as it generally provides higher-quality Wannier interpolation than just using PD with no frozen window.

\subsection{\label{sec:soc}Spin-orbit coupling}
SOC, as a relativistic effect, plays a significant role in systems involving heavy elements.
It can lift degeneracies~\cite{Dresselhaus2008} and open band gaps at certain $k$-points
where SOC plays a dominant role~\cite{Xiao2012},
leading to notable changes in the electronic properties.
Relativistic effects are described by the Dirac equation~\cite{Dirac1928},
whose solutions are four-component spinors. However, two of these components correspond to antimatter,
which is typically neglected in low-energy physics.
As a result, SOC is often approximated as a relativistic correction to the Schr\"{o}dinger equation~\cite{Bechstedt2015}, with wavefunctions described as two-component
spinors. The initial projectors for MLWFs should thus align with
the physics of SOC. Namely, one wants to use pseudo-atomic orbitals obtained from 
fully relativistic pseudopotentials.
Since currently the SSSP~\cite{Prandini2018} library does not offer a SOC version,
we performed our SOC calculations
using pseudopotentials from the \texttt{PseudoDojo} 0.4~\cite{VanSetten2018} library, containing norm-conserving fully relativistic pseudopotentials,
and conducted validation calculations using the \texttt{pslibrary}~\cite{DalCorso2014} 1.0.0 (PBE, PAW, fully relativistic).
For \texttt{pslibrary}, we selected the corresponding pseudopotentials based on the 
recommendations from Ref.~\cite{pslib_suggest}. We note that, for the \texttt{pslibrary}, some pseudopotentials for certain elements may either be missing or result in DFT calculations that fail to converge.
Consequently, for some elements,
we used older versions of \texttt{pslibrary} or substituted pseudopotentials from \texttt{PseudoDojo}.
Therefore, we refer to this set as a \texttt{modified-pslibrary} set in the following text;
the specific pseudopotential files used are detailed in Supplementary Table~\ref{sm-tab:mix_include}.

Due to SOC, the orbital quantum number $l$ and the spin quantum number $s$ are no longer good quantum numbers, and the system is instead described by the total angular momentum quantum number $j$.
Consequently, the projectors become $j$-dependent when SOC is considered.
To extend the workflow to SOC systems, we adjust the number of energy bands and projectors in the SOC system.
For the PDWF approach, since the projectors can be directly obtained from the pseudopotential files,
we modified the \texttt{pw2wannier90.x} code, part of \texttt{Q\textsc{uantum} ESPRESSO} (QE)~\cite{Giannozzi2009,Giannozzi2020}, for projecting
from plane-wave functions onto projectors accounting for their dependence on $j$.
Additionally, to enable the functionality of reading projectors from external files as implemented in Ref~\cite{Qiao2023},
we also implemented routines for reading $j$-dependent projectors from external files.

\begin{figure}[tb]
  \centering
  \includegraphics[width=8cm]{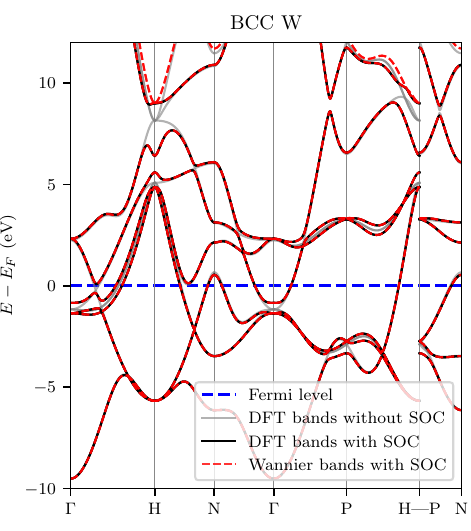}
  \caption{\label{fig:bands_bccw}
  \textbf{Electronic band structure of BCC tungsten}.
  The gray and black solid lines are the energy bands obtained
  directly from first-principles DFT calculations
  without and with SOC, respectively.
  The red dashed lines are the energy bands obtained from Wannier interpolation with SOC using our extended PDWF method.
  The Fermi level is marked as a horizontal blue dashed line.
  }
\end{figure}

The effect of SOC on the band structure of BCC tungsten, as well as the quality of the Wannier-interpolated bands obtained with our algorithm including SOC, are demonstrated in Fig. \ref{fig:bands_bccw}.
Some bands exhibit splitting due to SOC,
particularly along the $\Gamma-H$ $k$-path, where some band crossings transition into anti-crossings induced by SOC just below the Fermi level.
The Wannier-interpolated band structure obtained using our extension of the PDWF method displays a $\eta_2$ band distance of only 2.24~meV. Notably, even at approximately 8 eV above the Fermi level,
the Wannier-interpolated bands remain in good agreement with the DFT bands.
Thus, this Wannier tight-binding model is capable of accurately describing the electronic bands of tungsten
and is suitable for precise calculations of SOC-related properties.

To further validate our algorithm, we performed Wannierization calculations including SOC on the set of 200 chemically diverse materials extracted from the Materials Cloud 3D crystals database (MC3D)~\cite{mc3d} already used for benchmarking in Ref.~\cite{Qiao2023}.
173 among these materials contain elements with atomic numbers greater
than 20, where SOC effects become non-negligible. Therefore, this dataset is a also good test set for estimating the accuracy of Wannier interpolation with SOC, and where the few materials containing only light elements enable us to verify that the robustness of the PDWF approach is not disrupted when including SOC effects, even when these are negligible.

\begin{figure}[tb]
  \includegraphics[width=8.0cm]{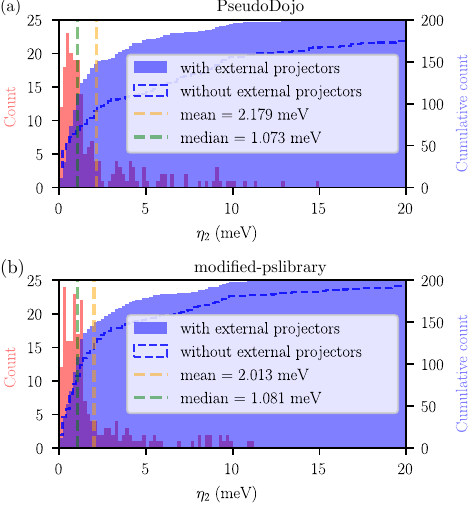}
  \caption{\label{fig:bd_soc_add_hyd}
  \textbf{Band distance $\eta_2$ of 200 systems with different pseudopotentials, including SOC effects.}
  Histogram (red) and cumulative histogram (blue) of the 
  band distance $\eta_2$ of 200 spin-orbit coupling systems 
  with different pseudopotentials sets: (a) the \texttt{PseudoDojo} library, and (b) the \texttt{modified-pslibrary} set (see main text).
  External hydrogenic atomic orbitals are introduced to the projectors to enhance
  the robustness of the PDWF method, making results obtained with the two pseudopotential libraries qualitatively very similar. As a comparison, the blue dashed lines are the cumulative histogram
  of $\eta_2$ without introducing such external projectors, exhibiting a lower success rate and a strong dependence on the pseudopotential library.
  The orange (green) vertical line is the mean (median) band distance $\eta_2$ of the 200 structures with external projectors;
  their values are shown in the legend of each panel.
  All 200 structures can be interpolated with a resulting $\eta_2\le 15$~meV for both pseudopotentials libraries, once hydrogenic atomic orbitals are introduced.
  }
\end{figure}

To compare the performance of different pseudopotential sets used to obtain PDWFs in the presence of SOC, we computed the band distance  $\eta_2$
using both the \texttt{PseudoDojo} and the \texttt{modified-pslibrary} sets described earlier. The cumulative histogram is shown with dashed lines in Fig.~\ref{fig:bd_soc_add_hyd}.
When using \texttt{PseudoDojo} there are 25 cases with band distance $\eta_2$ exceeding 20~meV, and
the median and mean $\eta_2$ are 2.873~meV and 10.136~meV, respectively.
When using the \texttt{modified-pslibrary} set, there are instead only 7 cases with $\eta_2$ larger than 20~meV, with median and mean $\eta_2$ being 1.411~meV and 4.597~meV, respectively.

We note that even though both sets of calculations involve the same set of materials, the results exhibit significant differences. Moreover, we still have few materials with large $\eta_2$ in both sets.
We discuss how to address and solve both these issues (dependence on the pseudopotentials used, and low-quality band interpolation) in the next section, thanks to the inclusion of selected additional hydrogenic projectors.
Nevertheless, we highlight that the quality of the results is already quite good, with the \texttt{modified-pslibrary} set (including SOC) achieving key metrics
(median and mean band distance) comparable to the data from Ref.~\cite{Qiao2023} (PDWF without SOC), and the statistical results of the \texttt{PseudoDojo} still
outperforming those obtained for HT calculations using the SCDM method~\cite{Vitale2020,Qiao2023}.

\subsection{\label{sec:add_hydrogenic}
  Improving Robustness by Adding Hydrogenic Projectors
}

In this section, we discuss the motivation and effectiveness of extending the projection space by
adding external projectors. 
We first observe that for about 3\% of systems (7/200 for mid-throughput calculations and 478/21737 in the case of the HT calculations in Ref.~\cite{Qiao2023}),
there remains a relatively large deviation between the Wannier interpolated bands and the DFT bands (band distance $\eta_2$ greater than 20~meV).
Analogously, our results using the PDWF method extended to SOC also exhibit similar trends, as discussed in the previous section.
Furthermore, the statistical results of band distance depend on the pseudopotential sets: as discussed earlier, when using \texttt{PseudoDojo}, 25/200 ($\sim$ 12.5\%) of the materials have $\eta_2\geq20$~meV,
while only 7/200 ($\sim$ 3.5\%) have $\eta_2\geq 20$~meV when using the \texttt{modified-pslibrary} set.
Since physical quantities should not depend on the pseudopotential choice, and since the implementation of the calculation of certain properties may require specific pseudopotentials (e.g., advanced properties might be implemented only for norm-conserving pseudopotentials), 
it is important to devise algorithms that are largely independent of the underlying pseudopotentials.
To address this, we aim to identify the causes of remaining discrepancies between DFT and Wannier-interpolated bands,
and design an appropriate algorithm that can be easily applied across various pseudopotential sets,
thereby enhancing the reliability and robustness of the PDWF method.

\begin{figure}[tb]
  \includegraphics[width=7.2cm]{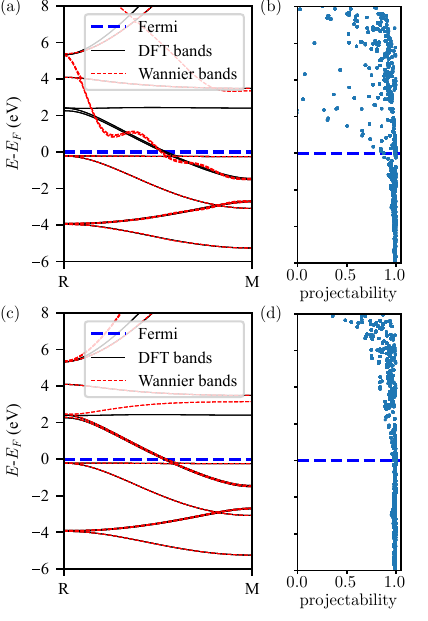}
  \caption{\label{fig:AlCo} \textbf{Effect of the introduction of additional hydrogenic projectors on the band structure and projectability of AlCo.}
  (a) DFT bands (black solid lines) compared with Wannier-interpolated bands (red dashed lines) 
  along the R--M path for AlCo without additional hydrogenic projectors, including only the orbitals from the pseudopotential files in the \texttt{PseudoDojo} library~\cite{VanSetten2018}:
  $3s$, $3p$, $3d$, $4s$ orbitals for cobalt, and
  $3s$, $3p$ orbitals for aluminum.
  (b) Projectability for all $k$-points for the system of panel (a). The projectability starts to decrease rapidly when the energy is larger than the Fermi level.
  (c) DFT bands compared with Wannier-interpolated bands when additional hydrogenic $4p$ orbitals are included for cobalt. (d) Projectability for all $k$-points for the system of panel (c). With the help of hydrogenic AOs, the projectability remains close to one up to a few eV above $E_F$.
  }
\end{figure}

We selected AlCo as an example, for which we computed PDWFs using the \texttt{PseudoDojo} set,
obtaining a fairly large $\eta_2=46.5$~meV. The primary source of the difference between DFT and Wannier bands is along the $\it k$-path from R $\left(1/2, 1/2, 1/2\right)$ to M $\left(1/2, 1/2, 0\right)$, as
shown in Fig.~\ref{fig:AlCo}(a).
Notably, there are significant oscillations in the Wannier-interpolated bands near the Fermi level.
To identify the source of the error, we examine the projectability of the Bloch states onto the trial PAOs within the
first Brillouin zone in Fig.~\ref{fig:AlCo}(b). The data shows that the projectability remains close to 1 below the Fermi level, but
gradually decreases at the Fermi level and above. Notably, already at 2~eV above the Fermi level there are states with essentially zero projectability (the states at the R point).
Tracing the $\it k$-path from R to M, the projectability increases
smoothly from 0, with approximately 0.3 projectability at the $k$-point $\left(1/2, 1/2, 1/3\right)$,
which is the neighborhood of R on the $\it k$-path.
However, since we employ the PD + ED algorithm, where ED freezes all bands below
$E_F$ + 2 eV, the disentanglement process will keep these bands unchanged.
Consequently, the almost-zero projectability at R leads to the exclusion of this band during the disentanglement,
replacing it with a new band constructed from linear combinations of higher-energy bands.
As a result, this forces a discontinuity (in reciprocal space) for this band, that is reflected in large oscillations of the resulting Wannier-interpolated bands.

Considering that we wish to maintain the accuracy of the interpolated bands within
the $E_F$ + 2 eV window, an approach to solve the issue described above for this system is to increase the projectability
of bands near the energy window, that can be achieved only by extending the projection space.
The most intuitive approach is thus to introduce external projectors
to expand the Hilbert space of the projections, thereby enhancing the overall projectability.
This was already considered in Ref~\cite{Qiao2023}, augmenting the number of projectors for silicon to include also $d$ states.

There are several approaches possible to obtain external projectors.
One approach consists in generating additional projectors
using the pseudopotential generation code, and releasing new pseudopotentials.
This approach is naturally the most physically accurate and can ensure maximal
orthogonality between projectors, while providing the most precise projectability.
However, it requires additional manual adaptation for different elements and pseudopotential libraries.
Another approach expands the projectors by referencing other libraries,
such as complementing missing projectors between \texttt{pslibrary} and \texttt{PseudoDojo}, or directly obtaining projectors
from third-party libraries, like extracting the desired projectors from the PAOs obtained from the \texttt{OpenMX} code~\cite{Ozaki2003,Qiao2023}.
However, these often yield projectors defined on different radial coordinates,
requiring separate projection calculations or interpolation to align the projectors 
on a common coordinate system. Moreover, this approach also introduces a higher
complexity, depending on several external libraries. Finally, there is no guarantee that the projectors obtained from different libraries are orthogonal to each other.
The approach that we will adopt in the following relies instead on adding hydrogenic atomic orbitals.
Standard hydrogenic-orbital projectors have been widely used in past applications, and have been
effectively applied in HT calculations~\cite{Gresch2018}.
Furthermore, since a hydrogenic AO is written using analytical expressions, it can be easily evaluated on the radial coordinates
of the original projectors.

\begin{table*}
  \caption{\label{tab:hydrogenic_radial_function}
  \textbf{Analytical expressions of hydrogenic AOs with different number of radial
  nodes ($n_r = n - 1$, where $n$ is the principal quantum number) and
  angular quantum numbers $l$.} Only the expressions for the values of $n_r$ and $l$ that are needed for covering the missing orbitals in the periodic table are reported.
  }
  \begin{tabular}{cccc}
  number of nodes ($n_r$) & 0 & 1 & 2\\ \hline
  $s$ ($l=0$) & $2\alpha^{3/2}\exp{(-\alpha r)}$ & 
  $\frac{1}{2\sqrt{2}}\alpha^{3/2}(2-\alpha r)\exp{(-\alpha r/2)}$&
  ...\\
  $p$ ($l=1$) & $\frac{1}{2\sqrt{6}}\alpha^{3/2}\alpha r\exp{(-\alpha r/2)}$ &
  $\frac{4}{81\sqrt{6}}\alpha^{3/2}(6\alpha r-\alpha^2 r^2)\exp{(-\alpha r/3)}$ &
  ...\\
  $d$ ($l=2$) & $\frac{4}{81\sqrt{30}}\alpha^{3/2}\alpha^2 r^2\exp{(-\alpha r/3)}$ &
  ... & ...\\
  \end{tabular}
\end{table*}

As a supplement to the standard hydrogenic approach, we have extended the radial function expressions
to accommodate different angular quantum numbers (since the radial part in general depends also on the angular quantum number $l$, in addition to the principal quantum number $n$). The various radial functions listed in 
Table~\ref{tab:hydrogenic_radial_function} can cover all the projectors needed for common elements.
The shape of these radial functions is controlled by the parameter $\alpha$. Therefore, in order to define which
hydrogenic AOs to use to expand missing projectors, we only need a table containing the minimal
required orbitals for each chemical element, together with their corresponding $\alpha$ values.
However, compared to projectors generated directly using pseudopotential generation codes,
using hydrogenic AO projectors inevitably introduces larger overlaps between AOs and the pseudopotential PAOs,
which impacts projectability. While intuitively this is not expected to be an issue for the added projectors, since these are are typically needed only to complete the Hilbert space for describing higher-energy bands, we conduct tests to evaluate
the effectiveness of this algorithm.

When considering the minimum required orbitals, we follow the principle of including one additional
higher AO for elements within the same period. For elements in the $n$-th period, if $n=1$, 
only the $1s$ orbital is considered. For $n=2$ or $3$, both $ns$ and $np$ orbitals are added to the
requirements list. For elements in higher periods, alkali and alkaline earth metals require $ns$ and 
$(n-1)d$ orbitals, transition metals need $ns$, $np$ and $(n-1)d$ orbitals, and elements from the boron
group to the noble gases require $ns$ and $np$.
According to these rules, we can list the minimal set of required orbitals for each element, which is also provided explicitly in the Supplementary Table~\ref{sm-tab:required-orbitals}.
Note that some of these orbitals might be redundant (i.e., having almost zero projectability on states below $E_F + 2$~eV). However, with the aim of maximizing the success rate for a fully automated algorithm, we include all of them. An analysis of the projectability of the individual orbitals could be performed as a post-processing step to determine which can be removed, if a minimal Wannier basis set is desired.

We stress that any additional hydrogenic projector should be orthogonal to the existing PAOs from the pseudopotential. In particular, if the pseudopotential PAOs already include an orbital with the same angular quantum number $l$ but smaller principal quantum number $n$, the additional projector should use a radial function that contains a node.
To obtain a table of appropriate $\alpha$ values, we employed two approaches.
For projectors with a node in the radial function,
we adjust $\alpha$ to ensure that the inner (PAOs) and outer (added hydrogenic AO for which we need to determine $\alpha$) projectors are orthogonal.
For projectors without nodes (i.e., when the pseudopotential does not already include an inner-shell projector with the same $l$), instead, we derive the $\alpha$ value
through fitting of our analytical expressions (Table~\ref{tab:hydrogenic_radial_function}) to the PAOs from the \texttt{OpenMX} code~\cite{Ozaki2003}. 
Because of the orthogonality condition to the underlying PAOs, the values of $\alpha$ will depend on the chosen pseudopotential library. Values of $\alpha$ for the pseudopotential libraries used in this work can be found in Supplementary Tables~\ref{sm-tab:alfa_sssp}, \ref{sm-tab:alfa_sssp_allhyd}, \ref{sm-tab:alfa_dojofr} and \ref{sm-tab:alfa_mixed}. 

Despite these precautions, the added orbitals are in general not orthogonal to the PAOs. Therefore, we apply a Gram--Schmidt orthonormalization~\cite{Leon2013} when external projectors are added.
More precisely, we first perform separately two L\"{o}wdin orthonormalizations for the PAOs from the pseudopotentials and
for the external hydrogenic projectors. Then, we fix the PAOs projectors and perform a further
Gram--Schmidt orthonormalization step on the hydrogenic orbitals only, ensuring that the full set (PAOs + additional hydrogenic orbitals) form an orthonormal set.
This procedure allows us to faithfully keep the PAOs from pseudopotentials unchanged, while at
the same time making sure that external projectors are always orthonormal to pseudopotential PAOs. Instead, a single L\"owdin orthogonalization of all orbitals would distort the PAOs, an undesirable effect since they are accurately describing the system chemistry.
For a detailed description of the orthonormalization procedure, see Supplementary Section~\ref{sm-sec:real_implement}.

The algorithms described above have been implemented and
will become available in the next releases of QE~\cite{Giannozzi2009,Giannozzi2020} (\texttt{pw2wannier90.x} code).
In addition, we can use the scripts in the AiiDA-Wannier90-Workflows repository~\cite{aww} (folder \texttt{dev/projectors})
to extract PAO information from the pseudopotentials and determine the minimal required additional projectors and the corresponding $\alpha$ values, according to the method described above. The script then generates the missing projectors and exports them as a \texttt{.dat} file,
which serves as an additional input for \texttt{pw2wannier90.x}.

Going back to the example of Fig.~\ref{fig:AlCo}, comparing the PAOs in the \texttt{PseudoDojo} with the minimum required AOs, we hypothesized that adding
an additional $4p$ AO for cobalt could enhance the overall projectability. The computational results
support this hypothesis: after introducing a $4p$ external projector, the minimum projectability near
$E_F$ + 2 eV increased to approximately 0.8, see Fig.~\ref{fig:AlCo}(d), which is sufficient to maintain continuity between adjacent
$\it k$-points using the PD+ED disentanglement approach. The spread of the cobalt WFs decreased from a range of
0.73--2.46 $\text{\r{A}}^2$ to 0.46--1.14$\text{\r{A}}^2$, confirming an increased smoothness of the wavefunctions in reciprocal space, thus resulting in more localized WFs and in highly accurate Wannier-interpolated bands (up to 2~eV above $E_F$), with $\eta_2 = 2.34$~meV, see Fig.~\ref{fig:AlCo}(c).

The computational results for the 200 materials with SOC, before and after introducing the external projectors,
are shown in Fig.~\ref{fig:bd_soc_add_hyd}. The statistical data shows significant improvements.
After adding external hydrogenic projectors to PAOs from \texttt{PseudoDojo} (\texttt{modified-pslibrary}),
the mean $\eta_2$ decreases from 10.136 (4.597)~meV to 2.179 (2.013)~meV, and the median $\eta_2$
drops from 2.873 (1.411)~meV to 1.073 (1.081)~meV.
Notably, the computational success rate increases from 87.5\% (96.5\%) to 100\%
when the \texttt{PseudoDojo} (\texttt{modified-pslibrary}) is used, with all systems having $\eta_2\le 20$~meV.
Therefore, by appropriately expanding the projection space, we both improve the robustness of the PDWF approach and strongly mitigate the pseudopotential dependence of the results.

\subsection{\label{sec:spin_unpol} Extended validation in spin-unpolarized systems}

In addition to the results on SOC systems mentioned above, we also conducted tests of our extended PDWF method with added hydrogenic projectors on non-SOC systems.
As shown in Fig.~\ref{fig:bd_compare}(e), the introduction of external projectors generally reduces the $\eta_2$ values,
with the mean $\eta_2$ decreasing from 4.231~meV to 2.002~meV and the median $\eta_2$
dropping from 1.597~meV to 1.153~meV with respect to the values of Ref.~\cite{Qiao2023}. Furthermore, also in this case we achieve a 100\% success rate, with $\eta_2\le 15$~meV for all materials, and 
the max distance $\eta_2^{max}$ also showing a significant reduction.
\begin{figure}[tb]
  \includegraphics[width=8.0cm]{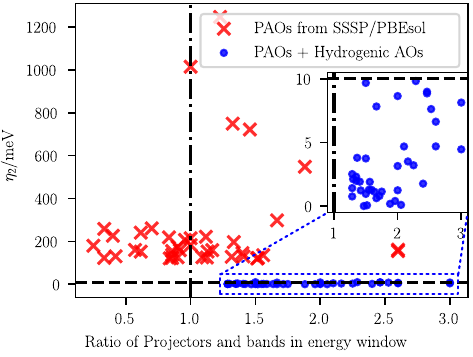}
  \caption{\label{fig:40_largest_eta} \textbf{Results of recalculating the 40 systems with largest band distance from 
  Ref.~\cite{Qiao2023}.} The red crosses are the original data directly from Ref.~\cite{Qiao2023}, whose projectors are
  PAOs from pseudopotentials in SSSP PBEsol Efficiency v1.1. The blue circles are the
  recalculated data where the same PAOs are complemented by additional hydrogenic AOs.
  The blue dashed rectangle area (which includes all blue circles) is zoomed in the inset, showing that after adding hydrogenic AOs, all 40 materials have a $\eta_2\le 10$~meV, thus demonstrating the effectiveness of our algorithm even in extremely challenging cases.
}
\end{figure}

We further test our extended algorithm on the 40 materials from Ref.~\cite{Qiao2023} that exhibit the largest $\eta_2$
among its 21,737 HT results. Each point in Fig.~\ref{fig:40_largest_eta} represents a structure,
where the $x$ coordinate is the ratio between the number of projectors (i.e., of Wannier functions) and the number of bands within the
energy window ($E_F$ + 2 eV), while the $y$ coordinate is
$\eta_2$. Since each projector can correspond to only one band state, and considering
that disentanglement involves a linear combination of several bands, when the number
of projectors is less than (or just slightly more than) the number of bands within the energy window, the disentanglement process does not have enough information from the initial projections, possibly resulting in a poor Wannier interpolation.
These results clearly indicate that the inaccuracy of the Wannier interpolated bands for many of 
these materials was due to the lack of a sufficient number of projectors. After adding external projectors (and enabling the guiding center setting during the
Wannierization process),
the $\eta_2$ for all systems was reduced to below 10~meV (blue circles in Fig.~\ref{fig:40_largest_eta}).
This demonstrates that the additional hydrogenic projectors can significantly enhance the robustness of the Wannier
interpolation also in non-SOC systems, even in extremely challenging cases.

To further illustrate how the various components of the extended PDWF method jointly contribute to the overall final robustness of the algorithm, we compare in Fig.~\ref{fig:bd_compare} the results obtained on the 200-structure set with different projection methods.
SOC is not included in these calculations, and the SSSP PBE Efficiency v1.1 pseudopotential library is used, to enable direct comparison with the results of Ref.~\cite{Qiao2023}.
Panel (a) of Fig.~\ref{fig:bd_compare} shows the results of the standard Souza--Marzari--Vanderbilt ED algorithm, using as starting projections the common analytical hydrogenic AOs (as, e.g., defined in the Wannier90 code~\cite{Mostofi2008,Pizzi2020}). The pseudopotential PAOs are used in this case solely to determine the angular character ($s$, $p$, $d$, ...) of the projectors to consider, but no further information is extracted from the pseudopotential projectors. Furthermore, the default values for $\alpha$ and for the projector analytical shapes are employed, as defined in the Wannier90 code. While the approach, that has been widely used in the literature, is able to provide a good Wannier interpolation for over 50\% of the systems, it still exhibits a significant number of failures, with a mean $\eta_2$ of 64.368~meV and a median $\eta_2$ of 7.561~meV. 

As discussed earlier, however, the
radial part of the hydrogenic orbitals should depend also on the angular quantum number $l$, meaning that different expressions must be applied for radial functions with $s$, $p$, or $d$ angular character (see Table~\ref{tab:hydrogenic_radial_function}).
Using these corrected radial functions as projectors, combined with optimized values of $\alpha$  (see Supplementary Table~\ref{sm-tab:alfa_sssp_allhyd}) obtained by fitting the corresponding PAOs radial functions (and complementing with fitting from OpenMX where PAOs are missing), results in panel (b) of Fig.~\ref{fig:bd_compare}.  The success rate significantly increases, and the mean (median) $\eta_2$ decreases to 27.478 (5.061)~meV, demonstrating that an improved choice of radial functions and $\alpha$ values already enhances the quality of the Wannier interpolation.

\begin{figure*}[tb]
  \includegraphics[width=16.5cm]{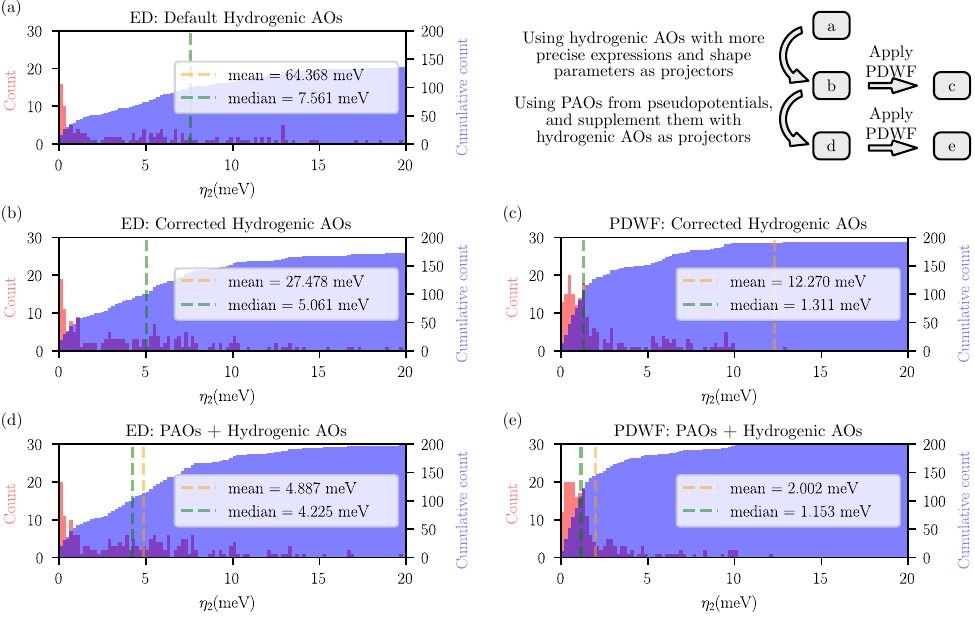}
  \caption{\label{fig:bd_compare}
  \textbf{Summary of the effect of several ingredients of the extended PDWF algorithm on the quality of Wannier interpolation.}
  Histogram (red) and cumulative histogram (blue) of band distance $\eta_2$ for our test set of
  200 structures,
  using different algorithms and projectors (all without SOC).
  All sets of calculations use the same number of projectors and same configurations of semi-core states.
  All frozen windows were set to $E_F + 2$~eV.
  (a) ED with the default all-hydrogenic AOs;
  (b) ED with the corrected all-hydrogenic AOs (see Table~\ref{tab:hydrogenic_radial_function});
  (c) PDWF with the corrected all-hydrogenic AOs;
  (d) ED with PAOs and external hydrogenic AOs;
  and (e) PDWF with PAOs and external hydrogenic AOs.
  The orange (green) vertical lines are the mean (median) band distance $\eta_2$,
  whose values are shown in the legend of each panel.
  The difference and the relation between each set of data are shown in the top right panel. Additional comparisons with further combinations of disentanglement method and starting projectors can be found in Supplementary Section~\ref{sm-sec:additional_compare}.
  }
\end{figure*}

Starting from panel (b), two possible directions can be taken to further improve the quality of the Wannier interpolation. The first is to use PDWFs instead of the ED algorithm, see panel (c) of Fig.~\ref{fig:bd_compare}. The mean (median) $\eta_2$ further reduces to 12.270 (1.311)~meV, demonstrating that the PDWF algorithm can provide a more accurate Wannier interpolation than the standard ED algorithm, even when using the same projectors. 
(As a note, the large mean $\eta_2$ values in panels (a-c) are due to few cases with very large $\eta_2$ that are not visible in the histogram as they are outside of the $x$-axis range.)
Alternatively, one can add external hydrogenic AOs to the PAOs, as shown in panel (d) of Fig.~\ref{fig:bd_compare}. The mean (median) $\eta_2$ also reduces, with respect to panel (b), to 4.887 (4.225)~meV, demonstrating that the addition of external hydrogenic AOs also plays an important role in improving the Wannierization robustness.

Finally, panel (e) of Fig.~\ref{fig:bd_compare} represents our improved algorithm, combining PDWF and the addition of hydrogenic AOs. 
The resulting Wannier interpolated bands exhibit the highest quality, with the lowest values for the mean (median) $\eta_2$ of 2.002 (1.153)~meV, and only such a combination achieves a 100\% success rate, with all systems having $\eta_2\le15$~meV. 

These results demonstrate that, although hydrogenic AOs are not as physically precise as the projectors from the pseudopotentials,
they are still essential as a complement to PAOs when these are not sufficient to cover all states up to $E_F + 2$~eV.
The additional advantage is that the present combined method essentially eliminates the dependence of PDWFs on pseudopotentials.
In fact, while different pseudopotentials may have different sets of PAOs, discrepancies are minimized by introducing hydrogenic functions to complete the projectors to the same list of orbitals (for a given chemical element), see SI Table~\ref{sm-tab:required-orbitals}.
We also stress that good interpolation quality can only be achieved if enough PAOs are already included into the pseudopotential (as these capture the detailed chemistry of the material close to the atoms), and we need to possibly add only a few more hydrogenic AOs to recover a large enough projectability for the high-energy bands in the relevant energy range.
Indeed, if no PAOs are available at all in the pseudopotentials, only considering hydrogenic AOs is equivalent to panel (b) of Fig.~\ref{fig:bd_compare}, showing that a much lower interpolation quality would be achieved.

The present method thus offers a flexible and straightforward approach to generate the needed additional projectors without requiring to execute and possibly modify the pseudopotential generation codes, making it a practical algorithm for enhancing the performance of PDWFs for high-throughput research.

\subsection{\label{sec:mag} Magnetization}
We finally discuss the extension of the PDWF workflows to magnetic systems.
In magnetic DFT calculations, one typically distinguishes between a collinear and non-collinear magnetic treatment. 
The former indicates that the spin can only be polarized along a given quantization axis (e.g., $z$), which is often appropriate to describe ferromagnetic and collinear antiferromagnetic materials, but can fail to accurately describe the magnetic structure of more complex systems where magnetic moments are not simply parallel or antiparallel,
such as non-collinear antiferromagnetic systems or spin spiral states.
In a collinear treatment, spinor wavefunctions have one of the two components being identically zero.
In QE, the code therefore considers collinear calculations by only storing the non-zero component of the spinor wavefunctions, effectively treating each of the two spin channels as spin-unpolarized calculations, but with a doubled set of $k-$points (one set for spin up, one for spin down). (Note that this is possible only in the absence of SOC, so that the Hamiltonian does not mix the two spin channels). Consequently, in our workflow design, we choose to separate the spin-up and spin-down calculations,
and later merge the band structures after the calculations are completed.

Instead, since formally the treatment of non-collinear spin systems is the same as that of SOC systems, because also in this case the wavefunctions are two-component spinors, the corresponding workflow structure is analogous to that of SOC calculations,
with the only additional requirement being a tool to incorporate the magnetic moment as an input.
Within the AiiDA-Wannier90-Workflows repository~\cite{aww} we have developed a new \texttt{MagneticStructureData} data plugin for AiiDA (available in the
\texttt{aiida\_wannier90\_workflows.data.structure} Python package) that processes magnetic moment structures and organizes input files.
In the following two subsections we briefly discuss the results of the verification of our workflows for collinear and non-collinear magnetic systems.

\subsubsection{\label{sec:col_mag} Collinear magnetic systems}

We performed calculations on 16 relevant magnetic systems, including collinear ferromagnetic body-centered cubic (BCC) iron,
collinear antiferromagnetic $\rm{Mn_2F_4}$, and non-collinear antiferromagnetic $L1_2$ $\rm{IrMn_3}$,
as well as other magnetic systems primarily composed of Fe, Co, Ni, and Mn elements.

\begin{figure}[tb]
  \includegraphics[width=8cm]{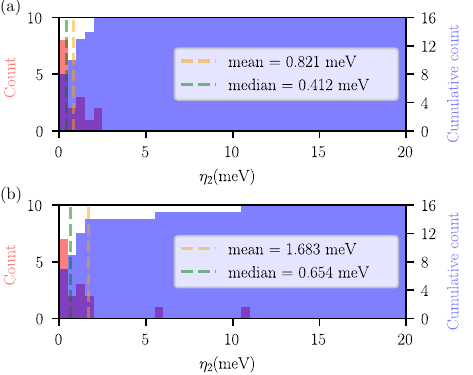}
  \caption{\label{fig:bd_cnc}
  \textbf{Results of the extended PDWF workflows for magnetic (collinear and non-collinear) systems.} Histogram (red) and cumulative histogram (blue) of the 
  band distance $\eta_2$ of (a) 16 collinear magnetic systems
  and (b) 16 non-collinear magnetic systems.
  The orange (green) vertical lines are the mean (median) band distance $\eta_2$,
  whose values are shown in the legend of each panel.
  In both cases, SSSP pseudopotentials and additional external hydrogenic projectors were used to obtain PDWFs.
  }
\end{figure}

In this set of calculations, the initial magnetic moments in DFT calculations are chosen to 
have the same magnitude as the $z$ component of the actual 3D magnetic structure.
Without applying strain, we allowed the DFT code
to converge to the ground-state magnetic moment structure within the framework of a collinear ferromagnetic system.
Our extended PDWF approach was used to accurately obtain the Wannier functions, incorporating external hydrogenic atomic orbitals into
the projectors to enhance robustness, as detailed earlier.
The calculated band distances within the energy window of $E_F+ 2$~eV are shown in Fig.~\ref{fig:bd_cnc}(a). Results on these systems are excellent, with very low mean and median $\eta_2$ (both below 1~meV).

\subsubsection{\label{sec:nocol_mag} Non-collinear magnetic systems}

Given the complexity of magnetic structures,
also in this case we selected only 16 magnetic systems, including both collinear and non-collinear configurations,
as well as ferromagnetic and antiferromagnetic systems.
The detailed list of materials, with their magnetic configurations, is provided in Supplementary Section~\ref{sm-sec:mag_struct}.
Using the PDWF algorithm and external hydrogenic AOs,
the calculated band distances within the energy window of $E_F+ 2$~eV  are shown in Fig.~\ref{fig:bd_cnc}(b).
Also in this case, the MLWFs generated by our extended PDWF method demonstrate an excellent performance in accurately describing electronic structures, with both mean and median $\eta_2$ below 2~meV, thus showcasing the effectiveness of our extended PDWF algorithm in treating magnetic systems.

\subsection{\label{sec:discussion} Discussion}
We present the extension of the PDWF algorithm to collinear and non-collinear magnetic systems, and to systems including SOC. 
Moreover, we further improve the robustness of the algorithm by defining an automated strategy for incorporating additional external (hydrogenic) projectors when the pseudopotentials do not include sufficient projectors to describe the lowest-energy unoccupied states.
Our extended algorithms are fully automated using the AiiDA workflow engine~\cite{Pizzi2016,Huber2020,Uhrin2021} and we make our implementation available open-source in the AiiDA-Wannier90-Workflows package~\cite{aww}.

To benchmark the algorithms, we calculate PDWFs for 
multiple systems (spin-unpolarized, magnetic, and SOC systems) and 
demonstrate that the present extended PDWF algorithm can reliably and automatically construct MLWF-based tight-binding models
for both magnetic and SOC systems. 
We find that essentially all deviations between Wannier-interpolated bands and DFT bands in earlier (spin-unpolarized) PDWF results~\cite{Qiao2023} can be attributed to the absence of all desirable projectors in the pseudopotentials.
This can lead to poor projectability of important low-energy empty states,
resulting in less localization of the corresponding Wannier functions and, in turn, in poorer interpolation of the electronic bands.
The addition of external hydrogenic projectors can improve significantly the accuracy of PDWFs and minimize the dependence of Wannier interpolation results on the specific set of PAOs defined in the chosen pseudopotentials, achieving
in our test set a 100\% success rate (defined as an average band distance $\eta_2\le20$~meV; in all systems in our test sets, the band distance remains below 15~meV)
when interpolating all bands up to the Fermi level + 2 eV.

Finally, we compare the results of the Wannierization procedure obtained with various types of projectors (only hydrogenic AOs, PAOs, PAOs + external hydrogenic AOs) and we elucidate the key contributions in the extended PDWF algorithm to the overall quality of the Wannier interpolation.
The present results underscore how PDWFs can be obtained in a systematic, robust and automated approach, enabling their straightforward and efficient application to several applications ranging from advanced property calculations~\cite{Marrazzo_2024,Liu2024} to high-throughput materials discovery projects~\cite{Bercx2025}.

\section{\label{sec:method} Methods}
\subsection{\label{sec:code} Code implementation}
We implemented projections with PAOs and hydrogenic AOs in the \texttt{pw2wannier90.x} executable, part of the
\texttt{Q\textsc{uantum} ESPRESSO} (QE)~\cite{Giannozzi2009,Giannozzi2020} package.
It can read user-provided custom projector files and calculate the projections of plane wave functions
onto the projectors. The implementation supports SOC systems, non-collinear and
collinear magnetic systems, as well as spin-unpolarized systems.
The projector files can be generated using a script in the AiiDA-Wannier90-Worflows package~\cite{aww} based on the rules defined in Sec.~\ref{sec:add_hydrogenic}.
Moreover, we implemented a strategy to perform orthogonalization of the projectors by first applying L\"{o}wdin orthogonalization~\cite{Lowdin1950} separately
to the PAOs and to the hydrogenic AOs, and then fixing the PAOs (ensuring that they remain intact) and performing a subsequent Gram--Schmidt orthogonalization~\cite{Leon2013} of
the entire set of projectors, as discussed in the main text.

\subsection{\label{sec:dft_calc} DFT calculations}
The DFT calculations are carried out using QE, with various pseudopotential libraries for different systems, as discussed in the main text. Specifically, and unless stated otherwise, for spin-unpolarized systems without SOC we use SSSP PBE Efficiency v1.1~\cite{Prandini2018}.
For SOC systems, we use the \texttt{PseudoDojo} PBE v1.4 (norm-conserving, fully relativistic) library~\cite{VanSetten2018} and the
pslibrary~\cite{DalCorso2014,pslib_suggest} (PAW, fully relativistic) for comparison and validation.
The HT calculations are managed with the AiiDA infrastructure~\cite{Pizzi2016,Huber2020,Uhrin2021}, which submits QE and \texttt{Wannier90}~\cite{Pizzi2020} calculations to remote clusters, parses
and stores the results into a database, while also orchestrating all sequences of simulations and workflows.
The automated AiiDA workflows are open-source and hosted on GitHub~\cite{aww}.
To attach magnetic information to the crystal structure, we extend the AiiDA \texttt{StructureData} plugin to a new \texttt{MagneticStructureData} class, defined as part of the 
AiiDA-Wannier90-Workflows package. In the future, we plan to replace this custom class with the new \texttt{StructureData} class defined in the
AiiDA-Atomistic~\cite{aatom} plugin.

\section*{\label{sec:acknowledgement} Acknowledgements}
This research was supported by the NCCR MARVEL, a National Centre of Competence in Research, funded by the Swiss National Science Foundation (grant number 205602).
YJ acknowledge support by the China Scholarship Council program.
JQ acknowledges support by the HORIZON-RIA 2D-PRINTABLE (proposal number: 101135196), and 
this work has received funding from the Swiss State Secretariat for Education, Research and Innovation (SERI).
NP and GP acknowledge support by the Swiss National Science Foundation (SNSF) Project Funding (grant 200021E\_206190 ``FISH4DIET'').
WZ acknowledge support by the National Key Research and Development Program of China (Grant No. 2022YFB4400200), National Natural Science Foundation of China (Grant Nos. T2394474, T2394470),
the Beijing Outstanding Young Scientist Program and Tencent Foundation through the XPLORER PRIZE.
We acknowledge access to Piz Daint or Alps at the Swiss National Supercomputing Centre, Switzerland under MARVEL's share with the project ID mr32.
We acknowledge fruitful discussions with Edward Baxter Linscott and Miki Bonacci.

\section*{Data availability}
All data generated in this work, as well as scripts to generate relevant plots, are available on the Materials Cloud Archive~\cite{Talirz2020} at \url{https://doi.org/10.24435/materialscloud:9g-ds}~\cite{MCA}. This entry also includes AiiDA~\cite{Huber2020} archive files with the full provenance of all simulations and data.

\clearpage
~
\onecolumngrid


\section*{Supplementary material}

# Supplementary Material for: Robust Wannierization including magnetization and spin-orbit coupling via projectability disentanglement

Yuhao Jiang, Junfeng Qiao, Nataliya Paulish, Weisheng Zhao, Nicola Marzari, and Giovanni Pizzi

## CONTENTS

# I. SOC-WEIGHTED BAND DISTANCE

When studying SOC systems, it is expected that the Wannier functions can accurately reproduce the plane wave functions as closely as possible within the SOC-strongly-influenced $k$-space. In terms of band distance, the accuracy near the SOC-induced band splitting should be maintained at a level comparable to other regions. As shown in an artificial schematics in Supplementary Fig. 1(a), an oscillating interpolated band may have a similar average band distance to that of an interpolated band with a small rigid shift of all bands. However, it is evident that the oscillating energy band corresponds to Wannier functions that are insufficient to accurately describe the system, because of the mismatch of the band splitting and of the band derivatives (which are an important ingredient, e.g., of transport calculations). Therefore, we designed a SOC-weighted band distance metric to emphasize the accuracy of the Wannier-interpolated bands near the SOC-induced splitting.

Similar to the definition of the band distance, we define the SOC-weighted band distance as

$$\eta_\nu^{SOC} = \sqrt{\frac{\sum_{n\mathbf{k}} \tilde{f}_{n\mathbf{k}}^{SOC} \left(\epsilon_{n\mathbf{k}}^{DFT} - \epsilon_{n\mathbf{k}}^{Wan}\right)^2}{\sum_{n\mathbf{k}} \tilde{f}_{n\mathbf{k}}^{SOC}}}, \qquad (1)$$

where $\tilde{f}_{n\mathbf{k}}^{SOC} = \sqrt{\tilde{f}_{n\mathbf{k}} |\tanh\left(\left(\epsilon_{n\mathbf{k}}^{SOC} - \epsilon_{n\mathbf{k}}^{noSOC}\right)/\sigma^{SOC}\right)|}$, $\tilde{f}_{n\mathbf{k}}$ being the effective Fermi-Dirac distribution. The tanh function is used to provide a weight to each band and $k-$point contribution, dependent on the splitting between the corresponding bands with and without SOC. We selected a $\sigma$ value of 50 meV, so that for instance a specific state with SOC splitting of 100 meV corresponds to a weight of 0.964, close to 1. In contrast, states without splitting or with very small splitting will have almost no weight.

In Supplementary Fig. 1(d), we can observe that the SOC-weighted band distance and the band distance are similar, proving that with our method the interpolation errors when SOC is included are not originating only SOC-specific errors, i.e., from regions where SOC has a large effect, but they are interpolation errors distributed across the BZ. The "standard" band distance can thus be safely used to describe the discrepancy between the Wannier-interpolated bands and the DFT bands also when SOC is included.

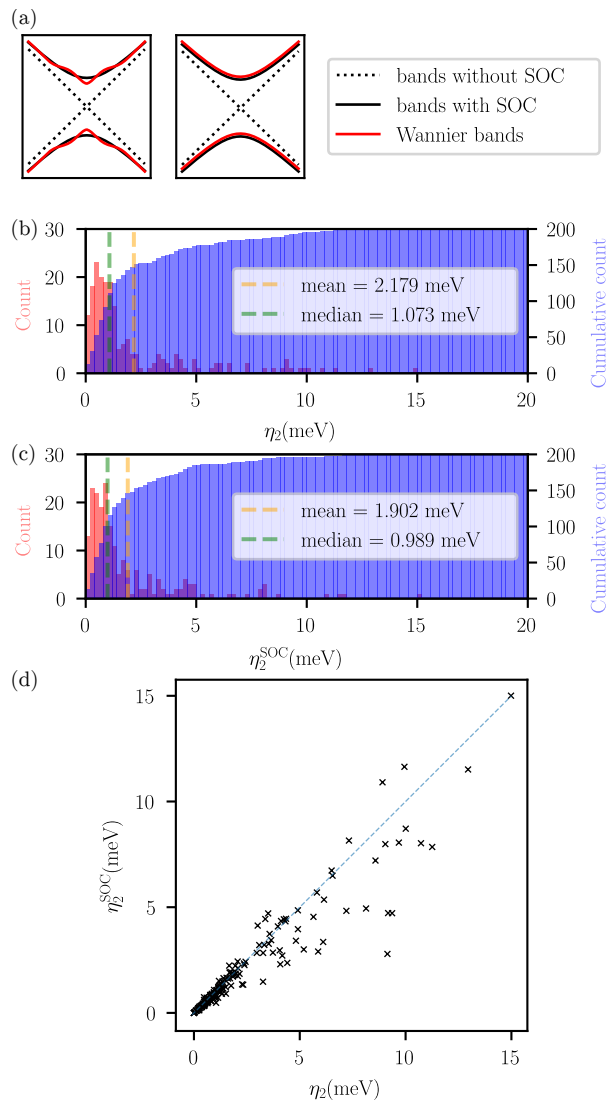

SUPPLEMENTARY FIG. 1. (a). Two sketches of Wannier interpolation of bands near a SOC-induced splitting. Note that the two sketches are only schematic representations and do not correspond to the band structure of any real system. While the average band distance in the two sketches is approximately similar, the left figure exhibits significant oscillations near the SOC splitting, which is detrimental for accurately predicting properties dominated by SOC. Histogram (red) and cumulative histogram (blue) of (b) the band distance $\eta_2$ and (c) the SOC-weighted band distance $\eta_2^{SOC}$. The orange (green) vertical line is the mean (median) band distance, whose value(s) are shown in the legend of each panel. `PseudoDojo` 0.4 and external hydrogenic projectors were used to perform a PDWF Wannierization. (d) Scatter plot of $\eta_2$ vs. $\eta_2^{SOC}$, showing that regardless of whether SOC-weighting is applied, the two band distances are approximately the same.

## II. BAND DISTANCE AFTER INTRODUCING EXTERNAL PROJECTORS

In Sec. II B of the main paper, we show that adding hydrogenic AOs can improve the performance of Wannier interpolation, especially in increasing the success rate in Supplementary Fig. 2. However, we observed that in regions with already very small band distance, the band distance might slightly increase after adding hydrogenic AOs. This effect is more pronounced when using `PseudoDojo`. To illustrate this, we have plotted the band distance $\eta_2$ before and after adding hydrogenic AOs in Supplementary Fig. 2. As a result, in the low-$\eta_2$ region, there are instances where the external projector actually increases the band distance (see red crosses in Supplementary Fig. 2). However, since even despite this small increase, the band distance in these systems remains very small, the resulting Wannier interpolation is still of high quality. We thus recommend to introduce by default the addition of external hydrogenic projectors, with the aim of improving the overall quality of Wannier interpolation for any input system in a general way. For systems negatively affected by this, a future workflow could incorporate a final *exclude projector* process to further optimize the results.

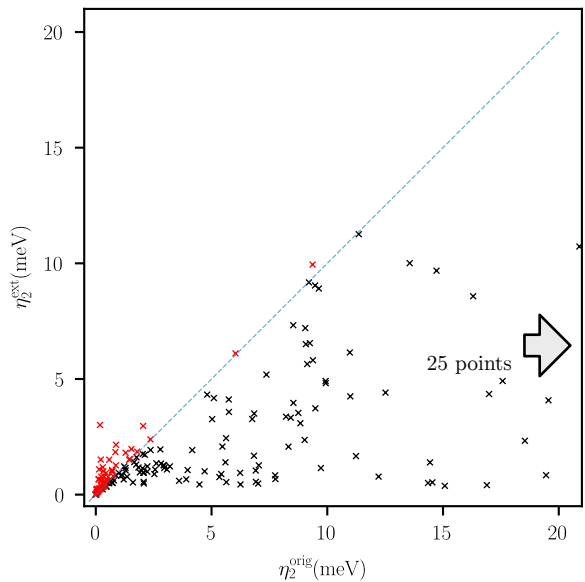

SUPPLEMENTARY FIG. 2. Scatter plot of the original band distance ($\eta_2^{\text{orig}}$) vs. the band distance after adding external projectors ($\eta_2^{\text{ext}}$) for materials calculated with `PseudoDojo`. The red crosses indicate systems for which the band distance increases after introducing external projectors (i.e., $\eta_2^{\text{ext}} > \eta_2^{\text{orig}}$). We note that for most of these points $\eta_2^{\text{ext}}$ remains $< 4$ meV, and in general the addition of external projectors has a minimal impact on the overall robustness. 25 structures with large $\eta_2^{\text{orig}}$ but low $\eta_2^{\text{ext}}$ ($< 10$ meV) are not shown in the figure.

## III. ADDITIONAL COMPARISONS BETWEEN APPROACHES FOR DISENTANGLEMENT AND INITIAL PROJECTIONS

We present in Supplementary Fig. 3 the comparison between PAO projectors only and PAO with external hydrogenic AO projectors. At variance with Fig. 5 in the main text, the number of projectors is not the same in all panels. The results of Supplementary Fig. 3 shows that, whether using ED or PDWF, the introduction of external hydrogenic AOs can significantly enhance the robustness and accuracy of the Wannier interpolation.

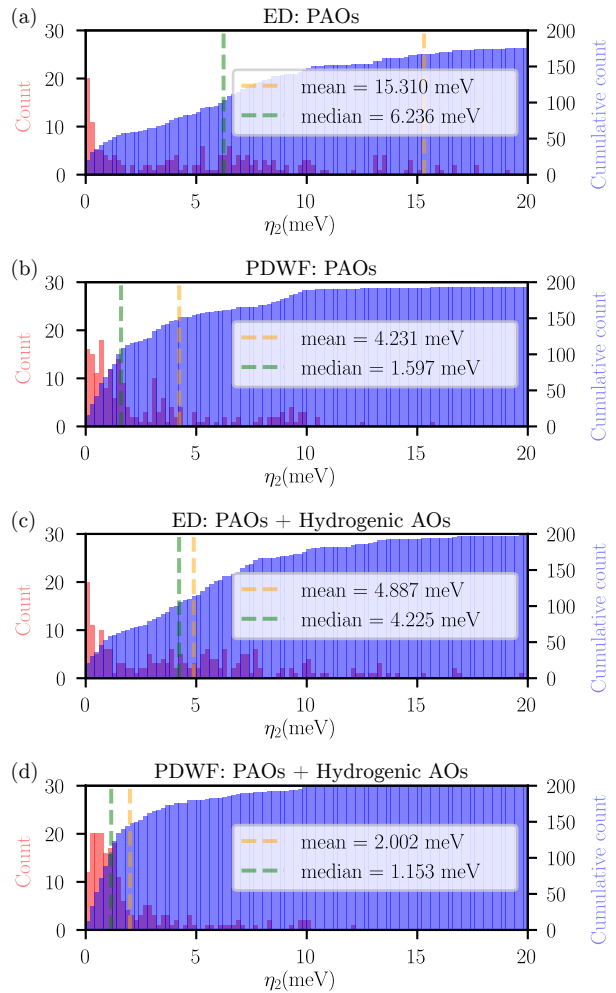

SUPPLEMENTARY FIG. 3. Histogram (red) and cumulative histogram (blue) of the band distance $\eta_2$ for the 200 structures used in Ref. [1] using different algorithms and projectors (without SOC). All frozen windows were set to $E_F + 2$ eV. (a) ED with PAOs; (b) PDWF with PAOs (same as Ref. [1]); (c) ED with PAOs and external hydrogenic AOs; and (d) PDWF with PAOs and external hydrogenic AOs. The orange (green) vertical line is the mean (median) band distance $\eta_2$, whose value(s) are shown in the legend of each panel.

## IV. ORTHONORMALIZATION STRATEGY

As discussed in the main text, using a Löwdin orthonormalization on the complete set of the projectors derived from the PAOs in the pseudopotential and from the external hydrogenic AOs is detrimental, as the (chemically accurate) PAOs in the pseudopotential get mixed and thus distorted in order to obtain orthogonality.

Instead, we recommend to apply Löwdin orthonormalization separately to the projectors derived from the PAOs in the pseudopotential and those from the external hydrogenic AOs. Then, we perform Gram–Schmidt orthonormalization on the hydrogenic AOs projectors only, using the PAO projectors as the reference basis. In our implementation, we perform the two Löwdin orthonormalization steps in a single matrix operation, and adopt the modified Gram–Schmidt algorithm for improved numerical precision. This chapter provides a brief overview of the underlying implementation details.

### A. Diagonalization of a block matrix

#### 1. Lemma

The eigen decomposition of a block diagonal matrix $A = \begin{bmatrix} A_1 & 0 \\ 0 & A_2 \end{bmatrix} = PDP^{-1}$ can always be written in the form of $P = \begin{bmatrix} P_1 & 0 \\ 0 & P_2 \end{bmatrix}$ and $D = \begin{bmatrix} D_1 \\ D_2 \end{bmatrix}$, where $D_1$ ($D_2$) and $P_1$ ($P_2$) are the eigenvalues and eigenvectors of $A_1$ ($A_2$), respectively, i.e., $A_1 = P_1 D_1 P_1^{-1}$.

#### 2. Proof

Suppose $A$, $A_1$, and $A_2$ have sizes $n \times n$, $p \times p$, and $q \times q$, respectively, and $p + q = n$. For any eigenvector $v$ of $A$, $Av = \lambda v$, where $\lambda \in \mathbb{Z}$ and $v \in \mathbb{Z}_n$. If we partition the $v$ into blocks, $v = \begin{bmatrix} v_1 \\ v_2 \end{bmatrix}$ where $v_1 \in \mathbb{Z}_p$ and $v_2 \in \mathbb{Z}_q$, we have

$$\begin{bmatrix} A_1 & 0 \\ 0 & A_2 \end{bmatrix} \begin{bmatrix} v_1 \\ v_2 \end{bmatrix} = \lambda \begin{bmatrix} v_1 \\ v_2 \end{bmatrix} \tag{2}$$

thus

$$\begin{bmatrix} A_1 v_1 \\ A_2 v_2 \end{bmatrix} = \lambda \begin{bmatrix} v_1 \\ v_2 \end{bmatrix}. \tag{3}$$

Therefore, $v_1$ and $v_2$ are respectively the eigenvectors of $A_1$ and $A_2$. Now, suppose $v$ is the eigenvector of $A$ and has normal length $||v|| = 1$, and the $v_1$ of $v$ is already an eigenvector of $A_1$ and has normal length $||v_1|| = 1$, then $v_2$ of $v$ must be 0, since $||v|| = ||v_1|| + ||v_2|| \Rightarrow ||v_2|| = 0$. Therefore, $A$ has in total $p$ eigenvectors $\begin{bmatrix} v_1 \\ 0 \end{bmatrix}$ coming from

$A_1$, and $q$ eigenvectors $\begin{bmatrix} 0 \\ v_2 \end{bmatrix}$ coming from $A_2$, that are linearly independent between each other. Correspondingly, $D$ has $p$ eigenvalues from $A_1$, and $q$ eigenvalues from $A_2$. The columns of $P$ are these eigenvectors, and one can always construct a permutation matrix $R$ such that $A = (PR)(R^{-1}DR)(R^{-1}P^{-1})$ and $PR$ has the block diagonal structure. Replacing $P$ by $PR$, one can always reach the form of $P = \begin{bmatrix} P_1 & 0 \\ 0 & P_2 \end{bmatrix}$ and $D = \begin{bmatrix} D_1 \\ D_2 \end{bmatrix}$, where the columns of $P_1$ are the eigenvectors $v_1$ of $A_1$, the $D_1$ are the eigenvalues of $A_1$, and similarly for $P_2$ and $D_2$.

### B. Löwdin orthonormalization

Suppose we have a complex matrix $A \in \mathbb{Z}_{m \times n}$, the Löwdin orthonormalization converts the columns of $A$ into a set of orthonormal vectors that are as close as possible to the original columns. The columns of $A$ should be linearly independent, thus requiring $m \geq n$.

This is achieved in several steps:

1. compute the overlap matrix $O = A^\dagger A$, which is Hermitian by construction;

2. perform an eigen-decomposition of the overlap matrix, $O = PDP^{-1}$; the eigenvalues $D = \text{diag}(\lambda_1, \ldots, \lambda_n)$ are real;

3. take the inverse square root of $D$, and create a new diagonal matrix $S = \text{diag}(\frac{1}{\sqrt{\lambda_1}}, \ldots, \frac{1}{\sqrt{\lambda_n}})$, and a new matrix $U = PSP^{-1}$;

4. the Löwdin orthonormalized matrix is $B = AU$. $B$ is semi-unitary since $B^\dagger B = (PS^\dagger P^\dagger A^\dagger)(APSP^{-1}) = PS^\dagger P^\dagger OPSP^{-1} = PS^\dagger DSP^{-1} = PP^{-1} = 1_{n \times n}$.

#### 1. Lemma

A Löwdin orthonormalization of a block diagonal matrix is equivalent to two separate Löwdin orthonormalizations of the submatrices.

#### 2. Proof

Suppose we have a block diagonal matrix $A = \begin{bmatrix} A_1 & 0 \\ 0 & A_2 \end{bmatrix}$, then the overlap matrix is $O = \begin{bmatrix} A_1^\dagger A_1 & 0 \\ 0 & A_2^\dagger A_2 \end{bmatrix}$. The eigen decoposition of $O$ is $O = PDP^{-1}$. According to Lemma IV A 1, $P$ and $D$ are both block-diagonal and they are composed of the eigenvectors and eigenvalues of $A_1$ and $A_2$. Therefore, the construction of $U$ in step 3 is equivalent to two

separate constructions of $U_1$ and $U_2$ for $A_1$ and $A_2$, and the multiplication of $U_1$ and $U_2$ is applied separately to $A_1$ and $A_2$. Thus, Lemma IV B 1 holds.

### C. Gram–Schmidt orthonormalization

#### 1. Classical Gram–Schmidt algorithm

To transform a $n \times m$ matrix $A$ into a semi-unitary form ($A^\dagger A = I_n$), the Gram–Schmidt algorithm is a commonly used method. The column-wise classical Gram–Schmidt (GS) algorithm can be expressed as

---
**Algorithm 1** Classical Gram–Schmidt algorithm
---
$B = A$
**for** $j \leq n$ **do**
    **for** $i < j$ **do**
        $B[:,j] = B[:,j] - \langle B[:,i], A[:,j] \rangle B[:,i]$
    **end for**
    $B[:,j] = B[:,j]/\|B[:,j]\|$
**end for**
---

The GS algorithm ensures that each column vector is orthogonal to all preceding columns. As a result, the final matrix $B$ is semi-unitary.

#### 2. Modified Gram–Schmidt algorithm

Since we implement the GS algorithm through computer code, modern computing commonly employs the modified GS algorithm, which has lower numerical errors. The algorithm principle is as follows:

---
**Algorithm 2** Modified Gram–Schmidt algorithm
---
$B = A$
**for** $j \leq n$ **do**
    **for** $i < j$ **do**
        $B[:,j] = B[:,j] - \langle B[:,i], B[:,j] \rangle B[:,i]$
    **end for**
    $B[:,j] = B[:,j]/\|B[:,j]\|$
**end for**
---

Mathematically, the two algorithms are equivalent, but the modified GS algorithm exhibits better numerical orthogonality.

#### 3. Modified Gram–Schmidt implemented in `pw2wannier90.x`

All discussions in this section are based on the case of ultrasoft pseudopotentials (USPP). Namely, an Hermitian overlap operator $\hat{S}$ is required to achieve $\langle \psi | \hat{S} | \psi \rangle = 1$. For non-USPP cases, one can simply take $\hat{S} = \hat{I}$.

We have $n = n_P + n_H$ projectors, of which $n_P$ are from Pseudo Atomic Orbitals (PAOs) and $n_H$ from external hydrogenic AOs. The real projector matrix is

$$A = \begin{bmatrix} A_P & A_H \end{bmatrix}$$

and the overlap matrix is

$$O = A^\dagger \hat{S} A = \begin{bmatrix} A_P^\dagger \hat{S} A_P & A_H^\dagger \hat{S} A_P \\ A_P^\dagger \hat{S} A_H & A_H^\dagger \hat{S} A_H \end{bmatrix},$$

where $A_P$ and $A_H$ are the projector matrices of PAOs and hydrogenic AOs respectively. However, we do not use the conventional Löwdin orthonormalization on the full set of $n$ projectors, as the procedure would also change the PAOs. So we first set the $A_H^\dagger \hat{S} A_P$ and $A_P^\dagger \hat{S} A_H$ to 0 in the overlap matrix, and then perform a Löwdin orthonormalization, so that the matrix we orthonormalize can actually be written as

$$A = \begin{bmatrix} A_P & 0 \\ 0 & A_H \end{bmatrix}.$$

According to Lemma IV B 1, the orthonormalization is equivalent to orthonormalizing the two sets of projectors separately.

When the matrix is written in the real projector form, it becomes

$$A = \begin{bmatrix} O_P^{-1/2} A_P & O_H^{-1/2} A_H \end{bmatrix}$$

and

$$SA = \begin{bmatrix} O_P^{-1/2} S A_P & O_H^{-1/2} S A_H \end{bmatrix}.$$

The off-diagonal part of the overlap matrix $\langle S, SA \rangle$ is not 0. So, an extra GS process is required to orthonormalize the projector matrix, while still keeping the PAOs unchanged. We therefore consider the projectors from PAOs frozen, and thus we can calculate the matrix by

---
**Algorithm 3** Practical algorithm in `pw2wannier90.x`
---
$B = A$
$SB = SA$
**for** projector $j$ **not in** frozen projectors **do**
    **for** projector $i$ **in** frozen projectors **do**
        $B[:,j] = B[:,j] - \langle B[:,i], SB[:,j] \rangle B[:,i]$
        $SB[:,j] = SB[:,j] - \langle B[:,i], SB[:,j] \rangle SB[:,i]$
    **end for**
    $SB[:,j] = SB[:,j]/\langle B[:,j], SB[:,j] \rangle$
    **set** projector $j$ frozen
**end for**
---

thus obtaining the final projection matrix $SB$.

# V. MAGNETIC STRUCTURES

With the goal of selecting a representative set of materials with a diversity of magnetic moment configurations, we selected several small- to medium-scale collinear and non-collinear magnetic structures from the MAGNDATA database [2]. These are graphically shown in Supplementary Fig. 4 and listed in Supplementary Table I.

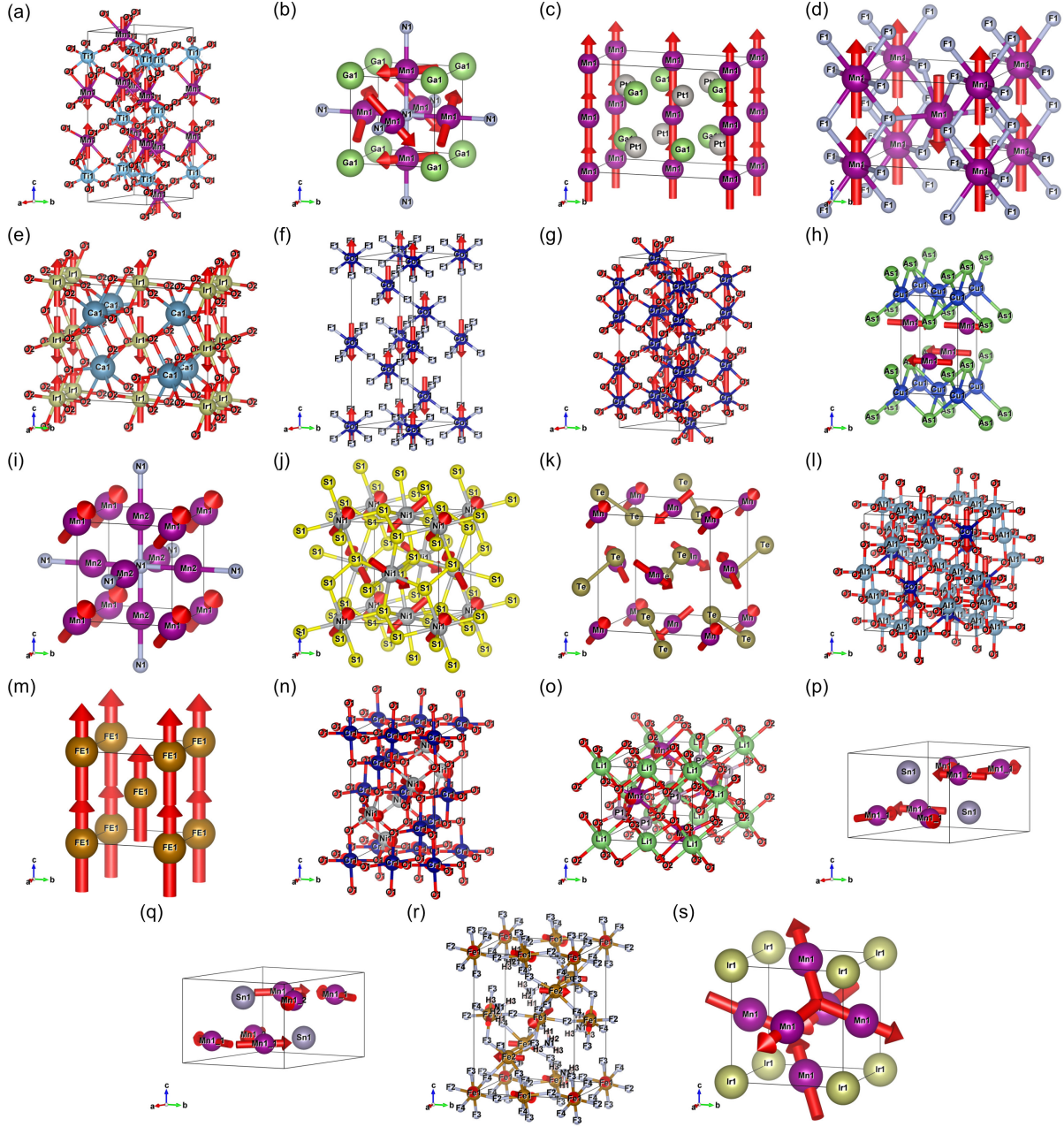

SUPPLEMENTARY FIG. 4. Visualization of the magnetic crystal structures used in collinear and non-collinear spin calculations, also listed in Supplementary Table I: (a) $Mn_2O_6Ti_2$, (b) $GaMn_3N$, (c) $Ga_2Mn_2Pt_2$, (d) $F_4Mn_2$, (e) $Ca_2Ir_2O_6$, (f) $Co_2F_6$, (g) $Cr_4O_6$, (h) $As_2Cu_2Mn_2$, (i) $Mn_4N$, (j) $Ni_4S_8$, (k) $Mn_4Te_8$, (l) $Al_4Co_2O_8$, (m) Fe, (n) $Cr_4Ni_2O_8$, (o) $Li_4Mn_4O_{16}P_4$, (p) $Mn_6Sn_2$, (q) $Mn_6Sn_2$, (r) $F_{24}Fe_8H_{16}N_4$, (s) $IrMn_3$. Magnetic moments are represented by red arrows.

SUPPLEMENTARY TABLE I. Magnetic structures used in collinear and non-collinear spin calculations.

| Label | Formula | Collinear calculation | Non-collinear calculation |
|---|---|:---:|:---:|
| a | $Mn_2O_6Ti_2$ | ✓ | ✓ |
| b | $GaMn_3N$ | ✓ | |
| c | $Ga_2Mn_2Pt_2$ | ✓ | ✓ |
| d | $F_4Mn_2$ | ✓ | ✓ |
| e | $Ca_2Ir_2O_6$ | ✓ | ✓ |
| f | $Co_2F_6$ | ✓ | ✓ |
| g | $Cr_4O_6$ | ✓ | ✓ |
| h | $As_2Cu_2Mn_2$ | ✓ | ✓ |
| i | $Mn_4N$ | ✓ | ✓ |
| j | $Ni_4S_8$ | ✓ | ✓ |
| k | $Mn_4Te_8$ | ✓ | ✓ |
| l | $Al_4Co_2O_8$ | ✓ | ✓ |
| m | $Fe$ | ✓ | ✓ |
| n | $Cr_4Ni_2O_8$ | | ✓ |
| o | $Li_4Mn_4O_{16}P_4$ | ✓ | ✓ |
| p | $Mn_6Sn_2$ | | ✓ |
| q | $Mn_6Sn_2$ | | ✓ |
| r | $F_{24}Fe_8H_{16}N_4$ | ✓ | |
| s | $IrMn_3$ | ✓ | |

## VI. REQUIRED PROJECTORS SET FOR A ROBUST PDWF WANNIERIZATION

In the main text, we employed a relatively aggressive strategy (see Section II B in the main text for detailed rules) to define a set of projectors that stabilize most PDWFs. These orbitals are listed in Supplementary Table II. We stress that, however, in practical applications this setting may result in an excessive number of projectors for certain structures. Thus, users are encouraged to select an appropriate projector set based on the specific system of interest.

SUPPLEMENTARY TABLE II: Required orbitals for a robust PDWF Wannierization.

| $Z$ | Element | Required orbitals |
|---|---|---|
| 1 | H | $1s$ |
| 2 | He | $1s$ |
| 3 | Li | $2s\ 2p$ |
| 4 | Be | $2s\ 2p$ |
| 5 | B | $2s\ 2p$ |
| 6 | C | $2s\ 2p$ |
| 7 | N | $2s\ 2p$ |
| 8 | O | $2s\ 2p$ |
| 9 | F | $2s\ 2p$ |
| 10 | Ne | $2s\ 2p$ |
| 11 | Na | $3s\ 3p$ |
| 12 | Mg | $3s\ 3p$ |
| 13 | Al | $3s\ 3p$ |
| 14 | Si | $3s\ 3p$ |
| 15 | P | $3s\ 3p$ |
| 16 | S | $3s\ 3p$ |
| 17 | Cl | $3s\ 3p$ |
| 18 | Ar | $3s\ 3p$ |
| 19 | K | $4s\ 3d$ |
| 20 | Ca | $4s\ 3d$ |
| 21 | Sc | $4s\ 4p\ 3d$ |
| 22 | Ti | $4s\ 4p\ 3d$ |
| 23 | V | $4s\ 4p\ 3d$ |
| 24 | Cr | $4s\ 4p\ 3d$ |
| 25 | Mn | $4s\ 4p\ 3d$ |
| 26 | Fe | $4s\ 4p\ 3d$ |
| 27 | Co | $4s\ 4p\ 3d$ |
| 28 | Ni | $4s\ 4p\ 3d$ |
| 29 | Cu | $4s\ 4p\ 3d$ |
| 30 | Zn | $4s\ 4p\ 3d$ |
| 31 | Ga | $4s\ 4p$ |
| 32 | Ge | $4s\ 4p$ |
| 33 | As | $4s\ 4p$ |
| 34 | Se | $4s\ 4p$ |
| 35 | Br | $4s\ 4p$ |
| 36 | Kr | $4s\ 4p$ |
| 37 | Rb | $5s\ 4d$ |
| 38 | Sr | $5s\ 4d$ |
| 39 | Y | $5s\ 5p\ 4d$ |
| 40 | Zr | $5s\ 5p\ 4d$ |
| 41 | Nb | $5s\ 5p\ 4d$ |
| 42 | Mo | $5s\ 5p\ 4d$ |
| 43 | Tc | $5s\ 5p\ 4d$ |
| 44 | Ru | $5s\ 5p\ 4d$ |
| 45 | Rh | $5s\ 5p\ 4d$ |
| 46 | Pd | $5s\ 5p\ 4d$ |

| $Z$ | Element | Required orbitals |
|---|---|---|
| 47 | Ag | $5s\ 5p\ 4d$ |
| 48 | Cd | $5s\ 5p\ 4d$ |
| 49 | In | $5s\ 5p$ |
| 50 | Sn | $5s\ 5p$ |
| 51 | Sb | $5s\ 5p$ |
| 52 | Te | $5s\ 5p$ |
| 53 | I | $5s\ 5p$ |
| 54 | Xe | $5s\ 5p$ |
| 55 | Cs | $6s\ 5d$ |
| 56 | Ba | $6s\ 5d$ |
| 72 | Hf | $6s\ 6p\ 5d$ |
| 73 | Ta | $6s\ 6p\ 5d$ |
| 74 | W | $6s\ 6p\ 5d$ |
| 75 | Re | $6s\ 6p\ 5d$ |
| 76 | Os | $6s\ 6p\ 5d$ |
| 77 | Ir | $6s\ 6p\ 5d$ |
| 78 | Pt | $6s\ 6p\ 5d$ |
| 79 | Au | $6s\ 6p\ 5d$ |
| 80 | Hg | $6s\ 6p\ 5d$ |
| 81 | Tl | $6s\ 6p$ |
| 82 | Pb | $6s\ 6p$ |
| 83 | Bi | $6s\ 6p$ |
| 84 | Po | $6s\ 6p$ |
| 85 | At | $6s\ 6p$ |
| 86 | Rn | $6s\ 6p$ |

## VII.  DETAILED INFORMATION ON THE PROJECTORS USED IN THIS WORK

Based on the radial expression of hydrogenic AOs (see Table I in main text for details), we obtained external projectors using different approaches, including orthonormalization fitting (detailed see Section II B in the main text) and all-hydrogenic fitting, which is the method to generate the projectors for "corrected hydrogenic AOs". The latter involves fitting the electron in the innermost shell to the pseudopotential PAOs while applying orthogonalization methods as much as possible to assign $\alpha$ to outer hydrogenic AOs. For missing AOs without inner shells, where therefore there is no orthogonality condition for the radial part to satisfy, we consider PAOs from OpenMX [3] and fit them accordingly to obtain the $\alpha$ values.

The values for the various pseudopotential libraries used in the main text, as well as using different orthogonalization fitting schemes, are reported in Supplementary Tables III, IV and V. The choice of $\alpha$ for the `modified-pslibrary` set (whose choice of pseudopotentials is reported in Supplementary Table VI) is reported in Supplementary Table VII.

In the Source column of the tables, "Pseudopotential" means that the orbital is simply taken from the PAO from the pseudopotential file, "Orthogonalization" means that the value is obtained by achieving orthogonality with the underlying PAO with same angular momentum $l$, while "OpenMX Fitting" or "Pseudopotential Fitting" indicate that the value has been obtained by fitting the corresponding atomic orbital from the OpenMX dataset [3] or from the pseudopotential PAOs. $\alpha$ values are in unit of bohr$^{-1}$.

Finally, to facilitate users in generating external projectors for other pseudopotential libraries, we provide in Supplementary Table VIII the $\alpha$ values for all hydrogenic AOs that we fitted in this work using `OpenMX`. This table allows users to use data directly from the table for projector generation. The recommended procedure to determine the $\alpha$ values for a new pseudopotential library is the following. First, use our script in the AiiDA-Wannier90-Workflows repository to generate external hydrogenic AOs based on the pseudopotential PAOs. Then, for missing projectors with nodes, employ orthogonalization fitting with the inner PAOs from pseudopotential to ensure greater accuracy. Finally, for any remaining projectors without nodes, use the value from Supplementary Table VIII.

SUPPLEMENTARY TABLE III: Values of $\alpha$ and minimal set of orbitals for the pseudopotentials of the SSSP PBE Efficiency v1.1 with orthonormalization fitting.

| Z | Element | Orbital | l | Source | $\alpha$(bohr$^{-1}$) |
|---|---------|---------|---|--------|------------------------|
| 1 | H | $1s$ | 0 | Pseudopotential | — |
| 2 | He | $1s$ | 0 | Pseudopotential | — |
| 3 | Li | $1s$ | 0 | Pseudopotential | — |
| | | $2s$ | 0 | Pseudopotential | — |
| | | $2p$ | 1 | Pseudopotential | — |
| 4 | Be | $1s$ | 0 | Pseudopotential | — |
| | | $2s$ | 0 | Pseudopotential | — |
| | | $2p$ | 1 | Pseudopotential | — |
| 5 | B | $2s$ | 0 | Pseudopotential | — |
| | | $2p$ | 1 | Pseudopotential | — |
| 6 | C | $2s$ | 0 | Pseudopotential | — |
| | | $2p$ | 1 | Pseudopotential | — |
| 7 | N | $2s$ | 0 | Pseudopotential | — |
| | | $2p$ | 1 | Pseudopotential | — |
| 8 | O | $2s$ | 0 | Pseudopotential | — |
| | | $2p$ | 1 | Pseudopotential | — |
| 9 | F | $2s$ | 0 | Pseudopotential | — |
| | | $2p$ | 1 | Pseudopotential | — |
| 10 | Ne | $2s$ | 0 | Pseudopotential | — |
| | | $2p$ | 1 | Pseudopotential | — |
| 11 | Na | $2s$ | 0 | Pseudopotential | — |
| | | $3s$ | 0 | Pseudopotential | — |
| | | $2p$ | 1 | Pseudopotential | — |
| | | $3p$ | 1 | Orthogonalization | 4.500 |
| 12 | Mg | $3s$ | 0 | Pseudopotential | — |
| | | $3p$ | 1 | Pseudopotential | — |
| 13 | Al | $3s$ | 0 | Pseudopotential | — |
| | | $3p$ | 1 | Pseudopotential | — |
| 14 | Si | $3s$ | 0 | Pseudopotential | — |
| | | $3p$ | 1 | Pseudopotential | — |
| 15 | P | $3s$ | 0 | Pseudopotential | — |
| | | $3p$ | 1 | Pseudopotential | — |
| 16 | S | $3s$ | 0 | Pseudopotential | — |
| | | $3p$ | 1 | Pseudopotential | — |
| 17 | Cl | $3s$ | 0 | Pseudopotential | — |
| | | $3p$ | 1 | Pseudopotential | — |
| 18 | Ar | $3s$ | 0 | Pseudopotential | — |
| | | $3p$ | 1 | Pseudopotential | — |
| 19 | K | $3s$ | 0 | Pseudopotential | — |
| | | $4s$ | 0 | Pseudopotential | — |
| | | $3p$ | 1 | Pseudopotential | — |
| | | $4p$ | 1 | Pseudopotential | — |
| | | $3d$ | 2 | OpenMX Fitting | 2.727 |
| 20 | Ca | $3s$ | 0 | Pseudopotential | — |
| | | $4s$ | 0 | Pseudopotential | — |
| | | $3p$ | 1 | Pseudopotential | — |
| | | $3d$ | 2 | Pseudopotential | — |
| 21 | Sc | $3s$ | 0 | Pseudopotential | — |
| | | $4s$ | 0 | Pseudopotential | — |
| | | $3p$ | 1 | Pseudopotential | — |
| | | $3d$ | 2 | Pseudopotential | — |
| | | $4p$ | 1 | Orthogonalization | 4.200 |
| 22 | Ti | $3s$ | 0 | Pseudopotential | — |
| | | $4s$ | 0 | Pseudopotential | — |
| | | $3p$ | 1 | Pseudopotential | — |
| | | $3d$ | 2 | Pseudopotential | — |
| | | $4p$ | 1 | Orthogonalization | 4.400 |
| 23 | V | $3s$ | 0 | Pseudopotential | — |
| | | $4s$ | 0 | Pseudopotential | — |
| | | $3p$ | 1 | Pseudopotential | — |
| | | $3d$ | 2 | Pseudopotential | — |
| | | $4p$ | 1 | Orthogonalization | 4.600 |
| 24 | Cr | $3s$ | 0 | Pseudopotential | — |
| | | $4s$ | 0 | Pseudopotential | — |
| | | $3p$ | 1 | Pseudopotential | — |
| | | $3d$ | 2 | Pseudopotential | — |
| | | $4p$ | 1 | Orthogonalization | 4.900 |
| 25 | Mn | $3s$ | 0 | Pseudopotential | — |
| | | $4s$ | 0 | Pseudopotential | — |
| | | $3p$ | 1 | Pseudopotential | — |
| | | $4p$ | 1 | Pseudopotential | — |
| | | $3d$ | 2 | Pseudopotential | — |
| 26 | Fe | $3s$ | 0 | Pseudopotential | — |
| | | $4s$ | 0 | Pseudopotential | — |
| | | $3p$ | 1 | Pseudopotential | — |
| | | $4p$ | 1 | Pseudopotential | — |
| | | $3d$ | 2 | Pseudopotential | — |
| 27 | Co | $3s$ | 0 | Pseudopotential | — |
| | | $4s$ | 0 | Pseudopotential | — |
| | | $3p$ | 1 | Pseudopotential | — |
| | | $4p$ | 1 | Pseudopotential | — |
| | | $3d$ | 2 | Pseudopotential | — |
| 28 | Ni | $3s$ | 0 | Pseudopotential | — |
| | | $4s$ | 0 | Pseudopotential | — |
| | | $3p$ | 1 | Pseudopotential | — |
| | | $4p$ | 1 | Pseudopotential | — |
| | | $3d$ | 2 | Pseudopotential | — |
| 29 | Cu | $3s$ | 0 | Pseudopotential | — |
| | | $4s$ | 0 | Pseudopotential | — |
| | | $3p$ | 1 | Pseudopotential | — |
| | | $4p$ | 1 | Pseudopotential | — |
| | | $3d$ | 2 | Pseudopotential | — |
| 30 | Zn | $3s$ | 0 | Pseudopotential | — |
| | | $4s$ | 0 | Pseudopotential | — |
| | | $3p$ | 1 | Pseudopotential | — |
| | | $4p$ | 1 | Pseudopotential | — |
| | | $3d$ | 2 | Pseudopotential | — |
| 31 | Ga | $4s$ | 0 | Pseudopotential | — |
| | | $4p$ | 1 | Pseudopotential | — |
| | | $3d$ | 2 | Pseudopotential | — |
| 32 | Ge | $4s$ | 0 | Pseudopotential | — |
| | | $4p$ | 1 | Pseudopotential | — |
| | | $3d$ | 2 | Pseudopotential | — |
| 33 | As | $4s$ | 0 | Pseudopotential | — |
| | | $4p$ | 1 | Pseudopotential | — |
| 34 | Se | $4s$ | 0 | Pseudopotential | — |
| | | $4p$ | 1 | Pseudopotential | — |
| 35 | Br | $4s$ | 0 | Pseudopotential | — |
| | | $4p$ | 1 | Pseudopotential | — |
| 36 | Kr | $4s$ | 0 | Pseudopotential | — |
| | | $4p$ | 1 | Pseudopotential | — |
| 37 | Rb | $4s$ | 0 | Pseudopotential | — |
| | | $5s$ | 0 | Pseudopotential | — |
| | | $4p$ | 1 | Pseudopotential | — |
| | | $4d$ | 2 | OpenMX Fitting | 2.197 |

| Z | Element | Orbital | l | Source | $\alpha(\text{bohr}^{-1})$ |
|---|---------|---------|---|--------|------------|
| | | 4s | 0 | Pseudopotential | — |
| | | 5s | 0 | Pseudopotential | — |
| 38 | Sr | 4p | 1 | Pseudopotential | — |
| | | 5p | 1 | Pseudopotential | — |
| | | 4d | 2 | Pseudopotential | — |
| | | 4s | 0 | Pseudopotential | — |
| | | 5s | 0 | Pseudopotential | — |
| 39 | Y | 4p | 1 | Pseudopotential | — |
| | | 5p | 1 | Pseudopotential | — |
| | | 4d | 2 | Pseudopotential | — |
| | | 4s | 0 | Pseudopotential | — |
| | | 5s | 0 | Pseudopotential | — |
| 40 | Zr | 4p | 1 | Pseudopotential | — |
| | | 5p | 1 | Pseudopotential | — |
| | | 4d | 2 | Pseudopotential | — |
| | | 4s | 0 | Pseudopotential | — |
| | | 5s | 0 | Pseudopotential | — |
| 41 | Nb | 4p | 1 | Pseudopotential | — |
| | | 4d | 2 | Pseudopotential | — |
| | | 5p | 1 | Orthogonalization | 4.100 |
| | | 4s | 0 | Pseudopotential | — |
| | | 5s | 0 | Pseudopotential | — |
| 42 | Mo | 4p | 1 | Pseudopotential | — |
| | | 4d | 2 | Pseudopotential | — |
| | | 5p | 1 | Orthogonalization | 4.300 |
| | | 4s | 0 | Pseudopotential | — |
| | | 5s | 0 | Pseudopotential | — |
| 43 | Tc | 4p | 1 | Pseudopotential | — |
| | | 4d | 2 | Pseudopotential | — |
| | | 5p | 1 | Orthogonalization | 4.400 |
| | | 4s | 0 | Pseudopotential | — |
| | | 5s | 0 | Pseudopotential | — |
| 44 | Ru | 4p | 1 | Pseudopotential | — |
| | | 4d | 2 | Pseudopotential | — |
| | | 5p | 1 | Orthogonalization | 4.600 |
| | | 4s | 0 | Pseudopotential | — |
| | | 5s | 0 | Pseudopotential | — |
| 45 | Rh | 4p | 1 | Pseudopotential | — |
| | | 4d | 2 | Pseudopotential | — |
| | | 5p | 1 | Orthogonalization | 4.800 |
| | | 4s | 0 | Pseudopotential | — |
| | | 5s | 0 | Pseudopotential | — |
| 46 | Pd | 4p | 1 | Pseudopotential | — |
| | | 4d | 2 | Pseudopotential | — |
| | | 5p | 1 | Orthogonalization | 5.000 |
| | | 4s | 0 | Pseudopotential | — |
| | | 5s | 0 | Pseudopotential | — |
| 47 | Ag | 4p | 1 | Pseudopotential | — |
| | | 4d | 2 | Pseudopotential | — |
| | | 5p | 1 | Orthogonalization | 5.100 |
| | | 5s | 0 | Pseudopotential | — |
| 48 | Cd | 5p | 1 | Pseudopotential | — |
| | | 4d | 2 | Pseudopotential | — |
| | | 5s | 0 | Pseudopotential | — |
| 49 | In | 5p | 1 | Pseudopotential | — |
| | | 4d | 2 | Pseudopotential | — |
| | | 5s | 0 | Pseudopotential | — |
| 50 | Sn | 5p | 1 | Pseudopotential | — |
| | | 4d | 2 | Pseudopotential | — |
| | | 5s | 0 | Pseudopotential | — |
| 51 | Sb | 5p | 1 | Pseudopotential | — |
| | | 4d | 2 | Pseudopotential | — |

| Z | Element | Orbital | l | Source | $\alpha(\text{bohr}^{-1})$ |
|---|---------|---------|---|--------|------------|
| 52 | Te | 5s | 0 | Pseudopotential | — |
| | | 5p | 1 | Pseudopotential | — |
| 53 | I | 5s | 0 | Pseudopotential | — |
| | | 5p | 1 | Pseudopotential | — |
| | | 5s | 0 | Pseudopotential | — |
| 54 | Xe | 5p | 1 | Pseudopotential | — |
| | | 4d | 2 | Pseudopotential | — |
| | | 5s | 0 | Pseudopotential | — |
| | | 6s | 0 | Pseudopotential | — |
| 55 | Cs | 5p | 1 | Pseudopotential | — |
| | | 6p | 1 | Pseudopotential | — |
| | | 5d | 2 | Pseudopotential | — |
| | | 5s | 0 | Pseudopotential | — |
| | | 6s | 0 | Pseudopotential | — |
| 56 | Ba | 5p | 1 | Pseudopotential | — |
| | | 5d | 2 | OpenMX Fitting | 3.039 |
| | | 5s | 0 | Pseudopotential | — |
| | | 6s | 0 | Pseudopotential | — |
| 72 | Hf | 5p | 1 | Pseudopotential | — |
| | | 5d | 2 | Pseudopotential | — |
| | | 6p | 1 | Orthogonalization | 3.900 |
| | | 5s | 0 | Pseudopotential | — |
| | | 6s | 0 | Pseudopotential | — |
| 73 | Ta | 5p | 1 | Pseudopotential | — |
| | | 6p | 1 | Pseudopotential | — |
| | | 5d | 2 | Pseudopotential | — |
| | | 5s | 0 | Pseudopotential | — |
| | | 6s | 0 | Pseudopotential | — |
| 74 | W | 5p | 1 | Pseudopotential | — |
| | | 6p | 1 | Pseudopotential | — |
| | | 5d | 2 | Pseudopotential | — |
| | | 5s | 0 | Pseudopotential | — |
| | | 6s | 0 | Pseudopotential | — |
| 75 | Re | 5p | 1 | Pseudopotential | — |
| | | 6p | 1 | Pseudopotential | — |
| | | 5d | 2 | Pseudopotential | — |
| | | 5s | 0 | Pseudopotential | — |
| | | 6s | 0 | Pseudopotential | — |
| 76 | Os | 5p | 1 | Pseudopotential | — |
| | | 6p | 1 | Pseudopotential | — |
| | | 5d | 2 | Pseudopotential | — |
| | | 6s | 0 | Pseudopotential | — |
| 77 | Ir | 5p | 1 | Pseudopotential | — |
| | | 6p | 1 | Pseudopotential | — |
| | | 5d | 2 | Pseudopotential | — |
| | | 6s | 0 | Pseudopotential | — |
| 78 | Pt | 5p | 1 | Pseudopotential | — |
| | | 6p | 1 | Pseudopotential | — |
| | | 5d | 2 | Pseudopotential | — |
| | | 5s | 0 | Pseudopotential | — |
| | | 6s | 0 | Pseudopotential | — |
| 79 | Au | 5p | 1 | Pseudopotential | — |
| | | 5d | 2 | Pseudopotential | — |
| | | 6p | 1 | Orthogonalization | 5.000 |
| | | 5s | 0 | Pseudopotential | — |
| | | 6s | 0 | Pseudopotential | — |
| 80 | Hg | 5p | 1 | Pseudopotential | — |
| | | 5d | 2 | Pseudopotential | — |
| | | 6p | 1 | Orthogonalization | 5.100 |
| | | 6s | 0 | Pseudopotential | — |
| 81 | Tl | 6p | 1 | Pseudopotential | — |
| | | 5d | 2 | Pseudopotential | — |

| $Z$ | Element | Orbital | $l$ | Source | $\alpha(\text{bohr}^{-1})$ |
|---|---|---|---|---|---|
| 82 | Pb | $6s$ | 0 | Pseudopotential | — |
|  |  | $6p$ | 1 | Pseudopotential | — |
|  |  | $5d$ | 2 | Pseudopotential | — |
| 83 | Bi | $6s$ | 0 | Pseudopotential | — |
|  |  | $6p$ | 1 | Pseudopotential | — |
|  |  | $5d$ | 2 | Pseudopotential | — |
| 84 | Po | $6s$ | 0 | Pseudopotential | — |
|  |  | $6p$ | 1 | Pseudopotential | — |
|  |  | $5d$ | 2 | Pseudopotential | — |
| 86 | Rn | $6s$ | 0 | Pseudopotential | — |
|  |  | $6p$ | 1 | Pseudopotential | — |
|  |  | $5d$ | 2 | Pseudopotential | — |

SUPPLEMENTARY TABLE IV: Values of $\alpha$ and minimal set of orbitals for the pseudopotentials of the SSSP PBE Efficiency v1.1 with all-hydrogenic fitting.

| $Z$ | Element | Orbital | $l$ | Source | $\alpha(\text{bohr}^{-1})$ |
|---|---|---|---|---|---|
| 1 | H | $1s$ | 0 | Pseudopotential Fitting | 0.910 |
| 2 | He | $1s$ | 0 | Pseudopotential Fitting | 1.522 |
| 3 | Li | $1s$ | 0 | Pseudopotential Fitting | 1.252 |
|  |  | $2s$ | 0 | Orthogonalization | 1.252 |
|  |  | $2p$ | 1 | Pseudopotential Fitting | 1.205 |
| 4 | Be | $1s$ | 0 | Pseudopotential Fitting | 0.702 |
|  |  | $2s$ | 0 | Orthogonalization | 0.702 |
|  |  | $2p$ | 1 | Pseudopotential Fitting | 1.612 |
| 5 | B | $2s$ | 0 | Pseudopotential Fitting | 0.614 |
|  |  | $2p$ | 1 | Pseudopotential Fitting | 2.283 |
| 6 | C | $2s$ | 0 | Pseudopotential Fitting | 0.863 |
|  |  | $2p$ | 1 | Pseudopotential Fitting | 2.834 |
| 7 | N | $2s$ | 0 | Pseudopotential Fitting | 1.109 |
|  |  | $2p$ | 1 | Pseudopotential Fitting | 3.195 |
| 8 | O | $2s$ | 0 | Pseudopotential Fitting | 1.287 |
|  |  | $2p$ | 1 | Pseudopotential Fitting | 3.564 |
| 9 | F | $2s$ | 0 | Pseudopotential Fitting | 1.403 |
|  |  | $2p$ | 1 | Pseudopotential Fitting | 3.476 |
| 10 | Ne | $2s$ | 0 | Pseudopotential Fitting | 1.565 |
|  |  | $2p$ | 1 | Pseudopotential Fitting | 4.326 |
| 11 | Na | $2s$ | 0 | Pseudopotential Fitting | 1.516 |
|  |  | $3s$ | 0 | Orthogonalization | 1.516 |
|  |  | $2p$ | 1 | Pseudopotential Fitting | 3.456 |
|  |  | $3p$ | 1 | Orthogonalization | 3.456 |
| 12 | Mg | $3s$ | 0 | Pseudopotential Fitting | 0.414 |
|  |  | $3p$ | 1 | Pseudopotential Fitting | 1.078 |
| 13 | Al | $3s$ | 0 | Pseudopotential Fitting | 0.478 |
|  |  | $3p$ | 1 | Pseudopotential Fitting | 1.412 |
| 14 | Si | $3s$ | 0 | Pseudopotential Fitting | 0.595 |
|  |  | $3p$ | 1 | Pseudopotential Fitting | 1.752 |
| 15 | P | $3s$ | 0 | Pseudopotential Fitting | 0.744 |
|  |  | $3p$ | 1 | Pseudopotential Fitting | 2.117 |
| 16 | S | $3s$ | 0 | Pseudopotential Fitting | 0.722 |
|  |  | $3p$ | 1 | Pseudopotential Fitting | 2.347 |
| 17 | Cl | $3s$ | 0 | Pseudopotential Fitting | 0.797 |
|  |  | $3p$ | 1 | Pseudopotential Fitting | 2.637 |
| 18 | Ar | $3s$ | 0 | Pseudopotential Fitting | 1.057 |
|  |  | $3p$ | 1 | Pseudopotential Fitting | 2.980 |
| 19 | K | $3s$ | 0 | Pseudopotential Fitting | 1.139 |
|  |  | $4s$ | 0 | Orthogonalization | 1.139 |
|  |  | $3p$ | 1 | Pseudopotential Fitting | 3.410 |
|  |  | $4p$ | 1 | Orthogonalization | 3.410 |
|  |  | $3d$ | 2 | OpenMX Fitting | 2.727 |
| 20 | Ca | $3s$ | 0 | Pseudopotential Fitting | 1.266 |
|  |  | $4s$ | 0 | Orthogonalization | 1.266 |
|  |  | $3p$ | 1 | Pseudopotential Fitting | 3.257 |
|  |  | $3d$ | 2 | Pseudopotential Fitting | 4.172 |
| 21 | Sc | $3s$ | 0 | Pseudopotential Fitting | 1.449 |
|  |  | $4s$ | 0 | Orthogonalization | 1.449 |
|  |  | $3p$ | 1 | Pseudopotential Fitting | 4.074 |
|  |  | $3d$ | 2 | Pseudopotential Fitting | 5.810 |
|  |  | $4p$ | 1 | Orthogonalization | 4.074 |
| 22 | Ti | $3s$ | 0 | Pseudopotential Fitting | 1.262 |
|  |  | $4s$ | 0 | Orthogonalization | 1.262 |
|  |  | $3p$ | 1 | Pseudopotential Fitting | 3.514 |
|  |  | $3d$ | 2 | Pseudopotential Fitting | 5.725 |
|  |  | $4p$ | 1 | Orthogonalization | 3.514 |

| Z | Element | Orbital | l | Source | α(bohr⁻¹) |
|---|---|---|---|---|---|
| 23 | V | 3s | 0 | Pseudopotential Fitting | 1.541 |
| | | 4s | 0 | Orthogonalization | 1.541 |
| | | 3p | 1 | Pseudopotential Fitting | 3.861 |
| | | 3d | 2 | Pseudopotential Fitting | 5.662 |
| | | 4p | 1 | Orthogonalization | 3.861 |
| 24 | Cr | 3s | 0 | Pseudopotential Fitting | 1.598 |
| | | 4s | 0 | Orthogonalization | 1.598 |
| | | 3p | 1 | Pseudopotential Fitting | 3.949 |
| | | 3d | 2 | Pseudopotential Fitting | 5.999 |
| | | 4p | 1 | Orthogonalization | 3.949 |
| 25 | Mn | 3s | 0 | Pseudopotential Fitting | 1.655 |
| | | 4s | 0 | Orthogonalization | 1.655 |
| | | 3p | 1 | Pseudopotential Fitting | 4.061 |
| | | 4p | 1 | Orthogonalization | 4.061 |
| | | 3d | 2 | Pseudopotential Fitting | 5.857 |
| 26 | Fe | 3s | 0 | Pseudopotential Fitting | 1.804 |
| | | 4s | 0 | Orthogonalization | 1.804 |
| | | 3p | 1 | Pseudopotential Fitting | 4.925 |
| | | 4p | 1 | Orthogonalization | 4.925 |
| | | 3d | 2 | Pseudopotential Fitting | 5.248 |
| 27 | Co | 3s | 0 | Pseudopotential Fitting | 1.656 |
| | | 4s | 0 | Orthogonalization | 1.656 |
| | | 3p | 1 | Pseudopotential Fitting | 3.888 |
| | | 4p | 1 | Orthogonalization | 3.888 |
| | | 3d | 2 | Pseudopotential Fitting | 6.165 |
| 28 | Ni | 3s | 0 | Pseudopotential Fitting | 1.746 |
| | | 4s | 0 | Orthogonalization | 1.746 |
| | | 3p | 1 | Pseudopotential Fitting | 3.868 |
| | | 4p | 1 | Orthogonalization | 3.868 |
| | | 3d | 2 | Pseudopotential Fitting | 6.521 |
| 29 | Cu | 3s | 0 | Pseudopotential Fitting | 1.674 |
| | | 4s | 0 | Orthogonalization | 1.674 |
| | | 3p | 1 | Pseudopotential Fitting | 3.477 |
| | | 4p | 1 | Orthogonalization | 3.477 |
| | | 3d | 2 | Pseudopotential Fitting | 6.327 |
| 30 | Zn | 3s | 0 | Pseudopotential Fitting | 1.676 |
| | | 4s | 0 | Orthogonalization | 1.676 |
| | | 3p | 1 | Pseudopotential Fitting | 3.762 |
| | | 4p | 1 | Orthogonalization | 3.762 |
| | | 3d | 2 | Pseudopotential Fitting | 5.849 |
| 31 | Ga | 4s | 0 | Pseudopotential Fitting | 0.558 |
| | | 4p | 1 | Pseudopotential Fitting | 1.484 |
| | | 3d | 2 | Pseudopotential Fitting | 5.166 |
| 32 | Ge | 4s | 0 | Pseudopotential Fitting | 0.607 |
| | | 4p | 1 | Pseudopotential Fitting | 1.669 |
| | | 3d | 2 | Pseudopotential Fitting | 1.332 |
| 33 | As | 4s | 0 | Pseudopotential Fitting | 0.710 |
| | | 4p | 1 | Pseudopotential Fitting | 1.984 |
| 34 | Se | 4s | 0 | Pseudopotential Fitting | 0.668 |
| | | 4p | 1 | Pseudopotential Fitting | 2.068 |
| 35 | Br | 4s | 0 | Pseudopotential Fitting | 0.713 |
| | | 4p | 1 | Pseudopotential Fitting | 2.267 |
| 36 | Kr | 4s | 0 | Pseudopotential Fitting | 0.985 |
| | | 4p | 1 | Pseudopotential Fitting | 2.566 |
| 37 | Rb | 4s | 0 | Pseudopotential Fitting | 1.099 |
| | | 5s | 0 | Orthogonalization | 1.099 |
| | | 4p | 1 | Pseudopotential Fitting | 2.894 |
| | | 4d | 2 | OpenMX Fitting | 2.197 |

| Z | Element | Orbital | l | Source | α(bohr⁻¹) |
|---|---|---|---|---|---|
| 38 | Sr | 4s | 0 | Pseudopotential Fitting | 1.064 |
| | | 5s | 0 | Orthogonalization | 1.064 |
| | | 4p | 1 | Pseudopotential Fitting | 3.045 |
| | | 5p | 1 | Orthogonalization | 3.045 |
| | | 4d | 2 | Pseudopotential Fitting | 3.121 |
| 39 | Y | 4s | 0 | Pseudopotential Fitting | 1.164 |
| | | 5s | 0 | Orthogonalization | 1.164 |
| | | 4p | 1 | Pseudopotential Fitting | 3.058 |
| | | 5p | 1 | Orthogonalization | 3.058 |
| | | 4d | 2 | Pseudopotential Fitting | 3.940 |
| 40 | Zr | 4s | 0 | Pseudopotential Fitting | 1.266 |
| | | 5s | 0 | Orthogonalization | 1.266 |
| | | 4p | 1 | Pseudopotential Fitting | 3.284 |
| | | 5p | 1 | Orthogonalization | 3.284 |
| | | 4d | 2 | Pseudopotential Fitting | 4.372 |
| 41 | Nb | 4s | 0 | Pseudopotential Fitting | 1.356 |
| | | 5s | 0 | Orthogonalization | 1.356 |
| | | 4p | 1 | Pseudopotential Fitting | 3.700 |
| | | 4d | 2 | Pseudopotential Fitting | 5.157 |
| | | 5p | 1 | Orthogonalization | 3.700 |
| 42 | Mo | 4s | 0 | Pseudopotential Fitting | 1.456 |
| | | 5s | 0 | Orthogonalization | 1.456 |
| | | 4p | 1 | Pseudopotential Fitting | 4.060 |
| | | 4d | 2 | Pseudopotential Fitting | 5.618 |
| | | 5p | 1 | Orthogonalization | 4.060 |
| 43 | Tc | 4s | 0 | Pseudopotential Fitting | 1.556 |
| | | 5s | 0 | Orthogonalization | 1.556 |
| | | 4p | 1 | Pseudopotential Fitting | 4.215 |
| | | 4d | 2 | Pseudopotential Fitting | 5.984 |
| | | 5p | 1 | Orthogonalization | 4.215 |
| 44 | Ru | 4s | 0 | Pseudopotential Fitting | 1.638 |
| | | 5s | 0 | Orthogonalization | 1.638 |
| | | 4p | 1 | Pseudopotential Fitting | 4.388 |
| | | 4d | 2 | Pseudopotential Fitting | 6.369 |
| | | 5p | 1 | Orthogonalization | 4.388 |
| 45 | Rh | 4s | 0 | Pseudopotential Fitting | 1.700 |
| | | 5s | 0 | Orthogonalization | 1.700 |
| | | 4p | 1 | Pseudopotential Fitting | 4.505 |
| | | 4d | 2 | Pseudopotential Fitting | 6.670 |
| | | 5p | 1 | Orthogonalization | 4.505 |
| 46 | Pd | 4s | 0 | Pseudopotential Fitting | 1.767 |
| | | 5s | 0 | Orthogonalization | 1.767 |
| | | 4p | 1 | Pseudopotential Fitting | 4.703 |
| | | 4d | 2 | Pseudopotential Fitting | 7.214 |
| | | 5p | 1 | Orthogonalization | 4.703 |
| 47 | Ag | 4s | 0 | Pseudopotential Fitting | 1.820 |
| | | 5s | 0 | Orthogonalization | 1.820 |
| | | 4p | 1 | Pseudopotential Fitting | 4.835 |
| | | 4d | 2 | Pseudopotential Fitting | 7.455 |
| | | 5p | 1 | Orthogonalization | 4.835 |
| 48 | Cd | 5s | 0 | Pseudopotential Fitting | 0.428 |
| | | 5p | 1 | Pseudopotential Fitting | 1.156 |
| | | 4d | 2 | Pseudopotential Fitting | 6.627 |
| 49 | In | 5s | 0 | Pseudopotential Fitting | 0.433 |
| | | 5p | 1 | Pseudopotential Fitting | 1.320 |
| | | 4d | 2 | Pseudopotential Fitting | 6.051 |
| 50 | Sn | 5s | 0 | Pseudopotential Fitting | 0.313 |
| | | 5p | 1 | Pseudopotential Fitting | 1.410 |
| | | 4d | 2 | Pseudopotential Fitting | 5.547 |
| 51 | Sb | 5s | 0 | Pseudopotential Fitting | 0.347 |
| | | 5p | 1 | Pseudopotential Fitting | 1.557 |
| | | 4d | 2 | Pseudopotential Fitting | 6.769 |

| Z | Element | Orbital | l | Source | α(bohr⁻¹) |
|---|---------|---------|---|--------|-----------|
| 52 | Te | 5s | 0 | Pseudopotential Fitting | 0.606 |
|   |   | 5p | 1 | Pseudopotential Fitting | 1.731 |
| 53 | I | 5s | 0 | Pseudopotential Fitting | 0.684 |
|   |   | 5p | 1 | Pseudopotential Fitting | 1.840 |
| 54 | Xe | 5s | 0 | Pseudopotential Fitting | 0.802 |
|   |   | 5p | 1 | Pseudopotential Fitting | 2.134 |
|   |   | 4d | 2 | Pseudopotential Fitting | 9.378 |
| 55 | Cs | 5s | 0 | Pseudopotential Fitting | 0.826 |
|   |   | 6s | 0 | Orthogonalization | 0.826 |
|   |   | 5p | 1 | Pseudopotential Fitting | 2.284 |
|   |   | 6p | 1 | Orthogonalization | 2.284 |
|   |   | 5d | 2 | Pseudopotential Fitting | 1.993 |
| 56 | Ba | 5s | 0 | Pseudopotential Fitting | 0.956 |
|   |   | 6s | 0 | Orthogonalization | 0.956 |
|   |   | 5p | 1 | Pseudopotential Fitting | 2.625 |
|   |   | 5d | 2 | OpenMX Fitting | 3.039 |
| 72 | Hf | 5s | 0 | Pseudopotential Fitting | 1.338 |
|   |   | 6s | 0 | Orthogonalization | 1.338 |
|   |   | 5p | 1 | Pseudopotential Fitting | 3.698 |
|   |   | 5d | 2 | Pseudopotential Fitting | 4.275 |
|   |   | 6p | 1 | Orthogonalization | 3.698 |
| 73 | Ta | 5s | 0 | Pseudopotential Fitting | 1.377 |
|   |   | 6s | 0 | Orthogonalization | 1.377 |
|   |   | 5p | 1 | Pseudopotential Fitting | 3.680 |
|   |   | 6p | 1 | Orthogonalization | 3.680 |
|   |   | 5d | 2 | Pseudopotential Fitting | 4.827 |
| 74 | W | 5s | 0 | Pseudopotential Fitting | 1.443 |
|   |   | 6s | 0 | Orthogonalization | 1.443 |
|   |   | 5p | 1 | Pseudopotential Fitting | 3.786 |
|   |   | 6p | 1 | Orthogonalization | 3.786 |
|   |   | 5d | 2 | Pseudopotential Fitting | 5.147 |
| 75 | Re | 5s | 0 | Pseudopotential Fitting | 1.473 |
|   |   | 6s | 0 | Orthogonalization | 1.473 |
|   |   | 5p | 1 | Pseudopotential Fitting | 3.874 |
|   |   | 6p | 1 | Orthogonalization | 3.874 |
|   |   | 5d | 2 | Pseudopotential Fitting | 5.504 |
| 76 | Os | 5s | 0 | Pseudopotential Fitting | 1.534 |
|   |   | 6s | 0 | Orthogonalization | 1.534 |
|   |   | 5p | 1 | Pseudopotential Fitting | 4.074 |
|   |   | 6p | 1 | Orthogonalization | 4.074 |
|   |   | 5d | 2 | Pseudopotential Fitting | 5.950 |
| 77 | Ir | 6s | 0 | Pseudopotential Fitting | 0.418 |
|   |   | 5p | 1 | Pseudopotential Fitting | 3.776 |
|   |   | 6p | 1 | Orthogonalization | 3.776 |
|   |   | 5d | 2 | Pseudopotential Fitting | 5.503 |
| 78 | Pt | 6s | 0 | Pseudopotential Fitting | 0.436 |
|   |   | 5p | 1 | Pseudopotential Fitting | 3.993 |
|   |   | 6p | 1 | Orthogonalization | 3.993 |
|   |   | 5d | 2 | Pseudopotential Fitting | 5.852 |
| 79 | Au | 5s | 0 | Pseudopotential Fitting | 1.840 |
|   |   | 6s | 0 | Orthogonalization | 1.840 |
|   |   | 5p | 1 | Pseudopotential Fitting | 4.682 |
|   |   | 5d | 2 | Pseudopotential Fitting | 6.697 |
|   |   | 6p | 1 | Orthogonalization | 4.682 |
| 80 | Hg | 5s | 0 | Pseudopotential Fitting | 1.891 |
|   |   | 6s | 0 | Orthogonalization | 1.891 |
|   |   | 5p | 1 | Pseudopotential Fitting | 4.840 |
|   |   | 5d | 2 | Pseudopotential Fitting | 6.962 |
|   |   | 6p | 1 | Orthogonalization | 4.840 |
| 81 | Tl | 6s | 0 | Pseudopotential Fitting | 0.548 |
|   |   | 6p | 1 | Pseudopotential Fitting | 1.219 |
|   |   | 5d | 2 | Pseudopotential Fitting | 4.322 |

| Z | Element | Orbital | l | Source | α(bohr⁻¹) |
|---|---------|---------|---|--------|-----------|
| 82 | Pb | 6s | 0 | Pseudopotential Fitting | 0.607 |
|   |   | 6p | 1 | Pseudopotential Fitting | 1.477 |
|   |   | 5d | 2 | Pseudopotential Fitting | 5.953 |
| 83 | Bi | 6s | 0 | Pseudopotential Fitting | 0.632 |
|   |   | 6p | 1 | Pseudopotential Fitting | 1.638 |
|   |   | 5d | 2 | Pseudopotential Fitting | 4.464 |
| 84 | Po | 6s | 0 | Pseudopotential Fitting | 0.622 |
|   |   | 6p | 1 | Pseudopotential Fitting | 1.658 |
|   |   | 5d | 2 | Pseudopotential Fitting | 7.897 |
| 86 | Rn | 6s | 0 | Pseudopotential Fitting | 0.615 |
|   |   | 6p | 1 | Pseudopotential Fitting | 1.915 |
|   |   | 5d | 2 | Pseudopotential Fitting | 8.335 |

SUPPLEMENTARY TABLE V: Values of $\alpha$ and minimal set of orbitals for the pseudopotentials of the PseudoDojo norm-conserving fully-relativistic v0.4 standard library with orthonormalization fitting.

| $Z$ | Element | Orbital | $l$ | $j$ | Source | $\alpha$(bohr$^{-1}$) |
|---|---|---|---|---|---|---|
| 1 | H | $1s$ | 0 | 0.5 | Pseudopotential | — |
| 2 | He | $1s$ | 0 | 0.5 | Pseudopotential | — |
| 3 | Li | $1s$ | 0 | 0.5 | Pseudopotential | — |
|   |   | $2s$ | 0 | 0.5 | Pseudopotential | — |
|   |   | $2p$ | 1 | 0.5 | OpenMX Fitting | 1.114 |
|   |   | $2p$ | 1 | 1.5 | OpenMX Fitting | 1.114 |
| 4 | Be | $1s$ | 0 | 0.5 | Pseudopotential | — |
|   |   | $2s$ | 0 | 0.5 | Pseudopotential | — |
|   |   | $2p$ | 1 | 0.5 | OpenMX Fitting | 1.834 |
|   |   | $2p$ | 1 | 1.5 | OpenMX Fitting | 1.834 |
| 5 | B | $2s$ | 0 | 0.5 | Pseudopotential | — |
|   |   | $2p$ | 1 | 0.5 | Pseudopotential | — |
|   |   | $2p$ | 1 | 1.5 | Pseudopotential | — |
| 6 | C | $2s$ | 0 | 0.5 | Pseudopotential | — |
|   |   | $2p$ | 1 | 0.5 | Pseudopotential | — |
|   |   | $2p$ | 1 | 1.5 | Pseudopotential | — |
| 7 | N | $2s$ | 0 | 0.5 | Pseudopotential | — |
|   |   | $2p$ | 1 | 0.5 | Pseudopotential | — |
|   |   | $2p$ | 1 | 1.5 | Pseudopotential | — |
| 8 | O | $2s$ | 0 | 0.5 | Pseudopotential | — |
|   |   | $2p$ | 1 | 0.5 | Pseudopotential | — |
|   |   | $2p$ | 1 | 1.5 | Pseudopotential | — |
| 9 | F | $2s$ | 0 | 0.5 | Pseudopotential | — |
|   |   | $2p$ | 1 | 0.5 | Pseudopotential | — |
|   |   | $2p$ | 1 | 1.5 | Pseudopotential | — |
| 10 | Ne | $2s$ | 0 | 0.5 | Pseudopotential | — |
|   |   | $2p$ | 1 | 0.5 | Pseudopotential | — |
|   |   | $2p$ | 1 | 1.5 | Pseudopotential | — |
| 11 | Na | $2s$ | 0 | 0.5 | Pseudopotential | — |
|   |   | $3s$ | 0 | 0.5 | Pseudopotential | — |
|   |   | $2p$ | 1 | 0.5 | Pseudopotential | — |
|   |   | $2p$ | 1 | 1.5 | Pseudopotential | — |
|   |   | $3p$ | 1 | 0.5 | Orthogonalization | 5.600 |
|   |   | $3p$ | 1 | 1.5 | Orthogonalization | 5.600 |
| 12 | Mg | $2s$ | 0 | 0.5 | Pseudopotential | — |
|   |   | $3s$ | 0 | 0.5 | Pseudopotential | — |
|   |   | $2p$ | 1 | 0.5 | Pseudopotential | — |
|   |   | $2p$ | 1 | 1.5 | Pseudopotential | — |
|   |   | $3p$ | 1 | 0.5 | Orthogonalization | 6.500 |
|   |   | $3p$ | 1 | 1.5 | Orthogonalization | 6.400 |
| 13 | Al | $3s$ | 0 | 0.5 | Pseudopotential | — |
|   |   | $3p$ | 1 | 0.5 | Pseudopotential | — |
|   |   | $3p$ | 1 | 1.5 | Pseudopotential | — |
| 14 | Si | $3s$ | 0 | 0.5 | Pseudopotential | — |
|   |   | $3p$ | 1 | 0.5 | Pseudopotential | — |
|   |   | $3p$ | 1 | 1.5 | Pseudopotential | — |
| 15 | P | $3s$ | 0 | 0.5 | Pseudopotential | — |
|   |   | $3p$ | 1 | 0.5 | Pseudopotential | — |
|   |   | $3p$ | 1 | 1.5 | Pseudopotential | — |
| 16 | S | $3s$ | 0 | 0.5 | Pseudopotential | — |
|   |   | $3p$ | 1 | 0.5 | Pseudopotential | — |
|   |   | $3p$ | 1 | 1.5 | Pseudopotential | — |
| 17 | Cl | $3s$ | 0 | 0.5 | Pseudopotential | — |
|   |   | $3p$ | 1 | 0.5 | Pseudopotential | — |
|   |   | $3p$ | 1 | 1.5 | Pseudopotential | — |
| 18 | Ar | $3s$ | 0 | 0.5 | Pseudopotential | — |
|   |   | $3p$ | 1 | 0.5 | Pseudopotential | — |
|   |   | $3p$ | 1 | 1.5 | Pseudopotential | — |

| $Z$ | Element | Orbital | $l$ | $j$ | Source | $\alpha$(bohr$^{-1}$) |
|---|---|---|---|---|---|---|
| 19 | K | $3s$ | 0 | 0.5 | Pseudopotential | — |
|   |   | $4s$ | 0 | 0.5 | Pseudopotential | — |
|   |   | $3p$ | 1 | 0.5 | Pseudopotential | — |
|   |   | $3p$ | 1 | 1.5 | Pseudopotential | — |
|   |   | $3d$ | 2 | 1.5 | OpenMX Fitting | 2.727 |
|   |   | $3d$ | 2 | 2.5 | OpenMX Fitting | 2.727 |
| 20 | Ca | $3s$ | 0 | 0.5 | Pseudopotential | — |
|   |   | $4s$ | 0 | 0.5 | Pseudopotential | — |
|   |   | $3p$ | 1 | 0.5 | Pseudopotential | — |
|   |   | $3p$ | 1 | 1.5 | Pseudopotential | — |
|   |   | $3d$ | 2 | 1.5 | OpenMX Fitting | 3.983 |
|   |   | $3d$ | 2 | 2.5 | OpenMX Fitting | 3.983 |
| 21 | Sc | $3s$ | 0 | 0.5 | Pseudopotential | — |
|   |   | $4s$ | 0 | 0.5 | Pseudopotential | — |
|   |   | $3p$ | 1 | 0.5 | Pseudopotential | — |
|   |   | $3p$ | 1 | 1.5 | Pseudopotential | — |
|   |   | $3d$ | 2 | 1.5 | Pseudopotential | — |
|   |   | $3d$ | 2 | 2.5 | Pseudopotential | — |
|   |   | $4p$ | 1 | 0.5 | Orthogonalization | 4.400 |
|   |   | $4p$ | 1 | 1.5 | Orthogonalization | 4.300 |
| 22 | Ti | $3s$ | 0 | 0.5 | Pseudopotential | — |
|   |   | $4s$ | 0 | 0.5 | Pseudopotential | — |
|   |   | $3p$ | 1 | 0.5 | Pseudopotential | — |
|   |   | $3p$ | 1 | 1.5 | Pseudopotential | — |
|   |   | $3d$ | 2 | 1.5 | Pseudopotential | — |
|   |   | $3d$ | 2 | 2.5 | Pseudopotential | — |
|   |   | $4p$ | 1 | 0.5 | Orthogonalization | 4.700 |
|   |   | $4p$ | 1 | 1.5 | Orthogonalization | 4.700 |
| 23 | V | $3s$ | 0 | 0.5 | Pseudopotential | — |
|   |   | $4s$ | 0 | 0.5 | Pseudopotential | — |
|   |   | $3p$ | 1 | 0.5 | Pseudopotential | — |
|   |   | $3p$ | 1 | 1.5 | Pseudopotential | — |
|   |   | $3d$ | 2 | 1.5 | Pseudopotential | — |
|   |   | $3d$ | 2 | 2.5 | Pseudopotential | — |
|   |   | $4p$ | 1 | 0.5 | Orthogonalization | 5.000 |
|   |   | $4p$ | 1 | 1.5 | Orthogonalization | 5.000 |
| 24 | Cr | $3s$ | 0 | 0.5 | Pseudopotential | — |
|   |   | $4s$ | 0 | 0.5 | Pseudopotential | — |
|   |   | $3p$ | 1 | 0.5 | Pseudopotential | — |
|   |   | $3p$ | 1 | 1.5 | Pseudopotential | — |
|   |   | $3d$ | 2 | 1.5 | Pseudopotential | — |
|   |   | $3d$ | 2 | 2.5 | Pseudopotential | — |
|   |   | $4p$ | 1 | 0.5 | Orthogonalization | 5.300 |
|   |   | $4p$ | 1 | 1.5 | Orthogonalization | 5.200 |
| 25 | Mn | $3s$ | 0 | 0.5 | Pseudopotential | — |
|   |   | $4s$ | 0 | 0.5 | Pseudopotential | — |
|   |   | $3p$ | 1 | 0.5 | Pseudopotential | — |
|   |   | $3p$ | 1 | 1.5 | Pseudopotential | — |
|   |   | $3d$ | 2 | 1.5 | Pseudopotential | — |
|   |   | $3d$ | 2 | 2.5 | Pseudopotential | — |
|   |   | $4p$ | 1 | 0.5 | Orthogonalization | 5.600 |
|   |   | $4p$ | 1 | 1.5 | Orthogonalization | 5.600 |
| 26 | Fe | $3s$ | 0 | 0.5 | Pseudopotential | — |
|   |   | $4s$ | 0 | 0.5 | Pseudopotential | — |
|   |   | $3p$ | 1 | 0.5 | Pseudopotential | — |
|   |   | $3p$ | 1 | 1.5 | Pseudopotential | — |
|   |   | $3d$ | 2 | 1.5 | Pseudopotential | — |
|   |   | $3d$ | 2 | 2.5 | Pseudopotential | — |
|   |   | $4p$ | 1 | 0.5 | Orthogonalization | 6.000 |
|   |   | $4p$ | 1 | 1.5 | Orthogonalization | 5.900 |

| $Z$ | Element | Orbital | $l$ | $j$ | Source | $\alpha(\text{bohr}^{-1})$ |
|---|---|---|---|---|---|---|
| | | $3s$ | 0 | 0.5 | Pseudopotential | — |
| | | $4s$ | 0 | 0.5 | Pseudopotential | — |
| | | $3p$ | 1 | 0.5 | Pseudopotential | — |
| 27 | Co | $3p$ | 1 | 1.5 | Pseudopotential | — |
| | | $3d$ | 2 | 1.5 | Pseudopotential | — |
| | | $3d$ | 2 | 2.5 | Pseudopotential | — |
| | | $4p$ | 1 | 0.5 | Orthogonalization | 6.300 |
| | | $4p$ | 1 | 1.5 | Orthogonalization | 6.200 |
| | | $3s$ | 0 | 0.5 | Pseudopotential | — |
| | | $4s$ | 0 | 0.5 | Pseudopotential | — |
| | | $3p$ | 1 | 0.5 | Pseudopotential | — |
| 28 | Ni | $3p$ | 1 | 1.5 | Pseudopotential | — |
| | | $3d$ | 2 | 1.5 | Pseudopotential | — |
| | | $3d$ | 2 | 2.5 | Pseudopotential | — |
| | | $4p$ | 1 | 0.5 | Orthogonalization | 6.500 |
| | | $4p$ | 1 | 1.5 | Orthogonalization | 6.500 |
| | | $3s$ | 0 | 0.5 | Pseudopotential | — |
| | | $4s$ | 0 | 0.5 | Pseudopotential | — |
| | | $3p$ | 1 | 0.5 | Pseudopotential | — |
| 29 | Cu | $3p$ | 1 | 1.5 | Pseudopotential | — |
| | | $3d$ | 2 | 1.5 | Pseudopotential | — |
| | | $3d$ | 2 | 2.5 | Pseudopotential | — |
| | | $4p$ | 1 | 0.5 | Orthogonalization | 6.700 |
| | | $4p$ | 1 | 1.5 | Orthogonalization | 6.600 |
| | | $3s$ | 0 | 0.5 | Pseudopotential | — |
| | | $4s$ | 0 | 0.5 | Pseudopotential | — |
| | | $3p$ | 1 | 0.5 | Pseudopotential | — |
| 30 | Zn | $3p$ | 1 | 1.5 | Pseudopotential | — |
| | | $3d$ | 2 | 1.5 | Pseudopotential | — |
| | | $3d$ | 2 | 2.5 | Pseudopotential | — |
| | | $4p$ | 1 | 0.5 | Orthogonalization | 6.900 |
| | | $4p$ | 1 | 1.5 | Orthogonalization | 6.800 |
| | | $4s$ | 0 | 0.5 | Pseudopotential | — |
| | | $4p$ | 1 | 0.5 | Pseudopotential | — |
| 31 | Ga | $4p$ | 1 | 1.5 | Pseudopotential | — |
| | | $3d$ | 2 | 1.5 | Pseudopotential | — |
| | | $3d$ | 2 | 2.5 | Pseudopotential | — |
| | | $4s$ | 0 | 0.5 | Pseudopotential | — |
| | | $4p$ | 1 | 0.5 | Pseudopotential | — |
| 32 | Ge | $4p$ | 1 | 1.5 | Pseudopotential | — |
| | | $3d$ | 2 | 1.5 | Pseudopotential | — |
| | | $3d$ | 2 | 2.5 | Pseudopotential | — |
| | | $4s$ | 0 | 0.5 | Pseudopotential | — |
| | | $4p$ | 1 | 0.5 | Pseudopotential | — |
| 33 | As | $4p$ | 1 | 1.5 | Pseudopotential | — |
| | | $3d$ | 2 | 1.5 | Pseudopotential | — |
| | | $3d$ | 2 | 2.5 | Pseudopotential | — |
| | | $4s$ | 0 | 0.5 | Pseudopotential | — |
| | | $4p$ | 1 | 0.5 | Pseudopotential | — |
| 34 | Se | $4p$ | 1 | 1.5 | Pseudopotential | — |
| | | $3d$ | 2 | 1.5 | Pseudopotential | — |
| | | $3d$ | 2 | 2.5 | Pseudopotential | — |
| | | $4s$ | 0 | 0.5 | Pseudopotential | — |
| 35 | Br | $4p$ | 1 | 0.5 | Pseudopotential | — |
| | | $4p$ | 1 | 1.5 | Pseudopotential | — |
| | | $4s$ | 0 | 0.5 | Pseudopotential | — |
| 36 | Kr | $4p$ | 1 | 0.5 | Pseudopotential | — |
| | | $4p$ | 1 | 1.5 | Pseudopotential | — |

| $Z$ | Element | Orbital | $l$ | $j$ | Source | $\alpha(\text{bohr}^{-1})$ |
|---|---|---|---|---|---|---|
| | | $4s$ | 0 | 0.5 | Pseudopotential | — |
| | | $5s$ | 0 | 0.5 | Pseudopotential | — |
| 37 | Rb | $4p$ | 1 | 0.5 | Pseudopotential | — |
| | | $4p$ | 1 | 1.5 | Pseudopotential | — |
| | | $4d$ | 2 | 1.5 | OpenMX Fitting | 2.197 |
| | | $4d$ | 2 | 2.5 | OpenMX Fitting | 2.197 |
| | | $4s$ | 0 | 0.5 | Pseudopotential | — |
| | | $5s$ | 0 | 0.5 | Pseudopotential | — |
| 38 | Sr | $4p$ | 1 | 0.5 | Pseudopotential | — |
| | | $4p$ | 1 | 1.5 | Pseudopotential | — |
| | | $4d$ | 2 | 1.5 | OpenMX Fitting | 4.400 |
| | | $4d$ | 2 | 2.5 | OpenMX Fitting | 4.400 |
| | | $4s$ | 0 | 0.5 | Pseudopotential | — |
| | | $5s$ | 0 | 0.5 | Pseudopotential | — |
| | | $4p$ | 1 | 0.5 | Pseudopotential | — |
| 39 | Y | $4p$ | 1 | 1.5 | Pseudopotential | — |
| | | $4d$ | 2 | 1.5 | Pseudopotential | — |
| | | $4d$ | 2 | 2.5 | Pseudopotential | — |
| | | $5p$ | 1 | 0.5 | Orthogonalization | 3.600 |
| | | $5p$ | 1 | 1.5 | Orthogonalization | 3.500 |
| | | $4s$ | 0 | 0.5 | Pseudopotential | — |
| | | $5s$ | 0 | 0.5 | Pseudopotential | — |
| | | $4p$ | 1 | 0.5 | Pseudopotential | — |
| 40 | Zr | $4p$ | 1 | 1.5 | Pseudopotential | — |
| | | $4d$ | 2 | 1.5 | Pseudopotential | — |
| | | $4d$ | 2 | 2.5 | Pseudopotential | — |
| | | $5p$ | 1 | 0.5 | Orthogonalization | 3.900 |
| | | $5p$ | 1 | 1.5 | Orthogonalization | 3.800 |
| | | $4s$ | 0 | 0.5 | Pseudopotential | — |
| | | $5s$ | 0 | 0.5 | Pseudopotential | — |
| | | $4p$ | 1 | 0.5 | Pseudopotential | — |
| 41 | Nb | $4p$ | 1 | 1.5 | Pseudopotential | — |
| | | $4d$ | 2 | 1.5 | Pseudopotential | — |
| | | $4d$ | 2 | 2.5 | Pseudopotential | — |
| | | $5p$ | 1 | 0.5 | Orthogonalization | 4.000 |
| | | $5p$ | 1 | 1.5 | Orthogonalization | 3.900 |
| | | $4s$ | 0 | 0.5 | Pseudopotential | — |
| | | $5s$ | 0 | 0.5 | Pseudopotential | — |
| | | $4p$ | 1 | 0.5 | Pseudopotential | — |
| 42 | Mo | $4p$ | 1 | 1.5 | Pseudopotential | — |
| | | $4d$ | 2 | 1.5 | Pseudopotential | — |
| | | $4d$ | 2 | 2.5 | Pseudopotential | — |
| | | $5p$ | 1 | 0.5 | Orthogonalization | 4.300 |
| | | $5p$ | 1 | 1.5 | Orthogonalization | 4.200 |
| | | $4s$ | 0 | 0.5 | Pseudopotential | — |
| | | $5s$ | 0 | 0.5 | Pseudopotential | — |
| | | $4p$ | 1 | 0.5 | Pseudopotential | — |
| 43 | Tc | $4p$ | 1 | 1.5 | Pseudopotential | — |
| | | $4d$ | 2 | 1.5 | Pseudopotential | — |
| | | $4d$ | 2 | 2.5 | Pseudopotential | — |
| | | $5p$ | 1 | 0.5 | Orthogonalization | 4.500 |
| | | $5p$ | 1 | 1.5 | Orthogonalization | 4.400 |
| | | $4s$ | 0 | 0.5 | Pseudopotential | — |
| | | $5s$ | 0 | 0.5 | Pseudopotential | — |
| | | $4p$ | 1 | 0.5 | Pseudopotential | — |
| 44 | Ru | $4p$ | 1 | 1.5 | Pseudopotential | — |
| | | $4d$ | 2 | 1.5 | Pseudopotential | — |
| | | $4d$ | 2 | 2.5 | Pseudopotential | — |
| | | $5p$ | 1 | 0.5 | Orthogonalization | 4.700 |
| | | $5p$ | 1 | 1.5 | Orthogonalization | 4.600 |

| $Z$ | Element | Orbital | $l$ | $j$ | Source | $\alpha(\text{bohr}^{-1})$ |
|---|---|---|---|---|---|---|
| | | 4s | 0 | 0.5 | Pseudopotential | — |
| | | 5s | 0 | 0.5 | Pseudopotential | — |
| | | 4p | 1 | 0.5 | Pseudopotential | — |
| 45 | Rh | 4p | 1 | 1.5 | Pseudopotential | — |
| | | 4d | 2 | 1.5 | Pseudopotential | — |
| | | 4d | 2 | 2.5 | Pseudopotential | — |
| | | 5p | 1 | 0.5 | Orthogonalization | 4.900 |
| | | 5p | 1 | 1.5 | Orthogonalization | 4.800 |
| | | 4s | 0 | 0.5 | Pseudopotential | — |
| | | 4p | 1 | 0.5 | Pseudopotential | — |
| | | 4p | 1 | 1.5 | Pseudopotential | — |
| 46 | Pd | 4d | 2 | 1.5 | Pseudopotential | — |
| | | 4d | 2 | 2.5 | Pseudopotential | — |
| | | 5s | 0 | 0.5 | Orthogonalization | 1.900 |
| | | 5p | 1 | 0.5 | Orthogonalization | 5.100 |
| | | 5p | 1 | 1.5 | Orthogonalization | 4.900 |
| | | 4s | 0 | 0.5 | Pseudopotential | — |
| | | 5s | 0 | 0.5 | Pseudopotential | — |
| | | 4p | 1 | 0.5 | Pseudopotential | — |
| 47 | Ag | 4p | 1 | 1.5 | Pseudopotential | — |
| | | 4d | 2 | 1.5 | Pseudopotential | — |
| | | 4d | 2 | 2.5 | Pseudopotential | — |
| | | 5p | 1 | 0.5 | Orthogonalization | 5.400 |
| | | 5p | 1 | 1.5 | Orthogonalization | 5.200 |
| | | 4s | 0 | 0.5 | Pseudopotential | — |
| | | 5s | 0 | 0.5 | Pseudopotential | — |
| | | 4p | 1 | 0.5 | Pseudopotential | — |
| 48 | Cd | 4p | 1 | 1.5 | Pseudopotential | — |
| | | 4d | 2 | 1.5 | Pseudopotential | — |
| | | 4d | 2 | 2.5 | Pseudopotential | — |
| | | 5p | 1 | 0.5 | Orthogonalization | 5.600 |
| | | 5p | 1 | 1.5 | Orthogonalization | 5.400 |
| | | 5s | 0 | 0.5 | Pseudopotential | — |
| | | 5p | 1 | 0.5 | Pseudopotential | — |
| 49 | In | 5p | 1 | 1.5 | Pseudopotential | — |
| | | 4d | 2 | 1.5 | Pseudopotential | — |
| | | 4d | 2 | 2.5 | Pseudopotential | — |
| | | 5s | 0 | 0.5 | Pseudopotential | — |
| | | 5p | 1 | 0.5 | Pseudopotential | — |
| 50 | Sn | 5p | 1 | 1.5 | Pseudopotential | — |
| | | 4d | 2 | 1.5 | Pseudopotential | — |
| | | 4d | 2 | 2.5 | Pseudopotential | — |
| | | 5s | 0 | 0.5 | Pseudopotential | — |
| | | 5p | 1 | 0.5 | Pseudopotential | — |
| 51 | Sb | 5p | 1 | 1.5 | Pseudopotential | — |
| | | 4d | 2 | 1.5 | Pseudopotential | — |
| | | 4d | 2 | 2.5 | Pseudopotential | — |
| | | 5s | 0 | 0.5 | Pseudopotential | — |
| | | 5p | 1 | 0.5 | Pseudopotential | — |
| 52 | Te | 5p | 1 | 1.5 | Pseudopotential | — |
| | | 4d | 2 | 1.5 | Pseudopotential | — |
| | | 4d | 2 | 2.5 | Pseudopotential | — |
| | | 5s | 0 | 0.5 | Pseudopotential | — |
| 53 | I | 5p | 1 | 0.5 | Pseudopotential | — |
| | | 5p | 1 | 1.5 | Pseudopotential | — |
| | | 5s | 0 | 0.5 | Pseudopotential | — |
| 54 | Xe | 5p | 1 | 0.5 | Pseudopotential | — |
| | | 5p | 1 | 1.5 | Pseudopotential | — |

| $Z$ | Element | Orbital | $l$ | $j$ | Source | $\alpha(\text{bohr}^{-1})$ |
|---|---|---|---|---|---|---|
| | | 5s | 0 | 0.5 | Pseudopotential | — |
| | | 6s | 0 | 0.5 | Pseudopotential | — |
| 55 | Cs | 5p | 1 | 0.5 | Pseudopotential | — |
| | | 5p | 1 | 1.5 | Pseudopotential | — |
| | | 5d | 2 | 1.5 | OpenMX Fitting | 3.566 |
| | | 5d | 2 | 2.5 | OpenMX Fitting | 3.566 |
| | | 5s | 0 | 0.5 | Pseudopotential | — |
| | | 6s | 0 | 0.5 | Pseudopotential | — |
| 56 | Ba | 5p | 1 | 0.5 | Pseudopotential | — |
| | | 5p | 1 | 1.5 | Pseudopotential | — |
| | | 5d | 2 | 1.5 | OpenMX Fitting | 3.039 |
| | | 5d | 2 | 2.5 | OpenMX Fitting | 3.039 |
| | | 5s | 0 | 0.5 | Pseudopotential | — |
| | | 6s | 0 | 0.5 | Pseudopotential | — |
| | | 5p | 1 | 0.5 | Pseudopotential | — |
| 72 | Hf | 5p | 1 | 1.5 | Pseudopotential | — |
| | | 5d | 2 | 1.5 | Pseudopotential | — |
| | | 5d | 2 | 2.5 | Pseudopotential | — |
| | | 6p | 1 | 0.5 | Orthogonalization | 4.200 |
| | | 6p | 1 | 1.5 | Orthogonalization | 3.800 |
| | | 5s | 0 | 0.5 | Pseudopotential | — |
| | | 6s | 0 | 0.5 | Pseudopotential | — |
| | | 5p | 1 | 0.5 | Pseudopotential | — |
| 73 | Ta | 5p | 1 | 1.5 | Pseudopotential | — |
| | | 5d | 2 | 1.5 | Pseudopotential | — |
| | | 5d | 2 | 2.5 | Pseudopotential | — |
| | | 6p | 1 | 0.5 | Orthogonalization | 4.300 |
| | | 6p | 1 | 1.5 | Orthogonalization | 4.000 |
| | | 5s | 0 | 0.5 | Pseudopotential | — |
| | | 6s | 0 | 0.5 | Pseudopotential | — |
| | | 5p | 1 | 0.5 | Pseudopotential | — |
| 74 | W | 5p | 1 | 1.5 | Pseudopotential | — |
| | | 5d | 2 | 1.5 | Pseudopotential | — |
| | | 5d | 2 | 2.5 | Pseudopotential | — |
| | | 6p | 1 | 0.5 | Orthogonalization | 4.500 |
| | | 6p | 1 | 1.5 | Orthogonalization | 4.100 |
| | | 5s | 0 | 0.5 | Pseudopotential | — |
| | | 6s | 0 | 0.5 | Pseudopotential | — |
| | | 5p | 1 | 0.5 | Pseudopotential | — |
| 75 | Re | 5p | 1 | 1.5 | Pseudopotential | — |
| | | 5d | 2 | 1.5 | Pseudopotential | — |
| | | 5d | 2 | 2.5 | Pseudopotential | — |
| | | 6p | 1 | 0.5 | Orthogonalization | 4.700 |
| | | 6p | 1 | 1.5 | Orthogonalization | 4.300 |
| | | 5s | 0 | 0.5 | Pseudopotential | — |
| | | 6s | 0 | 0.5 | Pseudopotential | — |
| | | 5p | 1 | 0.5 | Pseudopotential | — |
| 76 | Os | 5p | 1 | 1.5 | Pseudopotential | — |
| | | 5d | 2 | 1.5 | Pseudopotential | — |
| | | 5d | 2 | 2.5 | Pseudopotential | — |
| | | 6p | 1 | 0.5 | Orthogonalization | 4.800 |
| | | 6p | 1 | 1.5 | Orthogonalization | 4.400 |
| | | 5s | 0 | 0.5 | Pseudopotential | — |
| | | 6s | 0 | 0.5 | Pseudopotential | — |
| | | 5p | 1 | 0.5 | Pseudopotential | — |
| 77 | Ir | 5p | 1 | 1.5 | Pseudopotential | — |
| | | 5d | 2 | 1.5 | Pseudopotential | — |
| | | 5d | 2 | 2.5 | Pseudopotential | — |
| | | 6p | 1 | 0.5 | Orthogonalization | 5.000 |
| | | 6p | 1 | 1.5 | Orthogonalization | 4.600 |

| Z | Element | Orbital | l | j | Source | $\alpha(\text{bohr}^{-1})$ |
|---|---|---|---|---|---|---|
|  |  | 5s | 0 | 0.5 | Pseudopotential | — |
|  |  | 6s | 0 | 0.5 | Pseudopotential | — |
|  |  | 5p | 1 | 0.5 | Pseudopotential | — |
| 78 | Pt | 5p | 1 | 1.5 | Pseudopotential | — |
|  |  | 5d | 2 | 1.5 | Pseudopotential | — |
|  |  | 5d | 2 | 2.5 | Pseudopotential | — |
|  |  | 6p | 1 | 0.5 | Orthogonalization | 5.100 |
|  |  | 6p | 1 | 1.5 | Orthogonalization | 4.600 |
|  |  | 5s | 0 | 0.5 | Pseudopotential | — |
|  |  | 6s | 0 | 0.5 | Pseudopotential | — |
|  |  | 5p | 1 | 0.5 | Pseudopotential | — |
| 79 | Au | 5p | 1 | 1.5 | Pseudopotential | — |
|  |  | 5d | 2 | 1.5 | Pseudopotential | — |
|  |  | 5d | 2 | 2.5 | Pseudopotential | — |
|  |  | 6p | 1 | 0.5 | Orthogonalization | 5.300 |
|  |  | 6p | 1 | 1.5 | Orthogonalization | 4.800 |
|  |  | 5s | 0 | 0.5 | Pseudopotential | — |
|  |  | 6s | 0 | 0.5 | Pseudopotential | — |
|  |  | 5p | 1 | 0.5 | Pseudopotential | — |
| 80 | Hg | 5p | 1 | 1.5 | Pseudopotential | — |
|  |  | 5d | 2 | 1.5 | Pseudopotential | — |
|  |  | 5d | 2 | 2.5 | Pseudopotential | — |
|  |  | 6p | 1 | 0.5 | Orthogonalization | 5.500 |
|  |  | 6p | 1 | 1.5 | Orthogonalization | 5.000 |
|  |  | 6s | 0 | 0.5 | Pseudopotential | — |
|  |  | 6p | 1 | 0.5 | Pseudopotential | — |
| 81 | Tl | 6p | 1 | 1.5 | Pseudopotential | — |
|  |  | 5d | 2 | 1.5 | Pseudopotential | — |
|  |  | 5d | 2 | 2.5 | Pseudopotential | — |
|  |  | 6s | 0 | 0.5 | Pseudopotential | — |
|  |  | 6p | 1 | 0.5 | Pseudopotential | — |
| 82 | Pb | 6p | 1 | 1.5 | Pseudopotential | — |
|  |  | 5d | 2 | 1.5 | Pseudopotential | — |
|  |  | 5d | 2 | 2.5 | Pseudopotential | — |
|  |  | 6s | 0 | 0.5 | Pseudopotential | — |
|  |  | 6p | 1 | 0.5 | Pseudopotential | — |
| 83 | Bi | 6p | 1 | 1.5 | Pseudopotential | — |
|  |  | 5d | 2 | 1.5 | Pseudopotential | — |
|  |  | 5d | 2 | 2.5 | Pseudopotential | — |
|  |  | 6s | 0 | 0.5 | Pseudopotential | — |
|  |  | 6p | 1 | 0.5 | Pseudopotential | — |
| 84 | Po | 6p | 1 | 1.5 | Pseudopotential | — |
|  |  | 5d | 2 | 1.5 | Pseudopotential | — |
|  |  | 5d | 2 | 2.5 | Pseudopotential | — |
|  |  | 6s | 0 | 0.5 | Pseudopotential | — |
|  |  | 6p | 1 | 0.5 | Pseudopotential | — |
| 86 | Rn | 6p | 1 | 1.5 | Pseudopotential | — |
|  |  | 5d | 2 | 1.5 | Pseudopotential | — |
|  |  | 5d | 2 | 2.5 | Pseudopotential | — |

SUPPLEMENTARY TABLE VI: Source of the pseudopotentials in the `modified-pslibrary` set. We started by selecting pseudopotentials based on the suggested list [4] from pslibrary. However, during DFT calculations, we encountered convergence errors with certain pseudopotentials. Thus, some pseudopotentials were replaced with either older versions from pslibrary or alternatives from pseudoDojo, as reported in this table.

| Z | Elem. | Source | Filename |
|---|---|---|---|
| 1 | H | pslibrary | H.rel-pbe-kjpaw_psl.1.0.0.UPF |
| 2 | He | pslibrary | He.rel-pbe-kjpaw_psl.1.0.0.UPF |
| 3 | Li | pslibrary | Li.rel-pbe-sl-kjpaw_psl.1.0.0.UPF |
| 4 | Be | pslibrary | Be.rel-pbe-sl-kjpaw_psl.1.0.0.UPF |
| 5 | B | pslibrary | B.rel-pbe-n-kjpaw_psl.1.0.0.UPF |
| 6 | C | pslibrary | C.rel-pbe-n-kjpaw_psl.1.0.0.UPF |
| 7 | N | pslibrary | N.rel-pbe-n-kjpaw_psl.1.0.0.UPF |
| 8 | O | pslibrary | O.rel-pbe-n-kjpaw_psl.0.1.UPF |
| 9 | F | pslibrary | F.rel-pbe-n-kjpaw_psl.1.0.0.UPF |
| 10 | Ne | pslibrary | Ne.rel-pbe-n-kjpaw_psl.1.0.0.UPF |
| 11 | Na | pslibrary | Na.rel-pbe-spnl-kjpaw_psl.1.0.0.UPF |
| 12 | Mg | pslibrary | Mg.rel-pbe-spnl-kjpaw_psl.1.0.0.UPF |
| 13 | Al | pslibrary | Al.rel-pbe-nl-kjpaw_psl.1.0.0.UPF |
| 14 | Si | pslibrary | Si.rel-pbe-nl-kjpaw_psl.1.0.0.UPF |
| 15 | P | pslibrary | P.rel-pbe-nl-kjpaw_psl.1.0.0.UPF |
| 16 | S | pslibrary | S.rel-pbe-nl-kjpaw_psl.1.0.0.UPF |
| 17 | Cl | pslibrary | Cl.rel-pbe-nl-kjpaw_psl.1.0.0.UPF |
| 18 | Ar | pslibrary | Ar.rel-pbe-nl-kjpaw_psl.1.0.0.UPF |
| 19 | K | pslibrary | K.rel-pbe-spn-kjpaw_psl.1.0.0.UPF |
| 20 | Ca | pslibrary | Ca.rel-pbe-spn-kjpaw_psl.1.0.0.UPF |
| 21 | Sc | pslibrary | Sc.rel-pbe-spn-kjpaw_psl.1.0.0.UPF |
| 22 | Ti | pslibrary | Ti.rel-pbe-spn-kjpaw_psl.1.0.0.UPF |
| 23 | V | pslibrary | V.rel-pbe-spnl-kjpaw_psl.1.0.0.UPF |
| 24 | Cr | pslibrary | Cr.rel-pbe-spn-kjpaw_psl.1.0.0.UPF |
| 25 | Mn | pslibrary | Mn.rel-pbe-spn-kjpaw_psl.0.3.1.UPF |
| 26 | Fe | pslibrary | Fe.rel-pbe-n-kjpaw_psl.0.2.4.UPF |
| 27 | Co | pslibrary | Co.rel-pbe-spn-kjpaw_psl.0.3.1.UPF |
| 28 | Ni | pslibrary | Ni.rel-pbe-n-kjpaw_psl.0.1.UPF |
| 29 | Cu | pslibrary | Cu.rel-pbe-dn-kjpaw_psl.1.0.0.UPF |
| 30 | Zn | pslibrary | Zn.rel-pbe-dn-kjpaw_psl.1.0.0.UPF |
| 31 | Ga | pslibrary | Ga.rel-pbe-dnl-kjpaw_psl.1.0.0.UPF |
| 32 | Ge | pslibrary | Ge.rel-pbe-n-kjpaw_psl.1.0.0.UPF |
| 33 | As | pslibrary | As.rel-pbe-n-kjpaw_psl.1.0.0.UPF |
| 34 | Se | pslibrary | Se.rel-pbe-n-kjpaw_psl.1.0.0.UPF |
| 35 | Br | pslibrary | Br.rel-pbe-n-kjpaw_psl.1.0.0.UPF |
| 36 | Kr | pslibrary | Kr.rel-pbe-dn-kjpaw_psl.1.0.0.UPF |
| 37 | Rb | pslibrary | Rb.rel-pbe-spn-kjpaw_psl.1.0.0.UPF |
| 38 | Sr | pslibrary | Sr.rel-pbe-spn-kjpaw_psl.1.0.0.UPF |
| 39 | Y | pslibrary | Y.rel-pbe-spn-kjpaw_psl.1.0.0.UPF |
| 40 | Zr | pslibrary | Zr.rel-pbe-spn-kjpaw_psl.1.0.0.UPF |
| 41 | Nb | pslibrary | Nb.rel-pbe-spn-kjpaw_psl.1.0.0.UPF |
| 42 | Mo | pslibrary | Mo.rel-pbe-spn-kjpaw_psl.1.0.0.UPF |
| 43 | Tc | pslibrary | Tc.rel-pbe-spn-kjpaw_psl.0.3.0.UPF |
| 44 | Ru | pslibrary | Ru.rel-pbe-spn-kjpaw_psl.1.0.0.UPF |
| 45 | Rh | pseudoDojo | Rh.UPF |
| 46 | Pd | pslibrary | Pd.rel-pbe-n-kjpaw_psl.1.0.0.UPF |
| 47 | Ag | pslibrary | Ag.rel-pbe-n-kjpaw_psl.1.0.0.UPF |
| 48 | Cd | pslibrary | Cd.rel-pbe-dn-kjpaw_psl.0.3.1.UPF |
| 49 | In | pslibrary | In.rel-pbe-dn-kjpaw_psl.1.0.0.UPF |
| 50 | Sn | pslibrary | Sn.rel-pbe-dn-kjpaw_psl.1.0.0.UPF |
| 51 | Sb | pslibrary | Sb.rel-pbe-n-kjpaw_psl.1.0.0.UPF |
| 52 | Te | pslibrary | Te.rel-pbe-n-kjpaw_psl.1.0.0.UPF |
| 53 | I | pslibrary | I.rel-pbe-n-kjpaw_psl.1.0.0.UPF |
| 54 | Xe | pslibrary | Xe.rel-pbe-dn-kjpaw_psl.1.0.0.UPF |
| 55 | Cs | pseudoDojo | Cs.UPF |

| Z | Elem. | Source | Filename |
|---|---|---|---|
| 56 | Ba | pslibrary | Ba.rel-pbe-spn-kjpaw_psl.1.0.0.UPF |
| 72 | Hf | pslibrary | Hf.rel-pbe-spn-kjpaw_psl.1.0.0.UPF |
| 73 | Ta | pslibrary | Ta.rel-pbe-spn-kjpaw_psl.0.2.UPF |
| 74 | W | pslibrary | W.rel-pbe-spn-kjpaw_psl.1.0.1.UPF |
| 75 | Re | pslibrary | Re.rel-pbe-spn-kjpaw_psl.1.0.0.UPF |
| 76 | Os | pslibrary | Os.rel-pbe-spn-kjpaw_psl.1.0.0.UPF |
| 77 | Ir | pslibrary | Ir.rel-pbe-n-kjpaw_psl.0.2.3.UPF |
| 78 | Pt | pslibrary | Pt.rel-pbe-n-kjpaw_psl.1.0.0.UPF |
| 79 | Au | pslibrary | Au.rel-pbe-n-kjpaw_psl.1.0.0.UPF |
| 80 | Hg | pslibrary | Hg.rel-pbe-n-kjpaw_psl.1.0.0.UPF |
| 81 | Tl | pslibrary | Tl.rel-pbe-dn-kjpaw_psl.1.0.0.UPF |
| 82 | Pb | pslibrary | Pb.rel-pbe-dn-kjpaw_psl.0.2.2.UPF |
| 83 | Bi | pslibrary | Bi.rel-pbe-dn-kjpaw_psl.1.0.0.UPF |
| 84 | Po | pslibrary | Po.rel-pbe-dn-kjpaw_psl.1.0.0.UPF |
| 85 | At | pslibrary | At.rel-pbe-dn-kjpaw_psl.1.0.0.UPF |
| 86 | Rn | pslibrary | Rn.rel-pbe-dn-kjpaw_psl.1.0.0.UPF |
| 87 | Fr | pslibrary | Fr.rel-pbe-spdn-kjpaw_psl.1.0.0.UPF |
| 88 | Ra | pslibrary | Ra.rel-pbe-spdn-kjpaw_psl.1.0.0.UPF |

SUPPLEMENTARY TABLE VII: Values of $\alpha$ and minimal set of orbitals for the pseudopotentials of the `modified-pslibrary` set with orthonormalization fitting.

| Z | Element | Orbital | $l$ | $j$ | Source | $\alpha(\text{bohr}^{-1})$ |
|---|---|---|---|---|---|---|
| 1 | H | 1s | 0 | 0.5 | Pseudopotential | — |
| 2 | He | 1s | 0 | 0.5 | Pseudopotential | — |
| 3 | Li | 1s | 0 | 0.5 | Pseudopotential | — |
| | | 2s | 0 | 0.5 | Pseudopotential | — |
| | | 2p | 1 | 0.5 | Pseudopotential | — |
| | | 2p | 1 | 1.5 | Pseudopotential | — |
| 4 | Be | 1s | 0 | 0.5 | Pseudopotential | — |
| | | 2s | 0 | 0.5 | Pseudopotential | — |
| | | 2p | 1 | 0.5 | Pseudopotential | — |
| | | 2p | 1 | 1.5 | Pseudopotential | — |
| 5 | B | 2s | 0 | 0.5 | Pseudopotential | — |
| | | 2p | 1 | 0.5 | Pseudopotential | — |
| | | 2p | 1 | 1.5 | Pseudopotential | — |
| 6 | C | 2s | 0 | 0.5 | Pseudopotential | — |
| | | 2p | 1 | 0.5 | Pseudopotential | — |
| | | 2p | 1 | 1.5 | Pseudopotential | — |
| 7 | N | 2s | 0 | 0.5 | Pseudopotential | — |
| | | 2p | 1 | 0.5 | Pseudopotential | — |
| | | 2p | 1 | 1.5 | Pseudopotential | — |
| 8 | O | 2s | 0 | 0.5 | Pseudopotential | — |
| | | 2p | 1 | 0.5 | Pseudopotential | — |
| | | 2p | 1 | 1.5 | Pseudopotential | — |
| 9 | F | 2s | 0 | 0.5 | Pseudopotential | — |
| | | 2p | 1 | 0.5 | Pseudopotential | — |
| | | 2p | 1 | 1.5 | Pseudopotential | — |
| 10 | Ne | 2s | 0 | 0.5 | Pseudopotential | — |
| | | 2p | 1 | 0.5 | Pseudopotential | — |
| | | 2p | 1 | 1.5 | Pseudopotential | — |
| 11 | Na | 2s | 0 | 0.5 | Pseudopotential | — |
| | | 3s | 0 | 0.5 | Pseudopotential | — |
| | | 2p | 1 | 0.5 | Pseudopotential | — |
| | | 2p | 1 | 1.5 | Pseudopotential | — |
| | | 3p | 1 | 0.5 | Orthogonalization | 4.900 |
| | | 3p | 1 | 1.5 | Orthogonalization | 4.900 |
| 12 | Mg | 2s | 0 | 0.5 | Pseudopotential | — |
| | | 3s | 0 | 0.5 | Pseudopotential | — |
| | | 2p | 1 | 0.5 | Pseudopotential | — |
| | | 2p | 1 | 1.5 | Pseudopotential | — |
| | | 3p | 1 | 0.5 | Orthogonalization | 5.500 |
| | | 3p | 1 | 1.5 | Orthogonalization | 5.500 |
| 13 | Al | 3s | 0 | 0.5 | Pseudopotential | — |
| | | 3p | 1 | 0.5 | Pseudopotential | — |
| | | 3p | 1 | 1.5 | Pseudopotential | — |
| 14 | Si | 3s | 0 | 0.5 | Pseudopotential | — |
| | | 3p | 1 | 0.5 | Pseudopotential | — |
| | | 3p | 1 | 1.5 | Pseudopotential | — |
| 15 | P | 3s | 0 | 0.5 | Pseudopotential | — |
| | | 3p | 1 | 0.5 | Pseudopotential | — |
| | | 3p | 1 | 1.5 | Pseudopotential | — |
| 16 | S | 3s | 0 | 0.5 | Pseudopotential | — |
| | | 3p | 1 | 0.5 | Pseudopotential | — |
| | | 3p | 1 | 1.5 | Pseudopotential | — |
| 17 | Cl | 3s | 0 | 0.5 | Pseudopotential | — |
| | | 3p | 1 | 0.5 | Pseudopotential | — |
| | | 3p | 1 | 1.5 | Pseudopotential | — |
| 18 | Ar | 3s | 0 | 0.5 | Pseudopotential | — |
| | | 3p | 1 | 0.5 | Pseudopotential | — |
| | | 3p | 1 | 1.5 | Pseudopotential | — |

| Z | Element | Orbital | l | j | Source | $\alpha(\text{bohr}^{-1})$ |
|---|---|---|---|---|---|---|
| 19 | K | $3s$ | 0 | 0.5 | Pseudopotential | — |
| | | $4s$ | 0 | 0.5 | Pseudopotential | — |
| | | $3p$ | 1 | 0.5 | Pseudopotential | — |
| | | $4p$ | 1 | 0.5 | Pseudopotential | — |
| | | $3p$ | 1 | 1.5 | Pseudopotential | — |
| | | $4p$ | 1 | 1.5 | Pseudopotential | — |
| | | $3d$ | 2 | 1.5 | OpenMX Fitting | 2.727 |
| | | $3d$ | 2 | 2.5 | OpenMX Fitting | 2.727 |
| 20 | Ca | $3s$ | 0 | 0.5 | Pseudopotential | — |
| | | $4s$ | 0 | 0.5 | Pseudopotential | — |
| | | $3p$ | 1 | 0.5 | Pseudopotential | — |
| | | $4p$ | 1 | 0.5 | Pseudopotential | — |
| | | $3p$ | 1 | 1.5 | Pseudopotential | — |
| | | $4p$ | 1 | 1.5 | Pseudopotential | — |
| | | $3d$ | 2 | 1.5 | OpenMX Fitting | 3.983 |
| | | $3d$ | 2 | 2.5 | OpenMX Fitting | 3.983 |
| 21 | Sc | $3s$ | 0 | 0.5 | Pseudopotential | — |
| | | $4s$ | 0 | 0.5 | Pseudopotential | — |
| | | $3p$ | 1 | 0.5 | Pseudopotential | — |
| | | $3p$ | 1 | 1.5 | Pseudopotential | — |
| | | $3d$ | 2 | 1.5 | Pseudopotential | — |
| | | $3d$ | 2 | 2.5 | Pseudopotential | — |
| | | $4p$ | 1 | 0.5 | Orthogonalization | 4.400 |
| | | $4p$ | 1 | 1.5 | Orthogonalization | 4.400 |
| 22 | Ti | $3s$ | 0 | 0.5 | Pseudopotential | — |
| | | $4s$ | 0 | 0.5 | Pseudopotential | — |
| | | $3p$ | 1 | 0.5 | Pseudopotential | — |
| | | $3p$ | 1 | 1.5 | Pseudopotential | — |
| | | $3d$ | 2 | 1.5 | Pseudopotential | — |
| | | $3d$ | 2 | 2.5 | Pseudopotential | — |
| | | $4p$ | 1 | 0.5 | Orthogonalization | 4.600 |
| | | $4p$ | 1 | 1.5 | Orthogonalization | 4.500 |
| 23 | V | $3s$ | 0 | 0.5 | Pseudopotential | — |
| | | $4s$ | 0 | 0.5 | Pseudopotential | — |
| | | $3p$ | 1 | 0.5 | Pseudopotential | — |
| | | $3p$ | 1 | 1.5 | Pseudopotential | — |
| | | $3d$ | 2 | 1.5 | Pseudopotential | — |
| | | $3d$ | 2 | 2.5 | Pseudopotential | — |
| | | $4p$ | 1 | 0.5 | Orthogonalization | 5.000 |
| | | $4p$ | 1 | 1.5 | Orthogonalization | 4.900 |
| 24 | Cr | $3s$ | 0 | 0.5 | Pseudopotential | — |
| | | $4s$ | 0 | 0.5 | Pseudopotential | — |
| | | $3p$ | 1 | 0.5 | Pseudopotential | — |
| | | $3p$ | 1 | 1.5 | Pseudopotential | — |
| | | $3d$ | 2 | 1.5 | Pseudopotential | — |
| | | $3d$ | 2 | 2.5 | Pseudopotential | — |
| | | $4p$ | 1 | 0.5 | Orthogonalization | 5.200 |
| | | $4p$ | 1 | 1.5 | Orthogonalization | 5.200 |
| 25 | Mn | $3s$ | 0 | 0.5 | Pseudopotential | — |
| | | $4s$ | 0 | 0.5 | Pseudopotential | — |
| | | $3p$ | 1 | 0.5 | Pseudopotential | — |
| | | $3p$ | 1 | 1.5 | Pseudopotential | — |
| | | $3d$ | 2 | 1.5 | Pseudopotential | — |
| | | $3d$ | 2 | 2.5 | Pseudopotential | — |
| | | $4p$ | 1 | 0.5 | Orthogonalization | 4.700 |
| | | $4p$ | 1 | 1.5 | Orthogonalization | 4.700 |
| 26 | Fe | $4s$ | 0 | 0.5 | Pseudopotential | — |
| | | $4p$ | 1 | 0.5 | Pseudopotential | — |
| | | $4p$ | 1 | 1.5 | Pseudopotential | — |
| | | $3d$ | 2 | 1.5 | Pseudopotential | — |
| | | $3d$ | 2 | 2.5 | Pseudopotential | — |

| Z | Element | Orbital | l | j | Source | $\alpha(\text{bohr}^{-1})$ |
|---|---|---|---|---|---|---|
| 27 | Co | $3s$ | 0 | 0.5 | Pseudopotential | — |
| | | $4s$ | 0 | 0.5 | Pseudopotential | — |
| | | $3p$ | 1 | 0.5 | Pseudopotential | — |
| | | $3p$ | 1 | 1.5 | Pseudopotential | — |
| | | $3d$ | 2 | 1.5 | Pseudopotential | — |
| | | $3d$ | 2 | 2.5 | Pseudopotential | — |
| | | $4p$ | 1 | 0.5 | Orthogonalization | 5.900 |
| | | $4p$ | 1 | 1.5 | Orthogonalization | 5.900 |
| 28 | Ni | $4s$ | 0 | 0.5 | Pseudopotential | — |
| | | $4p$ | 1 | 0.5 | Pseudopotential | — |
| | | $4p$ | 1 | 1.5 | Pseudopotential | — |
| | | $3d$ | 2 | 1.5 | Pseudopotential | — |
| | | $3d$ | 2 | 2.5 | Pseudopotential | — |
| 29 | Cu | $4s$ | 0 | 0.5 | Pseudopotential | — |
| | | $4p$ | 1 | 0.5 | Pseudopotential | — |
| | | $4p$ | 1 | 1.5 | Pseudopotential | — |
| | | $3d$ | 2 | 1.5 | Pseudopotential | — |
| | | $3d$ | 2 | 2.5 | Pseudopotential | — |
| 30 | Zn | $4s$ | 0 | 0.5 | Pseudopotential | — |
| | | $4p$ | 1 | 0.5 | Pseudopotential | — |
| | | $4p$ | 1 | 1.5 | Pseudopotential | — |
| | | $3d$ | 2 | 1.5 | Pseudopotential | — |
| | | $3d$ | 2 | 2.5 | Pseudopotential | — |
| 31 | Ga | $4s$ | 0 | 0.5 | Pseudopotential | — |
| | | $4p$ | 1 | 0.5 | Pseudopotential | — |
| | | $4p$ | 1 | 1.5 | Pseudopotential | — |
| | | $3d$ | 2 | 1.5 | Pseudopotential | — |
| | | $3d$ | 2 | 2.5 | Pseudopotential | — |
| 32 | Ge | $4s$ | 0 | 0.5 | Pseudopotential | — |
| | | $4p$ | 1 | 0.5 | Pseudopotential | — |
| | | $4p$ | 1 | 1.5 | Pseudopotential | — |
| 33 | As | $4s$ | 0 | 0.5 | Pseudopotential | — |
| | | $4p$ | 1 | 0.5 | Pseudopotential | — |
| | | $4p$ | 1 | 1.5 | Pseudopotential | — |
| 34 | Se | $4s$ | 0 | 0.5 | Pseudopotential | — |
| | | $4p$ | 1 | 0.5 | Pseudopotential | — |
| | | $4p$ | 1 | 1.5 | Pseudopotential | — |
| 35 | Br | $4s$ | 0 | 0.5 | Pseudopotential | — |
| | | $4p$ | 1 | 0.5 | Pseudopotential | — |
| | | $4p$ | 1 | 1.5 | Pseudopotential | — |
| 36 | Kr | $4s$ | 0 | 0.5 | Pseudopotential | — |
| | | $4p$ | 1 | 0.5 | Pseudopotential | — |
| | | $4p$ | 1 | 1.5 | Pseudopotential | — |
| | | $3d$ | 2 | 1.5 | Pseudopotential | — |
| | | $3d$ | 2 | 2.5 | Pseudopotential | — |
| 37 | Rb | $4s$ | 0 | 0.5 | Pseudopotential | — |
| | | $5s$ | 0 | 0.5 | Pseudopotential | — |
| | | $4p$ | 1 | 0.5 | Pseudopotential | — |
| | | $5p$ | 1 | 0.5 | Pseudopotential | — |
| | | $4p$ | 1 | 1.5 | Pseudopotential | — |
| | | $5p$ | 1 | 1.5 | Pseudopotential | — |
| | | $4d$ | 2 | 1.5 | OpenMX Fitting | 2.197 |
| | | $4d$ | 2 | 2.5 | OpenMX Fitting | 2.197 |
| 38 | Sr | $4s$ | 0 | 0.5 | Pseudopotential | — |
| | | $5s$ | 0 | 0.5 | Pseudopotential | — |
| | | $4p$ | 1 | 0.5 | Pseudopotential | — |
| | | $5p$ | 1 | 0.5 | Pseudopotential | — |
| | | $4p$ | 1 | 1.5 | Pseudopotential | — |
| | | $5p$ | 1 | 1.5 | Pseudopotential | — |
| | | $4d$ | 2 | 1.5 | OpenMX Fitting | 4.400 |
| | | $4d$ | 2 | 2.5 | OpenMX Fitting | 4.400 |

| Z | Element | Orbital | l | j | Source | α(bohr⁻¹) |
|---|---------|---------|---|---|--------|-----------|
| 39 | Y | 4s | 0 | 0.5 | Pseudopotential | — |
| | | 5s | 0 | 0.5 | Pseudopotential | — |
| | | 4p | 1 | 0.5 | Pseudopotential | — |
| | | 5p | 1 | 0.5 | Pseudopotential | — |
| | | 4p | 1 | 1.5 | Pseudopotential | — |
| | | 5p | 1 | 1.5 | Pseudopotential | — |
| | | 4d | 2 | 1.5 | Pseudopotential | — |
| | | 4d | 2 | 2.5 | Pseudopotential | — |
| 40 | Zr | 4s | 0 | 0.5 | Pseudopotential | — |
| | | 5s | 0 | 0.5 | Pseudopotential | — |
| | | 4p | 1 | 0.5 | Pseudopotential | — |
| | | 5p | 1 | 0.5 | Pseudopotential | — |
| | | 4p | 1 | 1.5 | Pseudopotential | — |
| | | 5p | 1 | 1.5 | Pseudopotential | — |
| | | 4d | 2 | 1.5 | Pseudopotential | — |
| | | 4d | 2 | 2.5 | Pseudopotential | — |
| 41 | Nb | 4s | 0 | 0.5 | Pseudopotential | — |
| | | 5s | 0 | 0.5 | Pseudopotential | — |
| | | 4p | 1 | 0.5 | Pseudopotential | — |
| | | 5p | 1 | 0.5 | Pseudopotential | — |
| | | 4p | 1 | 1.5 | Pseudopotential | — |
| | | 5p | 1 | 1.5 | Pseudopotential | — |
| | | 4d | 2 | 1.5 | Pseudopotential | — |
| | | 4d | 2 | 2.5 | Pseudopotential | — |
| 42 | Mo | 4s | 0 | 0.5 | Pseudopotential | — |
| | | 5s | 0 | 0.5 | Pseudopotential | — |
| | | 4p | 1 | 0.5 | Pseudopotential | — |
| | | 4p | 1 | 1.5 | Pseudopotential | — |
| | | 4d | 2 | 1.5 | Pseudopotential | — |
| | | 4d | 2 | 2.5 | Pseudopotential | — |
| | | 5p | 1 | 0.5 | Orthogonalization | 4.500 |
| | | 5p | 1 | 1.5 | Orthogonalization | 4.400 |
| 43 | Tc | 4s | 0 | 0.5 | Pseudopotential | — |
| | | 5s | 0 | 0.5 | Pseudopotential | — |
| | | 4p | 1 | 0.5 | Pseudopotential | — |
| | | 5p | 1 | 0.5 | Pseudopotential | — |
| | | 4p | 1 | 1.5 | Pseudopotential | — |
| | | 4d | 2 | 1.5 | Pseudopotential | — |
| | | 4d | 2 | 2.5 | Pseudopotential | — |
| 44 | Ru | 4s | 0 | 0.5 | Pseudopotential | — |
| | | 5s | 0 | 0.5 | Pseudopotential | — |
| | | 4p | 1 | 0.5 | Pseudopotential | — |
| | | 4p | 1 | 1.5 | Pseudopotential | — |
| | | 4d | 2 | 1.5 | Pseudopotential | — |
| | | 4d | 2 | 2.5 | Pseudopotential | — |
| | | 5p | 1 | 0.5 | Orthogonalization | 4.900 |
| | | 5p | 1 | 1.5 | Orthogonalization | 4.800 |
| 45 | Rh | 4s | 0 | 0.5 | Pseudopotential | — |
| | | 5s | 0 | 0.5 | Pseudopotential | — |
| | | 4p | 1 | 0.5 | Pseudopotential | — |
| | | 4p | 1 | 1.5 | Pseudopotential | — |
| | | 4d | 2 | 1.5 | Pseudopotential | — |
| | | 4d | 2 | 2.5 | Pseudopotential | — |
| | | 5p | 1 | 0.5 | Orthogonalization | 4.900 |
| | | 5p | 1 | 1.5 | Orthogonalization | 4.800 |
| 46 | Pd | 5s | 0 | 0.5 | Pseudopotential | — |
| | | 5p | 1 | 0.5 | Pseudopotential | — |
| | | 5p | 1 | 1.5 | Pseudopotential | — |
| | | 4d | 2 | 1.5 | Pseudopotential | — |
| | | 4d | 2 | 2.5 | Pseudopotential | — |
| 47 | Ag | 5s | 0 | 0.5 | Pseudopotential | — |
| | | 5p | 1 | 0.5 | Pseudopotential | — |
| | | 5p | 1 | 1.5 | Pseudopotential | — |
| | | 4d | 2 | 1.5 | Pseudopotential | — |
| | | 4d | 2 | 2.5 | Pseudopotential | — |
| 48 | Cd | 5s | 0 | 0.5 | Pseudopotential | — |
| | | 5p | 1 | 0.5 | Pseudopotential | — |
| | | 5p | 1 | 1.5 | Pseudopotential | — |
| | | 4d | 2 | 1.5 | Pseudopotential | — |
| | | 4d | 2 | 2.5 | Pseudopotential | — |
| 49 | In | 5s | 0 | 0.5 | Pseudopotential | — |
| | | 5p | 1 | 0.5 | Pseudopotential | — |
| | | 5p | 1 | 1.5 | Pseudopotential | — |
| | | 4d | 2 | 1.5 | Pseudopotential | — |
| | | 4d | 2 | 2.5 | Pseudopotential | — |
| 50 | Sn | 5s | 0 | 0.5 | Pseudopotential | — |
| | | 5p | 1 | 0.5 | Pseudopotential | — |
| | | 5p | 1 | 1.5 | Pseudopotential | — |
| | | 4d | 2 | 1.5 | Pseudopotential | — |
| | | 4d | 2 | 2.5 | Pseudopotential | — |
| 51 | Sb | 5s | 0 | 0.5 | Pseudopotential | — |
| | | 5p | 1 | 0.5 | Pseudopotential | — |
| | | 5p | 1 | 1.5 | Pseudopotential | — |
| 52 | Te | 5s | 0 | 0.5 | Pseudopotential | — |
| | | 5p | 1 | 0.5 | Pseudopotential | — |
| | | 5p | 1 | 1.5 | Pseudopotential | — |
| 53 | I | 5s | 0 | 0.5 | Pseudopotential | — |
| | | 5p | 1 | 0.5 | Pseudopotential | — |
| | | 5p | 1 | 1.5 | Pseudopotential | — |
| 54 | Xe | 5s | 0 | 0.5 | Pseudopotential | — |
| | | 5p | 1 | 0.5 | Pseudopotential | — |
| | | 5p | 1 | 1.5 | Pseudopotential | — |
| | | 4d | 2 | 1.5 | Pseudopotential | — |
| | | 4d | 2 | 2.5 | Pseudopotential | — |
| 55 | Cs | 5s | 0 | 0.5 | Pseudopotential | — |
| | | 6s | 0 | 0.5 | Pseudopotential | — |
| | | 5p | 1 | 0.5 | Pseudopotential | — |
| | | 5p | 1 | 1.5 | Pseudopotential | — |
| | | 5d | 2 | 1.5 | OpenMX Fitting | 3.566 |
| | | 5d | 2 | 2.5 | OpenMX Fitting | 3.566 |
| 56 | Ba | 5s | 0 | 0.5 | Pseudopotential | — |
| | | 6s | 0 | 0.5 | Pseudopotential | — |
| | | 5p | 1 | 0.5 | Pseudopotential | — |
| | | 5p | 1 | 1.5 | Pseudopotential | — |
| | | 5d | 2 | 1.5 | OpenMX Fitting | 3.039 |
| | | 5d | 2 | 2.5 | OpenMX Fitting | 3.039 |
| 72 | Hf | 5s | 0 | 0.5 | Pseudopotential | — |
| | | 6s | 0 | 0.5 | Pseudopotential | — |
| | | 5p | 1 | 0.5 | Pseudopotential | — |
| | | 5p | 1 | 1.5 | Pseudopotential | — |
| | | 5d | 2 | 1.5 | Pseudopotential | — |
| | | 5d | 2 | 2.5 | Pseudopotential | — |
| | | 6p | 1 | 0.5 | Orthogonalization | 4.400 |
| | | 6p | 1 | 1.5 | Orthogonalization | 3.900 |
| 73 | Ta | 5s | 0 | 0.5 | Pseudopotential | — |
| | | 6s | 0 | 0.5 | Pseudopotential | — |
| | | 5p | 1 | 0.5 | Pseudopotential | — |
| | | 6p | 1 | 0.5 | Pseudopotential | — |
| | | 5p | 1 | 1.5 | Pseudopotential | — |
| | | 6p | 1 | 1.5 | Pseudopotential | — |
| | | 5d | 2 | 1.5 | Pseudopotential | — |
| | | 5d | 2 | 2.5 | Pseudopotential | — |

| Z | Element | Orbital | l | j | Source | α(bohr⁻¹) |
|---|---------|---------|---|---|--------|-----------|
| 74 W | | 5s | 0 | 0.5 | Pseudopotential | — |
| | | 6s | 0 | 0.5 | Pseudopotential | — |
| | | 5p | 1 | 0.5 | Pseudopotential | — |
| | | 6p | 1 | 0.5 | Pseudopotential | — |
| | | 5p | 1 | 1.5 | Pseudopotential | — |
| | | 6p | 1 | 1.5 | Pseudopotential | — |
| | | 5d | 2 | 1.5 | Pseudopotential | — |
| | | 5d | 2 | 2.5 | Pseudopotential | — |
| 75 Re | | 5s | 0 | 0.5 | Pseudopotential | — |
| | | 6s | 0 | 0.5 | Pseudopotential | — |
| | | 5p | 1 | 0.5 | Pseudopotential | — |
| | | 5p | 1 | 1.5 | Pseudopotential | — |
| | | 5d | 2 | 1.5 | Pseudopotential | — |
| | | 5d | 2 | 2.5 | Pseudopotential | — |
| | | 6p | 1 | 0.5 | Orthogonalization | 4.900 |
| | | 6p | 1 | 1.5 | Orthogonalization | 4.500 |
| 76 Os | | 5s | 0 | 0.5 | Pseudopotential | — |
| | | 6s | 0 | 0.5 | Pseudopotential | — |
| | | 5p | 1 | 0.5 | Pseudopotential | — |
| | | 5p | 1 | 1.5 | Pseudopotential | — |
| | | 5d | 2 | 1.5 | Pseudopotential | — |
| | | 5d | 2 | 2.5 | Pseudopotential | — |
| | | 6p | 1 | 0.5 | Orthogonalization | 5.100 |
| | | 6p | 1 | 1.5 | Orthogonalization | 4.600 |
| 77 Ir | | 6s | 0 | 0.5 | Pseudopotential | — |
| | | 6p | 1 | 0.5 | Pseudopotential | — |
| | | 6p | 1 | 1.5 | Pseudopotential | — |
| | | 5d | 2 | 1.5 | Pseudopotential | — |
| | | 5d | 2 | 2.5 | Pseudopotential | — |
| 78 Pt | | 6s | 0 | 0.5 | Pseudopotential | — |
| | | 6p | 1 | 0.5 | Pseudopotential | — |
| | | 6p | 1 | 1.5 | Pseudopotential | — |
| | | 5d | 2 | 1.5 | Pseudopotential | — |
| | | 5d | 2 | 2.5 | Pseudopotential | — |
| 79 Au | | 6s | 0 | 0.5 | Pseudopotential | — |
| | | 6p | 1 | 0.5 | Pseudopotential | — |
| | | 6p | 1 | 1.5 | Pseudopotential | — |
| | | 5d | 2 | 1.5 | Pseudopotential | — |
| | | 5d | 2 | 2.5 | Pseudopotential | — |
| 80 Hg | | 6s | 0 | 0.5 | Pseudopotential | — |
| | | 6p | 1 | 0.5 | Pseudopotential | — |
| | | 6p | 1 | 1.5 | Pseudopotential | — |
| | | 5d | 2 | 1.5 | Pseudopotential | — |
| | | 5d | 2 | 2.5 | Pseudopotential | — |
| 81 Tl | | 6s | 0 | 0.5 | Pseudopotential | — |
| | | 6p | 1 | 0.5 | Pseudopotential | — |
| | | 6p | 1 | 1.5 | Pseudopotential | — |
| | | 5d | 2 | 1.5 | Pseudopotential | — |
| | | 5d | 2 | 2.5 | Pseudopotential | — |
| 82 Pb | | 6s | 0 | 0.5 | Pseudopotential | — |
| | | 6p | 1 | 0.5 | Pseudopotential | — |
| | | 6p | 1 | 1.5 | Pseudopotential | — |
| | | 5d | 2 | 1.5 | Pseudopotential | — |
| | | 5d | 2 | 2.5 | Pseudopotential | — |
| 83 Bi | | 6s | 0 | 0.5 | Pseudopotential | — |
| | | 6p | 1 | 0.5 | Pseudopotential | — |
| | | 6p | 1 | 1.5 | Pseudopotential | — |
| | | 5d | 2 | 1.5 | Pseudopotential | — |
| | | 5d | 2 | 2.5 | Pseudopotential | — |

| Z | Element | Orbital | l | j | Source | α(bohr⁻¹) |
|---|---------|---------|---|---|--------|-----------|
| 84 Po | | 6s | 0 | 0.5 | Pseudopotential | — |
| | | 6p | 1 | 0.5 | Pseudopotential | — |
| | | 6p | 1 | 1.5 | Pseudopotential | — |
| | | 5d | 2 | 1.5 | Pseudopotential | — |
| | | 5d | 2 | 2.5 | Pseudopotential | — |
| 85 At | | 6s | 0 | 0.5 | Pseudopotential | — |
| | | 6p | 1 | 0.5 | Pseudopotential | — |
| | | 6p | 1 | 1.5 | Pseudopotential | — |
| | | 5d | 2 | 1.5 | Pseudopotential | — |
| | | 5d | 2 | 2.5 | Pseudopotential | — |
| 86 Rn | | 6s | 0 | 0.5 | Pseudopotential | — |
| | | 6p | 1 | 0.5 | Pseudopotential | — |
| | | 6p | 1 | 1.5 | Pseudopotential | — |
| | | 5d | 2 | 1.5 | Pseudopotential | — |
| | | 5d | 2 | 2.5 | Pseudopotential | — |
| 87 Fr | | 6s | 0 | 0.5 | Pseudopotential | — |
| | | 7s | 0 | 0.5 | Pseudopotential | — |
| | | 6p | 1 | 0.5 | Pseudopotential | — |
| | | 6p | 1 | 1.5 | Pseudopotential | — |
| | | 5d | 2 | 1.5 | Pseudopotential | — |
| | | 5d | 2 | 2.5 | Pseudopotential | — |
| 88 Ra | | 6s | 0 | 0.5 | Pseudopotential | — |
| | | 7s | 0 | 0.5 | Pseudopotential | — |
| | | 6p | 1 | 0.5 | Pseudopotential | — |
| | | 6p | 1 | 1.5 | Pseudopotential | — |
| | | 5d | 2 | 1.5 | Pseudopotential | — |
| | | 5d | 2 | 2.5 | Pseudopotential | — |

SUPPLEMENTARY TABLE VIII: Values of $\alpha$ and minimal set of orbitals for projectors from `OpenMX` fitting. We note that we only report values for those orbitals for which a value was fitted from OpenMX in this work (i.e., for orbitals that were not included in the PAOs for at least one pseudopotential family, and that did not have underlying orbitals with the same angular character, so that no orthogonality condition on the radial part could be enforced).

| $Z$ | Element | Filename | Orbital | $l$ | $\alpha(\text{bohr}^{-1})$ |
|---|---|---|---|---|---|
| 1 | H | H5.0.pao | 1s | 0 | 1.075 |
| 2 | He | He10.0.pao | 1s | 0 | 1.529 |
| 3 | Li | Li12.0.pao | 2s | 0 | 2.520 |
| | | | 2p | 1 | 1.114 |
| 4 | Be | Be8.0.pao | 2s | 0 | 0.649 |
| | | | 2p | 1 | 1.834 |
| 5 | B | B9.0.pao | 2s | 0 | 0.833 |
| | | | 2p | 1 | 2.143 |
| 6 | C | C6.0.pao | 2s | 0 | 1.000 |
| | | | 2p | 1 | 2.684 |
| 7 | N | N6.0p.pao | 2s | 0 | 1.227 |
| | | | 2p | 1 | 3.228 |
| 8 | O | O6.0.pao | 2s | 0 | 1.412 |
| | | | 2p | 1 | 3.755 |
| 9 | F | F7.0.pao | 2s | 0 | 1.608 |
| | | | 2p | 1 | 4.230 |
| 10 | Ne | Ne11.0.pao | 2s | 0 | 1.797 |
| | | | 2p | 1 | 4.748 |
| 11 | Na | Na11.0.pao | 3s | 0 | 2.072 |
| | | | 3p | 1 | 5.830 |
| 12 | Mg | Mg7.0.pao | 3s | 0 | 0.587 |
| | | | 3p | 1 | 6.876 |
| 13 | Al | Al7.0.pao | 3s | 0 | 0.682 |
| | | | 3p | 1 | 1.768 |
| 14 | Si | Si8.0.pao | 3s | 0 | 0.765 |
| | | | 3p | 1 | 1.876 |
| 15 | P | P8.0.pao | 3s | 0 | 0.886 |
| | | | 3p | 1 | 2.191 |
| 16 | S | S9.0.pao | 3s | 0 | 0.992 |
| | | | 3p | 1 | 2.432 |
| 17 | Cl | Cl7.0.pao | 3s | 0 | 1.096 |
| | | | 3p | 1 | 2.733 |
| 18 | Ar | Ar11.0.pao | 3s | 0 | 1.197 |
| | | | 3p | 1 | 3.001 |
| 19 | K | K12.0.pao | 4s | 0 | 1.336 |
| | | | 3d | 2 | 2.727 |
| 20 | Ca | Ca9.0.pao | 4s | 0 | 1.474 |
| | | | 3d | 2 | 3.983 |
| 21 | Sc | Sc11.0.pao | 4s | 0 | 1.589 |
| | | | 4p | 1 | 4.353 |
| | | | 3d | 2 | 5.200 |
| 22 | Ti | Ti11.0.pao | 4s | 0 | 1.698 |
| | | | 4p | 1 | 4.628 |
| | | | 3d | 2 | 5.246 |
| 23 | V | V8.0.pao | 4s | 0 | 1.796 |
| | | | 4p | 1 | 4.925 |
| | | | 3d | 2 | 5.774 |
| 24 | Cr | Cr6.0.pao | 4s | 0 | 1.902 |
| | | | 4p | 1 | 5.222 |
| | | | 3d | 2 | 6.565 |
| 25 | Mn | Mn8.0.pao | 4s | 0 | 1.999 |
| | | | 4p | 1 | 5.603 |
| | | | 3d | 2 | 8.113 |
| 26 | Fe | Fe6.0S.pao | 4s | 0 | 0.641 |
| | | | 4p | 1 | 5.882 |
| | | | 3d | 2 | 8.140 |
| 27 | Co | Co8.0H.pao | 4s | 0 | 2.195 |
| | | | 4p | 1 | 6.180 |
| | | | 3d | 2 | 8.453 |
| 28 | Ni | Ni10.0H.pao | 4s | 0 | 2.321 |
| | | | 4p | 1 | 6.481 |
| | | | 3d | 2 | 8.893 |
| 29 | Cu | Cu8.0S.pao | 4s | 0 | 0.601 |
| | | | 4p | 1 | 2.149 |
| | | | 3d | 2 | 9.296 |
| 30 | Zn | Zn10.0S.pao | 4s | 0 | 0.628 |
| | | | 4p | 1 | 1.361 |
| | | | 3d | 2 | 10.651 |
| 31 | Ga | Ga7.0.pao | 4s | 0 | 0.729 |
| | | | 4p | 1 | 1.781 |
| 32 | Ge | Ge9.0.pao | 4s | 0 | 0.798 |
| | | | 4p | 1 | 1.815 |
| 33 | As | As9.0.pao | 4s | 0 | 0.875 |
| | | | 4p | 1 | 2.035 |
| 34 | Se | Se7.0.pao | 4s | 0 | 0.947 |
| | | | 4p | 1 | 2.272 |
| 35 | Br | Br7.0.pao | 4s | 0 | 1.018 |
| | | | 4p | 1 | 2.447 |
| 36 | Kr | Kr10.0.pao | 4s | 0 | 1.087 |
| | | | 4p | 1 | 2.612 |
| 37 | Rb | Rb13.0.pao | 5s | 0 | 1.182 |
| | | | 4d | 2 | 2.197 |
| 38 | Sr | Sr12.0.pao | 5s | 0 | 1.275 |
| | | | 4d | 2 | 4.400 |
| 39 | Y | Y10.0.pao | 5s | 0 | 1.348 |
| | | | 5p | 1 | 3.560 |
| | | | 4d | 2 | 3.758 |
| 40 | Zr | Zr9.0.pao | 5s | 0 | 1.430 |
| | | | 5p | 1 | 3.760 |
| | | | 4d | 2 | 4.062 |
| 41 | Nb | Nb9.0.pao | 5s | 0 | 1.505 |
| | | | 5p | 1 | 3.981 |
| | | | 4d | 2 | 4.590 |
| 42 | Mo | Mo7.0.pao | 5s | 0 | 1.567 |
| | | | 5p | 1 | 4.190 |
| | | | 4d | 2 | 5.150 |
| 43 | Tc | Tc11.0.pao | 5s | 0 | 1.641 |
| | | | 5p | 1 | 4.411 |
| | | | 4d | 2 | 5.645 |
| 44 | Ru | Ru7.0.pao | 5s | 0 | 0.569 |
| | | | 5p | 1 | 4.599 |
| | | | 4d | 2 | 5.920 |
| 45 | Rh | Rh9.0.pao | 5s | 0 | 0.534 |
| | | | 5p | 1 | 4.800 |
| | | | 4d | 2 | 6.305 |
| 46 | Pd | Pd11.0.pao | 5s | 0 | 0.522 |
| | | | 5p | 1 | 4.966 |
| | | | 4d | 2 | 6.254 |
| 47 | Ag | Ag9.0.pao | 5s | 0 | 0.547 |
| | | | 5p | 1 | 5.191 |
| | | | 4d | 2 | 7.107 |
| 48 | Cd | Cd7.0.pao | 5s | 0 | 0.621 |
| | | | 5p | 1 | 1.815 |
| | | | 4d | 2 | 7.835 |

| $Z$ | Element | Filename | Orbital | $l$ | $\alpha(\text{bohr}^{-1})$ |
|---|---|---|---|---|---|
| 49 | In | In7.0.pao | 5s | 0 | 0.668 |
| | | | 5p | 1 | 1.684 |
| 50 | Sn | Sn9.0.pao | 5s | 0 | 0.713 |
| | | | 5p | 1 | 1.652 |
| 51 | Sb | Sb11.0.pao | 5s | 0 | 0.771 |
| | | | 5p | 1 | 1.791 |
| 52 | Te | Te9.0.pao | 5s | 0 | 0.825 |
| | | | 5p | 1 | 1.946 |
| 53 | I | I7.0.pao | 5s | 0 | 0.879 |
| | | | 5p | 1 | 2.144 |
| 54 | Xe | Xe13.0.pao | 5s | 0 | 0.930 |
| | | | 5p | 1 | 2.230 |
| 55 | Cs | Cs12.0.pao | 6s | 0 | 1.001 |
| | | | 5d | 2 | 3.566 |
| 56 | Ba | Ba10.0.pao | 6s | 0 | 1.069 |
| | | | 5d | 2 | 3.039 |
| 72 | Hf | Hf11.0.pao | 6s | 0 | 1.544 |
| | | | 6p | 1 | 3.885 |
| | | | 5d | 2 | 4.081 |
| 73 | Ta | Ta11.0.pao | 6s | 0 | 1.593 |
| | | | 6p | 1 | 4.070 |
| | | | 5d | 2 | 4.551 |
| 74 | W | W7.0.pao | 6s | 0 | 0.577 |
| | | | 6p | 1 | 4.137 |
| | | | 5d | 2 | 4.789 |
| 75 | Re | Re7.0.pao | 6s | 0 | 1.698 |
| | | | 6p | 1 | 4.338 |
| | | | 5d | 2 | 5.392 |

| $Z$ | Element | Filename | Orbital | $l$ | $\alpha(\text{bohr}^{-1})$ |
|---|---|---|---|---|---|
| 76 | Os | Os7.0.pao | 6s | 0 | 0.600 |
| | | | 6p | 1 | 4.490 |
| | | | 5d | 2 | 5.623 |
| 77 | Ir | Ir11.0.pao | 6s | 0 | 0.576 |
| | | | 6p | 1 | 4.615 |
| | | | 5d | 2 | 5.716 |
| 78 | Pt | Pt9.0.pao | 6s | 0 | 0.590 |
| | | | 6p | 1 | 4.758 |
| | | | 5d | 2 | 5.988 |
| 79 | Au | Au11.0.pao | 6s | 0 | 0.589 |
| | | | 6p | 1 | 4.873 |
| | | | 5d | 2 | 6.301 |
| 80 | Hg | Hg8.0.pao | 6s | 0 | 0.638 |
| | | | 6p | 1 | 5.073 |
| | | | 5d | 2 | 6.837 |
| 81 | Tl | Tl8.0.pao | 6s | 0 | 0.689 |
| | | | 6p | 1 | 5.245 |
| 82 | Pb | Pb8.0.pao | 6s | 0 | 0.742 |
| | | | 6p | 1 | 1.666 |
| 83 | Bi | Bi8.0.pao | 6s | 0 | 0.792 |
| | | | 6p | 1 | 1.786 |
| 84 | Po | Po10.0.pao | 6s | 0 | 0.839 |
| | | | 6p | 1 | 1.850 |
| 86 | Rn | Rn13.0.pao | 6s | 0 | 0.934 |
| | | | 6p | 1 | 2.091 |



\begin{thebibliography}{56}%
\makeatletter
\providecommand \@ifxundefined [1]{%
 \@ifx{#1\undefined}
}%
\providecommand \@ifnum [1]{%
 \ifnum #1\expandafter \@firstoftwo
 \else \expandafter \@secondoftwo
 \fi
}%
\providecommand \@ifx [1]{%
 \ifx #1\expandafter \@firstoftwo
 \else \expandafter \@secondoftwo
 \fi
}%
\providecommand \natexlab [1]{#1}%
\providecommand \enquote  [1]{``#1''}%
\providecommand \bibnamefont  [1]{#1}%
\providecommand \bibfnamefont [1]{#1}%
\providecommand \citenamefont [1]{#1}%
\providecommand \href@noop [0]{\@secondoftwo}%
\providecommand \href [0]{\begingroup \@sanitize@url \@href}%
\providecommand \@href[1]{\@@startlink{#1}\@@href}%
\providecommand \@@href[1]{\endgroup#1\@@endlink}%
\providecommand \@sanitize@url [0]{\catcode `\\12\catcode `\$12\catcode `\&12\catcode `\#12\catcode `\^12\catcode `\_12\catcode `\%12\relax}%
\providecommand \@@startlink[1]{}%
\providecommand \@@endlink[0]{}%
\providecommand \url  [0]{\begingroup\@sanitize@url \@url }%
\providecommand \@url [1]{\endgroup\@href {#1}{\urlprefix }}%
\providecommand \urlprefix  [0]{URL }%
\providecommand \Eprint [0]{\href }%
\providecommand \doibase [0]{https://doi.org/}%
\providecommand \selectlanguage [0]{\@gobble}%
\providecommand \bibinfo  [0]{\@secondoftwo}%
\providecommand \bibfield  [0]{\@secondoftwo}%
\providecommand \translation [1]{[#1]}%
\providecommand \BibitemOpen [0]{}%
\providecommand \bibitemStop [0]{}%
\providecommand \bibitemNoStop [0]{.\EOS\space}%
\providecommand \EOS [0]{\spacefactor3000\relax}%
\providecommand \BibitemShut  [1]{\csname bibitem#1\endcsname}%
\let\auto@bib@innerbib\@empty
%</preamble>
\bibitem [{\citenamefont {Hohenberg}\ and\ \citenamefont {Kohn}(1964)}]{Hohenberg1964}%
  \BibitemOpen
  \bibfield  {author} {\bibinfo {author} {\bibfnamefont {P.}~\bibnamefont {Hohenberg}}\ and\ \bibinfo {author} {\bibfnamefont {W.}~\bibnamefont {Kohn}},\ }\bibfield  {title} {\bibinfo {title} {{Inhomogeneous Electron Gas}},\ }\href {https://doi.org/10.1103/PhysRev.136.B864} {\bibfield  {journal} {\bibinfo  {journal} {Physical Review}\ }\textbf {\bibinfo {volume} {136}},\ \bibinfo {pages} {B864} (\bibinfo {year} {1964})}\BibitemShut {NoStop}%
\bibitem [{\citenamefont {Nagaosa}\ \emph {et~al.}(2010)\citenamefont {Nagaosa}, \citenamefont {Sinova}, \citenamefont {Onoda}, \citenamefont {MacDonald},\ and\ \citenamefont {Ong}}]{Nagaosa2010}%
  \BibitemOpen
  \bibfield  {author} {\bibinfo {author} {\bibfnamefont {N.}~\bibnamefont {Nagaosa}}, \bibinfo {author} {\bibfnamefont {J.}~\bibnamefont {Sinova}}, \bibinfo {author} {\bibfnamefont {S.}~\bibnamefont {Onoda}}, \bibinfo {author} {\bibfnamefont {A.~H.}\ \bibnamefont {MacDonald}},\ and\ \bibinfo {author} {\bibfnamefont {N.~P.}\ \bibnamefont {Ong}},\ }\bibfield  {title} {\bibinfo {title} {{Anomalous Hall effect}},\ }\href {https://doi.org/10.1103/RevModPhys.82.1539} {\bibfield  {journal} {\bibinfo  {journal} {Reviews of Modern Physics}\ }\textbf {\bibinfo {volume} {82}},\ \bibinfo {pages} {1539} (\bibinfo {year} {2010})}\BibitemShut {NoStop}%
\bibitem [{\citenamefont {Sinova}\ \emph {et~al.}(2015)\citenamefont {Sinova}, \citenamefont {Valenzuela}, \citenamefont {Wunderlich}, \citenamefont {Back},\ and\ \citenamefont {Jungwirth}}]{Sinova2015}%
  \BibitemOpen
  \bibfield  {author} {\bibinfo {author} {\bibfnamefont {J.}~\bibnamefont {Sinova}}, \bibinfo {author} {\bibfnamefont {S.~O.}\ \bibnamefont {Valenzuela}}, \bibinfo {author} {\bibfnamefont {J.}~\bibnamefont {Wunderlich}}, \bibinfo {author} {\bibfnamefont {C.~H.}\ \bibnamefont {Back}},\ and\ \bibinfo {author} {\bibfnamefont {T.}~\bibnamefont {Jungwirth}},\ }\bibfield  {title} {\bibinfo {title} {{Spin Hall effects}},\ }\href {https://doi.org/10.1103/RevModPhys.87.1213} {\bibfield  {journal} {\bibinfo  {journal} {Reviews of Modern Physics}\ }\textbf {\bibinfo {volume} {87}},\ \bibinfo {pages} {1213} (\bibinfo {year} {2015})}\BibitemShut {NoStop}%
\bibitem [{\citenamefont {Derunova}\ \emph {et~al.}(2019)\citenamefont {Derunova}, \citenamefont {Sun}, \citenamefont {Felser}, \citenamefont {Parkin}, \citenamefont {Yan},\ and\ \citenamefont {Ali}}]{Derunova2019}%
  \BibitemOpen
  \bibfield  {author} {\bibinfo {author} {\bibfnamefont {E.}~\bibnamefont {Derunova}}, \bibinfo {author} {\bibfnamefont {Y.}~\bibnamefont {Sun}}, \bibinfo {author} {\bibfnamefont {C.}~\bibnamefont {Felser}}, \bibinfo {author} {\bibfnamefont {S.~S.~P.}\ \bibnamefont {Parkin}}, \bibinfo {author} {\bibfnamefont {B.}~\bibnamefont {Yan}},\ and\ \bibinfo {author} {\bibfnamefont {M.~N.}\ \bibnamefont {Ali}},\ }\bibfield  {title} {\bibinfo {title} {{Giant intrinsic spin Hall effect in $\rm{W_3 Ta}$ and other A15 superconductors}},\ }\href {https://doi.org/10.1126/sciadv.aav8575} {\bibfield  {journal} {\bibinfo  {journal} {Science Advances}\ }\textbf {\bibinfo {volume} {5}},\ \bibinfo {pages} {1} (\bibinfo {year} {2019})}\BibitemShut {NoStop}%
\bibitem [{\citenamefont {Yao}\ \emph {et~al.}(2004)\citenamefont {Yao}, \citenamefont {Kleinman}, \citenamefont {MacDonald}, \citenamefont {Sinova}, \citenamefont {Jungwirth}, \citenamefont {sheng Wang}, \citenamefont {Wang},\ and\ \citenamefont {Niu}}]{Yao2004}%
  \BibitemOpen
  \bibfield  {author} {\bibinfo {author} {\bibfnamefont {Y.}~\bibnamefont {Yao}}, \bibinfo {author} {\bibfnamefont {L.}~\bibnamefont {Kleinman}}, \bibinfo {author} {\bibfnamefont {A.~H.}\ \bibnamefont {MacDonald}}, \bibinfo {author} {\bibfnamefont {J.}~\bibnamefont {Sinova}}, \bibinfo {author} {\bibfnamefont {T.}~\bibnamefont {Jungwirth}}, \bibinfo {author} {\bibfnamefont {D.}~\bibnamefont {sheng Wang}}, \bibinfo {author} {\bibfnamefont {E.}~\bibnamefont {Wang}},\ and\ \bibinfo {author} {\bibfnamefont {Q.}~\bibnamefont {Niu}},\ }\bibfield  {title} {\bibinfo {title} {{First Principles Calculation of Anomalous Hall Conductivity in Ferromagnetic bcc Fe}},\ }\href {https://doi.org/10.1103/PhysRevLett.92.037204} {\bibfield  {journal} {\bibinfo  {journal} {Physical Review Letters}\ }\textbf {\bibinfo {volume} {92}},\ \bibinfo {pages} {4} (\bibinfo {year} {2004})}\BibitemShut {NoStop}%
\bibitem [{\citenamefont {Guo}\ \emph {et~al.}(2008)\citenamefont {Guo}, \citenamefont {Murakami}, \citenamefont {Chen},\ and\ \citenamefont {Nagaosa}}]{Guo2008}%
  \BibitemOpen
  \bibfield  {author} {\bibinfo {author} {\bibfnamefont {G.~Y.}\ \bibnamefont {Guo}}, \bibinfo {author} {\bibfnamefont {S.}~\bibnamefont {Murakami}}, \bibinfo {author} {\bibfnamefont {T.-W.}\ \bibnamefont {Chen}},\ and\ \bibinfo {author} {\bibfnamefont {N.}~\bibnamefont {Nagaosa}},\ }\bibfield  {title} {\bibinfo {title} {{Intrinsic Spin Hall Effect in Platinum: First-Principles Calculations}},\ }\href {https://doi.org/10.1103/PhysRevLett.100.096401} {\bibfield  {journal} {\bibinfo  {journal} {Physical Review Letters}\ }\textbf {\bibinfo {volume} {100}},\ \bibinfo {pages} {096401} (\bibinfo {year} {2008})}\BibitemShut {NoStop}%
\bibitem [{\citenamefont {Kohn}\ and\ \citenamefont {Sham}(1965)}]{Kohn1965}%
  \BibitemOpen
  \bibfield  {author} {\bibinfo {author} {\bibfnamefont {W.}~\bibnamefont {Kohn}}\ and\ \bibinfo {author} {\bibfnamefont {L.~J.}\ \bibnamefont {Sham}},\ }\bibfield  {title} {\bibinfo {title} {{Self-Consistent Equations Including Exchange and Correlation Effects}},\ }\href {https://doi.org/10.1103/PhysRev.140.A1133} {\bibfield  {journal} {\bibinfo  {journal} {Physical Review}\ }\textbf {\bibinfo {volume} {140}},\ \bibinfo {pages} {A1133} (\bibinfo {year} {1965})}\BibitemShut {NoStop}%
\bibitem [{\citenamefont {Marzari}\ \emph {et~al.}(2012)\citenamefont {Marzari}, \citenamefont {Mostofi}, \citenamefont {Yates}, \citenamefont {Souza},\ and\ \citenamefont {Vanderbilt}}]{Marzari2012}%
  \BibitemOpen
  \bibfield  {author} {\bibinfo {author} {\bibfnamefont {N.}~\bibnamefont {Marzari}}, \bibinfo {author} {\bibfnamefont {A.~A.}\ \bibnamefont {Mostofi}}, \bibinfo {author} {\bibfnamefont {J.~R.}\ \bibnamefont {Yates}}, \bibinfo {author} {\bibfnamefont {I.}~\bibnamefont {Souza}},\ and\ \bibinfo {author} {\bibfnamefont {D.}~\bibnamefont {Vanderbilt}},\ }\bibfield  {title} {\bibinfo {title} {{Maximally localized Wannier functions: Theory and applications}},\ }\href {https://doi.org/10.1103/RevModPhys.84.1419} {\bibfield  {journal} {\bibinfo  {journal} {Reviews of Modern Physics}\ }\textbf {\bibinfo {volume} {84}},\ \bibinfo {pages} {1419} (\bibinfo {year} {2012})}\BibitemShut {NoStop}%
\bibitem [{\citenamefont {Marzari}\ and\ \citenamefont {Vanderbilt}(1997)}]{Marzari1997}%
  \BibitemOpen
  \bibfield  {author} {\bibinfo {author} {\bibfnamefont {N.}~\bibnamefont {Marzari}}\ and\ \bibinfo {author} {\bibfnamefont {D.}~\bibnamefont {Vanderbilt}},\ }\bibfield  {title} {\bibinfo {title} {{Maximally localized generalized Wannier functions for composite energy bands}},\ }\href {https://doi.org/10.1103/PhysRevB.56.12847} {\bibfield  {journal} {\bibinfo  {journal} {Physical Review B - Condensed Matter and Materials Physics}\ }\textbf {\bibinfo {volume} {56}},\ \bibinfo {pages} {12847} (\bibinfo {year} {1997})}\BibitemShut {NoStop}%
\bibitem [{\citenamefont {Marrazzo}\ \emph {et~al.}(2024)\citenamefont {Marrazzo}, \citenamefont {Beck}, \citenamefont {Margine}, \citenamefont {Marzari}, \citenamefont {Mostofi}, \citenamefont {Qiao}, \citenamefont {Souza}, \citenamefont {Tsirkin}, \citenamefont {Yates},\ and\ \citenamefont {Pizzi}}]{Marrazzo_2024}%
  \BibitemOpen
  \bibfield  {author} {\bibinfo {author} {\bibfnamefont {A.}~\bibnamefont {Marrazzo}}, \bibinfo {author} {\bibfnamefont {S.}~\bibnamefont {Beck}}, \bibinfo {author} {\bibfnamefont {E.~R.}\ \bibnamefont {Margine}}, \bibinfo {author} {\bibfnamefont {N.}~\bibnamefont {Marzari}}, \bibinfo {author} {\bibfnamefont {A.~A.}\ \bibnamefont {Mostofi}}, \bibinfo {author} {\bibfnamefont {J.}~\bibnamefont {Qiao}}, \bibinfo {author} {\bibfnamefont {I.}~\bibnamefont {Souza}}, \bibinfo {author} {\bibfnamefont {S.~S.}\ \bibnamefont {Tsirkin}}, \bibinfo {author} {\bibfnamefont {J.~R.}\ \bibnamefont {Yates}},\ and\ \bibinfo {author} {\bibfnamefont {G.}~\bibnamefont {Pizzi}},\ }\bibfield  {title} {\bibinfo {title} {Wannier-function software ecosystem for materials simulations},\ }\href@noop {} {\bibfield  {journal} {\bibinfo  {journal} {Reviews of Modern Physics}\ }\textbf {\bibinfo {volume} {96}},\ \bibinfo {pages} {045008} (\bibinfo {year} {2024})}\BibitemShut {NoStop}%
\bibitem [{\citenamefont {Zhang}\ \emph {et~al.}(2009)\citenamefont {Zhang}, \citenamefont {Liu}, \citenamefont {Qi}, \citenamefont {Deng}, \citenamefont {Dai}, \citenamefont {Zhang},\ and\ \citenamefont {Fang}}]{Zhang2009}%
  \BibitemOpen
  \bibfield  {author} {\bibinfo {author} {\bibfnamefont {H.~J.}\ \bibnamefont {Zhang}}, \bibinfo {author} {\bibfnamefont {C.~X.}\ \bibnamefont {Liu}}, \bibinfo {author} {\bibfnamefont {X.~L.}\ \bibnamefont {Qi}}, \bibinfo {author} {\bibfnamefont {X.~Y.}\ \bibnamefont {Deng}}, \bibinfo {author} {\bibfnamefont {X.}~\bibnamefont {Dai}}, \bibinfo {author} {\bibfnamefont {S.~C.}\ \bibnamefont {Zhang}},\ and\ \bibinfo {author} {\bibfnamefont {Z.}~\bibnamefont {Fang}},\ }\bibfield  {title} {\bibinfo {title} {{Electronic structures and surface states of the topological insulator $\rm{Bi_{1-x}Sb_x}$}},\ }\href {https://doi.org/10.1103/PhysRevB.80.085307} {\bibfield  {journal} {\bibinfo  {journal} {Physical Review B - Condensed Matter and Materials Physics}\ }\textbf {\bibinfo {volume} {80}},\ \bibinfo {pages} {1} (\bibinfo {year} {2009})}\BibitemShut {NoStop}%
\bibitem [{\citenamefont {Pizzi}\ \emph {et~al.}(2014)\citenamefont {Pizzi}, \citenamefont {Volja}, \citenamefont {Kozinsky}, \citenamefont {Fornari},\ and\ \citenamefont {Marzari}}]{Pizzi_2014}%
  \BibitemOpen
  \bibfield  {author} {\bibinfo {author} {\bibfnamefont {G.}~\bibnamefont {Pizzi}}, \bibinfo {author} {\bibfnamefont {D.}~\bibnamefont {Volja}}, \bibinfo {author} {\bibfnamefont {B.}~\bibnamefont {Kozinsky}}, \bibinfo {author} {\bibfnamefont {M.}~\bibnamefont {Fornari}},\ and\ \bibinfo {author} {\bibfnamefont {N.}~\bibnamefont {Marzari}},\ }\bibfield  {title} {\bibinfo {title} {Boltzwann: A code for the evaluation of thermoelectric and electronic transport properties with a maximally-localized wannier functions basis},\ }\href {https://doi.org/https://doi.org/10.1016/j.cpc.2013.09.015} {\bibfield  {journal} {\bibinfo  {journal} {Computer Physics Communications}\ }\textbf {\bibinfo {volume} {185}},\ \bibinfo {pages} {422} (\bibinfo {year} {2014})}\BibitemShut {NoStop}%
\bibitem [{\citenamefont {Wang}\ \emph {et~al.}(2006)\citenamefont {Wang}, \citenamefont {Yates}, \citenamefont {Souza},\ and\ \citenamefont {Vanderbilt}}]{Wang2006}%
  \BibitemOpen
  \bibfield  {author} {\bibinfo {author} {\bibfnamefont {X.}~\bibnamefont {Wang}}, \bibinfo {author} {\bibfnamefont {J.~R.}\ \bibnamefont {Yates}}, \bibinfo {author} {\bibfnamefont {I.}~\bibnamefont {Souza}},\ and\ \bibinfo {author} {\bibfnamefont {D.}~\bibnamefont {Vanderbilt}},\ }\bibfield  {title} {\bibinfo {title} {{Ab initio calculation of the anomalous Hall conductivity by Wannier interpolation}},\ }\href {https://doi.org/10.1103/PhysRevB.74.195118} {\bibfield  {journal} {\bibinfo  {journal} {Physical Review B - Condensed Matter and Materials Physics}\ }\textbf {\bibinfo {volume} {74}},\ \bibinfo {pages} {1} (\bibinfo {year} {2006})}\BibitemShut {NoStop}%
\bibitem [{\citenamefont {Thonhauser}\ \emph {et~al.}(2005)\citenamefont {Thonhauser}, \citenamefont {Ceresoli}, \citenamefont {Vanderbilt},\ and\ \citenamefont {Resta}}]{Thonhauser2005}%
  \BibitemOpen
  \bibfield  {author} {\bibinfo {author} {\bibfnamefont {T.}~\bibnamefont {Thonhauser}}, \bibinfo {author} {\bibfnamefont {D.}~\bibnamefont {Ceresoli}}, \bibinfo {author} {\bibfnamefont {D.}~\bibnamefont {Vanderbilt}},\ and\ \bibinfo {author} {\bibfnamefont {R.}~\bibnamefont {Resta}},\ }\bibfield  {title} {\bibinfo {title} {{Orbital magnetization in periodic insulators}},\ }\href {https://doi.org/10.1103/PhysRevLett.95.137205} {\bibfield  {journal} {\bibinfo  {journal} {Physical Review Letters}\ }\textbf {\bibinfo {volume} {95}},\ \bibinfo {pages} {1} (\bibinfo {year} {2005})}\BibitemShut {NoStop}%
\bibitem [{\citenamefont {Qiao}\ \emph {et~al.}(2018)\citenamefont {Qiao}, \citenamefont {Zhou}, \citenamefont {Yuan},\ and\ \citenamefont {Zhao}}]{Qiao2018}%
  \BibitemOpen
  \bibfield  {author} {\bibinfo {author} {\bibfnamefont {J.}~\bibnamefont {Qiao}}, \bibinfo {author} {\bibfnamefont {J.}~\bibnamefont {Zhou}}, \bibinfo {author} {\bibfnamefont {Z.}~\bibnamefont {Yuan}},\ and\ \bibinfo {author} {\bibfnamefont {W.}~\bibnamefont {Zhao}},\ }\bibfield  {title} {\bibinfo {title} {{Calculation of intrinsic spin Hall conductivity by Wannier interpolation}},\ }\href {https://doi.org/10.1103/PhysRevB.98.214402} {\bibfield  {journal} {\bibinfo  {journal} {Physical Review B}\ }\textbf {\bibinfo {volume} {98}},\ \bibinfo {pages} {1} (\bibinfo {year} {2018})}\BibitemShut {NoStop}%
\bibitem [{\citenamefont {Ryoo}\ \emph {et~al.}(2019)\citenamefont {Ryoo}, \citenamefont {Park},\ and\ \citenamefont {Souza}}]{Ryoo2019}%
  \BibitemOpen
  \bibfield  {author} {\bibinfo {author} {\bibfnamefont {J.~H.}\ \bibnamefont {Ryoo}}, \bibinfo {author} {\bibfnamefont {C.~H.}\ \bibnamefont {Park}},\ and\ \bibinfo {author} {\bibfnamefont {I.}~\bibnamefont {Souza}},\ }\bibfield  {title} {\bibinfo {title} {{Computation of intrinsic spin Hall conductivities from first principles using maximally localized Wannier functions}},\ }\href {https://doi.org/10.1103/PhysRevB.99.235113} {\bibfield  {journal} {\bibinfo  {journal} {Physical Review B}\ }\textbf {\bibinfo {volume} {99}},\ \bibinfo {pages} {235113} (\bibinfo {year} {2019})}\BibitemShut {NoStop}%
\bibitem [{\citenamefont {Souza}\ \emph {et~al.}(2001)\citenamefont {Souza}, \citenamefont {Marzari},\ and\ \citenamefont {Vanderbilt}}]{Souza2002}%
  \BibitemOpen
  \bibfield  {author} {\bibinfo {author} {\bibfnamefont {I.}~\bibnamefont {Souza}}, \bibinfo {author} {\bibfnamefont {N.}~\bibnamefont {Marzari}},\ and\ \bibinfo {author} {\bibfnamefont {D.}~\bibnamefont {Vanderbilt}},\ }\bibfield  {title} {\bibinfo {title} {{Maximally localized Wannier functions for entangled energy bands}},\ }\href {https://doi.org/10.1103/PhysRevB.65.035109} {\bibfield  {journal} {\bibinfo  {journal} {Physical Review B}\ }\textbf {\bibinfo {volume} {65}},\ \bibinfo {pages} {035109} (\bibinfo {year} {2001})}\BibitemShut {NoStop}%
\bibitem [{\citenamefont {L{\"{o}}wdin}(1950)}]{Lowdin1950}%
  \BibitemOpen
  \bibfield  {author} {\bibinfo {author} {\bibfnamefont {P.-O.}\ \bibnamefont {L{\"{o}}wdin}},\ }\bibfield  {title} {\bibinfo {title} {{On the Non-Orthogonality Problem Connected with the Use of Atomic Wave Functions in the Theory of Molecules and Crystals}},\ }\href {https://doi.org/10.1063/1.1747632} {\bibfield  {journal} {\bibinfo  {journal} {The Journal of Chemical Physics}\ }\textbf {\bibinfo {volume} {18}},\ \bibinfo {pages} {365} (\bibinfo {year} {1950})}\BibitemShut {NoStop}%
\bibitem [{\citenamefont {Gresch}\ \emph {et~al.}(2018)\citenamefont {Gresch}, \citenamefont {Wu}, \citenamefont {Winkler}, \citenamefont {H{\"{a}}uselmann}, \citenamefont {Troyer},\ and\ \citenamefont {Soluyanov}}]{Gresch2018}%
  \BibitemOpen
  \bibfield  {author} {\bibinfo {author} {\bibfnamefont {D.}~\bibnamefont {Gresch}}, \bibinfo {author} {\bibfnamefont {Q.}~\bibnamefont {Wu}}, \bibinfo {author} {\bibfnamefont {G.~W.}\ \bibnamefont {Winkler}}, \bibinfo {author} {\bibfnamefont {R.}~\bibnamefont {H{\"{a}}uselmann}}, \bibinfo {author} {\bibfnamefont {M.}~\bibnamefont {Troyer}},\ and\ \bibinfo {author} {\bibfnamefont {A.~A.}\ \bibnamefont {Soluyanov}},\ }\bibfield  {title} {\bibinfo {title} {{Automated construction of symmetrized Wannier-like tight-binding models from ab initio calculations}},\ }\href {https://doi.org/10.1103/PhysRevMaterials.2.103805} {\bibfield  {journal} {\bibinfo  {journal} {Physical Review Materials}\ }\textbf {\bibinfo {volume} {2}},\ \bibinfo {pages} {103805} (\bibinfo {year} {2018})}\BibitemShut {NoStop}%
\bibitem [{\citenamefont {Vitale}\ \emph {et~al.}(2020)\citenamefont {Vitale}, \citenamefont {Pizzi}, \citenamefont {Marrazzo}, \citenamefont {Yates}, \citenamefont {Marzari},\ and\ \citenamefont {Mostofi}}]{Vitale2020}%
  \BibitemOpen
  \bibfield  {author} {\bibinfo {author} {\bibfnamefont {V.}~\bibnamefont {Vitale}}, \bibinfo {author} {\bibfnamefont {G.}~\bibnamefont {Pizzi}}, \bibinfo {author} {\bibfnamefont {A.}~\bibnamefont {Marrazzo}}, \bibinfo {author} {\bibfnamefont {J.~R.}\ \bibnamefont {Yates}}, \bibinfo {author} {\bibfnamefont {N.}~\bibnamefont {Marzari}},\ and\ \bibinfo {author} {\bibfnamefont {A.~A.}\ \bibnamefont {Mostofi}},\ }\bibfield  {title} {\bibinfo {title} {{Automated high-throughput Wannierisation}},\ }\href {https://doi.org/10.1038/s41524-020-0312-y} {\bibfield  {journal} {\bibinfo  {journal} {npj Computational Materials}\ }\textbf {\bibinfo {volume} {6}},\ \bibinfo {pages} {66} (\bibinfo {year} {2020})}\BibitemShut {NoStop}%
\bibitem [{\citenamefont {Qiao}\ \emph {et~al.}(2023{\natexlab{a}})\citenamefont {Qiao}, \citenamefont {Pizzi},\ and\ \citenamefont {Marzari}}]{Qiao2023}%
  \BibitemOpen
  \bibfield  {author} {\bibinfo {author} {\bibfnamefont {J.}~\bibnamefont {Qiao}}, \bibinfo {author} {\bibfnamefont {G.}~\bibnamefont {Pizzi}},\ and\ \bibinfo {author} {\bibfnamefont {N.}~\bibnamefont {Marzari}},\ }\bibfield  {title} {\bibinfo {title} {{Projectability disentanglement for accurate and automated electronic-structure Hamiltonians}},\ }\href {https://doi.org/10.1038/s41524-023-01146-w} {\bibfield  {journal} {\bibinfo  {journal} {npj Computational Materials}\ }\textbf {\bibinfo {volume} {9}},\ \bibinfo {pages} {208} (\bibinfo {year} {2023}{\natexlab{a}})}\BibitemShut {NoStop}%
\bibitem [{\citenamefont {Qiao}\ \emph {et~al.}(2023{\natexlab{b}})\citenamefont {Qiao}, \citenamefont {Pizzi},\ and\ \citenamefont {Marzari}}]{Qiao2023a}%
  \BibitemOpen
  \bibfield  {author} {\bibinfo {author} {\bibfnamefont {J.}~\bibnamefont {Qiao}}, \bibinfo {author} {\bibfnamefont {G.}~\bibnamefont {Pizzi}},\ and\ \bibinfo {author} {\bibfnamefont {N.}~\bibnamefont {Marzari}},\ }\bibfield  {title} {\bibinfo {title} {Automated mixing of maximally localized wannier functions into target manifolds},\ }\href {https://doi.org/10.1038/s41524-023-01147-9} {\bibfield  {journal} {\bibinfo  {journal} {npj Computational Materials}\ }\textbf {\bibinfo {volume} {9}},\ \bibinfo {pages} {206} (\bibinfo {year} {2023}{\natexlab{b}})}\BibitemShut {NoStop}%
\bibitem [{\citenamefont {Agapito}\ \emph {et~al.}(2016)\citenamefont {Agapito}, \citenamefont {Ismail-Beigi}, \citenamefont {Curtarolo}, \citenamefont {Fornari},\ and\ \citenamefont {Nardelli}}]{Agapito2016}%
  \BibitemOpen
  \bibfield  {author} {\bibinfo {author} {\bibfnamefont {L.~A.}\ \bibnamefont {Agapito}}, \bibinfo {author} {\bibfnamefont {S.}~\bibnamefont {Ismail-Beigi}}, \bibinfo {author} {\bibfnamefont {S.}~\bibnamefont {Curtarolo}}, \bibinfo {author} {\bibfnamefont {M.}~\bibnamefont {Fornari}},\ and\ \bibinfo {author} {\bibfnamefont {M.~B.}\ \bibnamefont {Nardelli}},\ }\bibfield  {title} {\bibinfo {title} {{Accurate tight-binding Hamiltonian matrices from ab initio calculations: Minimal basis sets}},\ }\href {https://doi.org/10.1103/PhysRevB.93.035104} {\bibfield  {journal} {\bibinfo  {journal} {Physical Review B}\ }\textbf {\bibinfo {volume} {93}},\ \bibinfo {pages} {035104} (\bibinfo {year} {2016})}\BibitemShut {NoStop}%
\bibitem [{\citenamefont {Baltz}\ \emph {et~al.}(2018)\citenamefont {Baltz}, \citenamefont {Manchon}, \citenamefont {Tsoi}, \citenamefont {Moriyama}, \citenamefont {Ono},\ and\ \citenamefont {Tserkovnyak}}]{Baltz2018}%
  \BibitemOpen
  \bibfield  {author} {\bibinfo {author} {\bibfnamefont {V.}~\bibnamefont {Baltz}}, \bibinfo {author} {\bibfnamefont {A.}~\bibnamefont {Manchon}}, \bibinfo {author} {\bibfnamefont {M.}~\bibnamefont {Tsoi}}, \bibinfo {author} {\bibfnamefont {T.}~\bibnamefont {Moriyama}}, \bibinfo {author} {\bibfnamefont {T.}~\bibnamefont {Ono}},\ and\ \bibinfo {author} {\bibfnamefont {Y.}~\bibnamefont {Tserkovnyak}},\ }\bibfield  {title} {\bibinfo {title} {{Antiferromagnetic spintronics}},\ }\href {https://doi.org/10.1103/RevModPhys.90.015005} {\bibfield  {journal} {\bibinfo  {journal} {Reviews of Modern Physics}\ }\textbf {\bibinfo {volume} {90}},\ \bibinfo {pages} {015005} (\bibinfo {year} {2018})}\BibitemShut {NoStop}%
\bibitem [{\citenamefont {Xiong}\ \emph {et~al.}(2022)\citenamefont {Xiong}, \citenamefont {Jiang}, \citenamefont {Shi}, \citenamefont {Du}, \citenamefont {Yao}, \citenamefont {Guo}, \citenamefont {Zhu}, \citenamefont {Cao}, \citenamefont {Peng}, \citenamefont {Cai}, \citenamefont {Zhu},\ and\ \citenamefont {Zhao}}]{Xiong2022}%
  \BibitemOpen
  \bibfield  {author} {\bibinfo {author} {\bibfnamefont {D.}~\bibnamefont {Xiong}}, \bibinfo {author} {\bibfnamefont {Y.}~\bibnamefont {Jiang}}, \bibinfo {author} {\bibfnamefont {K.}~\bibnamefont {Shi}}, \bibinfo {author} {\bibfnamefont {A.}~\bibnamefont {Du}}, \bibinfo {author} {\bibfnamefont {Y.}~\bibnamefont {Yao}}, \bibinfo {author} {\bibfnamefont {Z.}~\bibnamefont {Guo}}, \bibinfo {author} {\bibfnamefont {D.}~\bibnamefont {Zhu}}, \bibinfo {author} {\bibfnamefont {K.}~\bibnamefont {Cao}}, \bibinfo {author} {\bibfnamefont {S.}~\bibnamefont {Peng}}, \bibinfo {author} {\bibfnamefont {W.}~\bibnamefont {Cai}}, \bibinfo {author} {\bibfnamefont {D.}~\bibnamefont {Zhu}},\ and\ \bibinfo {author} {\bibfnamefont {W.}~\bibnamefont {Zhao}},\ }\bibfield  {title} {\bibinfo {title} {{Antiferromagnetic spintronics: An overview and outlook}},\ }\href {https://doi.org/10.1016/j.fmre.2022.03.016} {\bibfield  {journal} {\bibinfo  {journal} {Fundamental Research}\ }\textbf {\bibinfo {volume} {2}},\ \bibinfo {pages} {522} (\bibinfo {year} {2022})}\BibitemShut {NoStop}%
\bibitem [{\citenamefont {Kim}\ \emph {et~al.}(2022)\citenamefont {Kim}, \citenamefont {Beach}, \citenamefont {Lee}, \citenamefont {Ono}, \citenamefont {Rasing},\ and\ \citenamefont {Yang}}]{Kim2022}%
  \BibitemOpen
  \bibfield  {author} {\bibinfo {author} {\bibfnamefont {S.~K.}\ \bibnamefont {Kim}}, \bibinfo {author} {\bibfnamefont {G.~S.~D.}\ \bibnamefont {Beach}}, \bibinfo {author} {\bibfnamefont {K.-J.}\ \bibnamefont {Lee}}, \bibinfo {author} {\bibfnamefont {T.}~\bibnamefont {Ono}}, \bibinfo {author} {\bibfnamefont {T.}~\bibnamefont {Rasing}},\ and\ \bibinfo {author} {\bibfnamefont {H.}~\bibnamefont {Yang}},\ }\bibfield  {title} {\bibinfo {title} {{Ferrimagnetic spintronics}},\ }\href {https://doi.org/10.1038/s41563-021-01139-4} {\bibfield  {journal} {\bibinfo  {journal} {Nature Materials}\ }\textbf {\bibinfo {volume} {21}},\ \bibinfo {pages} {24} (\bibinfo {year} {2022})}\BibitemShut {NoStop}%
\bibitem [{\citenamefont {Brinker}\ \emph {et~al.}(2020)\citenamefont {Brinker}, \citenamefont {{dos Santos Dias}},\ and\ \citenamefont {Lounis}}]{Brinker2020}%
  \BibitemOpen
  \bibfield  {author} {\bibinfo {author} {\bibfnamefont {S.}~\bibnamefont {Brinker}}, \bibinfo {author} {\bibfnamefont {M.}~\bibnamefont {{dos Santos Dias}}},\ and\ \bibinfo {author} {\bibfnamefont {S.}~\bibnamefont {Lounis}},\ }\bibfield  {title} {\bibinfo {title} {{Prospecting chiral multisite interactions in prototypical magnetic systems}},\ }\href {https://doi.org/10.1103/PhysRevResearch.2.033240} {\bibfield  {journal} {\bibinfo  {journal} {Physical Review Research}\ }\textbf {\bibinfo {volume} {2}},\ \bibinfo {pages} {033240} (\bibinfo {year} {2020})}\BibitemShut {NoStop}%
\bibitem [{\citenamefont {Ham}\ \emph {et~al.}(2021)\citenamefont {Ham}, \citenamefont {Pradipto}, \citenamefont {Yakushiji}, \citenamefont {Kim}, \citenamefont {Rhim}, \citenamefont {Nakamura}, \citenamefont {Shiota}, \citenamefont {Kim},\ and\ \citenamefont {Ono}}]{Ham2021}%
  \BibitemOpen
  \bibfield  {author} {\bibinfo {author} {\bibfnamefont {W.~S.}\ \bibnamefont {Ham}}, \bibinfo {author} {\bibfnamefont {A.-M.}\ \bibnamefont {Pradipto}}, \bibinfo {author} {\bibfnamefont {K.}~\bibnamefont {Yakushiji}}, \bibinfo {author} {\bibfnamefont {K.}~\bibnamefont {Kim}}, \bibinfo {author} {\bibfnamefont {S.~H.}\ \bibnamefont {Rhim}}, \bibinfo {author} {\bibfnamefont {K.}~\bibnamefont {Nakamura}}, \bibinfo {author} {\bibfnamefont {Y.}~\bibnamefont {Shiota}}, \bibinfo {author} {\bibfnamefont {S.}~\bibnamefont {Kim}},\ and\ \bibinfo {author} {\bibfnamefont {T.}~\bibnamefont {Ono}},\ }\bibfield  {title} {\bibinfo {title} {{Dzyaloshinskii–Moriya interaction in noncentrosymmetric superlattices}},\ }\href {https://doi.org/10.1038/s41524-021-00592-8} {\bibfield  {journal} {\bibinfo  {journal} {npj Computational Materials}\ }\textbf {\bibinfo {volume} {7}},\ \bibinfo {pages} {129} (\bibinfo {year} {2021})}\BibitemShut {NoStop}%
\bibitem [{\citenamefont {Armitage}\ \emph {et~al.}(2018)\citenamefont {Armitage}, \citenamefont {Mele},\ and\ \citenamefont {Vishwanath}}]{Armitage2018}%
  \BibitemOpen
  \bibfield  {author} {\bibinfo {author} {\bibfnamefont {N.~P.}\ \bibnamefont {Armitage}}, \bibinfo {author} {\bibfnamefont {E.~J.}\ \bibnamefont {Mele}},\ and\ \bibinfo {author} {\bibfnamefont {A.}~\bibnamefont {Vishwanath}},\ }\bibfield  {title} {\bibinfo {title} {{Weyl and Dirac semimetals in three-dimensional solids}},\ }\href {https://doi.org/10.1103/RevModPhys.90.015001} {\bibfield  {journal} {\bibinfo  {journal} {Reviews of Modern Physics}\ }\textbf {\bibinfo {volume} {90}},\ \bibinfo {pages} {015001} (\bibinfo {year} {2018})}\BibitemShut {NoStop}%
\bibitem [{\citenamefont {Hasan}\ and\ \citenamefont {Kane}(2010)}]{Hasan2010}%
  \BibitemOpen
  \bibfield  {author} {\bibinfo {author} {\bibfnamefont {M.~Z.}\ \bibnamefont {Hasan}}\ and\ \bibinfo {author} {\bibfnamefont {C.~L.}\ \bibnamefont {Kane}},\ }\bibfield  {title} {\bibinfo {title} {{Colloquium : Topological insulators}},\ }\href {https://doi.org/10.1103/RevModPhys.82.3045} {\bibfield  {journal} {\bibinfo  {journal} {Reviews of Modern Physics}\ }\textbf {\bibinfo {volume} {82}},\ \bibinfo {pages} {3045} (\bibinfo {year} {2010})}\BibitemShut {NoStop}%
\bibitem [{\citenamefont {Zhou}\ \emph {et~al.}(2024)\citenamefont {Zhou}, \citenamefont {{dos Santos Dias}}, \citenamefont {Zhang}, \citenamefont {Zhao},\ and\ \citenamefont {Lounis}}]{Zhou2024}%
  \BibitemOpen
  \bibfield  {author} {\bibinfo {author} {\bibfnamefont {H.}~\bibnamefont {Zhou}}, \bibinfo {author} {\bibfnamefont {M.}~\bibnamefont {{dos Santos Dias}}}, \bibinfo {author} {\bibfnamefont {Y.}~\bibnamefont {Zhang}}, \bibinfo {author} {\bibfnamefont {W.}~\bibnamefont {Zhao}},\ and\ \bibinfo {author} {\bibfnamefont {S.}~\bibnamefont {Lounis}},\ }\bibfield  {title} {\bibinfo {title} {{Kagomerization of transition metal monolayers induced by two-dimensional hexagonal boron nitride}},\ }\href {https://doi.org/10.1038/s41467-024-48973-z} {\bibfield  {journal} {\bibinfo  {journal} {Nature Communications}\ }\textbf {\bibinfo {volume} {15}},\ \bibinfo {pages} {4854} (\bibinfo {year} {2024})}\BibitemShut {NoStop}%
\bibitem [{\citenamefont {Manchon}\ \emph {et~al.}(2015)\citenamefont {Manchon}, \citenamefont {Koo}, \citenamefont {Nitta}, \citenamefont {Frolov},\ and\ \citenamefont {Duine}}]{Manchon2015}%
  \BibitemOpen
  \bibfield  {author} {\bibinfo {author} {\bibfnamefont {A.}~\bibnamefont {Manchon}}, \bibinfo {author} {\bibfnamefont {H.~C.}\ \bibnamefont {Koo}}, \bibinfo {author} {\bibfnamefont {J.}~\bibnamefont {Nitta}}, \bibinfo {author} {\bibfnamefont {S.~M.}\ \bibnamefont {Frolov}},\ and\ \bibinfo {author} {\bibfnamefont {R.~A.}\ \bibnamefont {Duine}},\ }\bibfield  {title} {\bibinfo {title} {{New perspectives for Rashba spin-orbit coupling}},\ }\href {https://doi.org/10.1038/nmat4360} {\bibfield  {journal} {\bibinfo  {journal} {Nature Materials}\ }\textbf {\bibinfo {volume} {14}},\ \bibinfo {pages} {871} (\bibinfo {year} {2015})}\BibitemShut {NoStop}%
\bibitem [{\citenamefont {Prandini}\ \emph {et~al.}(2018)\citenamefont {Prandini}, \citenamefont {Marrazzo}, \citenamefont {Castelli}, \citenamefont {Mounet},\ and\ \citenamefont {Marzari}}]{Prandini2018}%
  \BibitemOpen
  \bibfield  {author} {\bibinfo {author} {\bibfnamefont {G.}~\bibnamefont {Prandini}}, \bibinfo {author} {\bibfnamefont {A.}~\bibnamefont {Marrazzo}}, \bibinfo {author} {\bibfnamefont {I.~E.}\ \bibnamefont {Castelli}}, \bibinfo {author} {\bibfnamefont {N.}~\bibnamefont {Mounet}},\ and\ \bibinfo {author} {\bibfnamefont {N.}~\bibnamefont {Marzari}},\ }\bibfield  {title} {\bibinfo {title} {{Precision and efficiency in solid-state pseudopotential calculations}},\ }\bibfield  {journal} {\bibinfo  {journal} {npj Computational Materials}\ }\textbf {\bibinfo {volume} {4}},\ \href {https://doi.org/10.1038/s41524-018-0127-2} {10.1038/s41524-018-0127-2} (\bibinfo {year} {2018})\BibitemShut {NoStop}%
\bibitem [{\citenamefont {Dresselhaus}\ \emph {et~al.}(2008)\citenamefont {Dresselhaus}, \citenamefont {Dresselhaus},\ and\ \citenamefont {Jorio}}]{Dresselhaus2008}%
  \BibitemOpen
  \bibfield  {author} {\bibinfo {author} {\bibfnamefont {M.~S.}\ \bibnamefont {Dresselhaus}}, \bibinfo {author} {\bibfnamefont {G.}~\bibnamefont {Dresselhaus}},\ and\ \bibinfo {author} {\bibfnamefont {A.}~\bibnamefont {Jorio}},\ }\href {https://doi.org/10.1007/978-3-540-32899-5} {\emph {\bibinfo {title} {Group Theory}}}\ (\bibinfo  {publisher} {Springer Berlin Heidelberg},\ \bibinfo {address} {Berlin, Heidelberg},\ \bibinfo {year} {2008})\ pp.\ \bibinfo {pages} {1--582}\BibitemShut {NoStop}%
\bibitem [{\citenamefont {Xiao}\ \emph {et~al.}(2011)\citenamefont {Xiao}, \citenamefont {Liu}, \citenamefont {Feng}, \citenamefont {Xu},\ and\ \citenamefont {Yao}}]{Xiao2012}%
  \BibitemOpen
  \bibfield  {author} {\bibinfo {author} {\bibfnamefont {D.}~\bibnamefont {Xiao}}, \bibinfo {author} {\bibfnamefont {G.-B.}\ \bibnamefont {Liu}}, \bibinfo {author} {\bibfnamefont {W.}~\bibnamefont {Feng}}, \bibinfo {author} {\bibfnamefont {X.}~\bibnamefont {Xu}},\ and\ \bibinfo {author} {\bibfnamefont {W.}~\bibnamefont {Yao}},\ }\bibfield  {title} {\bibinfo {title} {{Coupled spin and valley physics in monolayers of MoS2 and other group-VI dichalcogenides}},\ }\href {https://doi.org/10.1103/PhysRevLett.108.196802} {\bibfield  {journal} {\bibinfo  {journal} {Physical Review Letters}\ }\textbf {\bibinfo {volume} {108}},\ \bibinfo {pages} {196802} (\bibinfo {year} {2011})}\BibitemShut {NoStop}%
\bibitem [{\citenamefont {Dirac}(1928)}]{Dirac1928}%
  \BibitemOpen
  \bibfield  {author} {\bibinfo {author} {\bibfnamefont {P.~A.~M.}\ \bibnamefont {Dirac}},\ }\bibfield  {title} {\bibinfo {title} {{The quantum theory of the electron}},\ }\href {https://doi.org/10.1098/rspa.1928.0023} {\bibfield  {journal} {\bibinfo  {journal} {Proceedings of the Royal Society of London. Series A, Containing Papers of a Mathematical and Physical Character}\ }\textbf {\bibinfo {volume} {117}},\ \bibinfo {pages} {610} (\bibinfo {year} {1928})}\BibitemShut {NoStop}%
\bibitem [{\citenamefont {Bechstedt}(2015)}]{Bechstedt2015}%
  \BibitemOpen
  \bibfield  {author} {\bibinfo {author} {\bibfnamefont {F.}~\bibnamefont {Bechstedt}},\ }\href {https://doi.org/10.1007/978-3-662-44593-8} {\emph {\bibinfo {title} {Many-Body Approach to Electronic Excitations}}},\ \bibinfo {series} {Springer Series in Solid-State Sciences}, Vol.\ \bibinfo {volume} {181}\ (\bibinfo  {publisher} {Springer Berlin Heidelberg},\ \bibinfo {address} {Berlin, Heidelberg},\ \bibinfo {year} {2015})\ p.\ \bibinfo {pages} {457}\BibitemShut {NoStop}%
\bibitem [{\citenamefont {van Setten}\ \emph {et~al.}(2018)\citenamefont {van Setten}, \citenamefont {Giantomassi}, \citenamefont {Bousquet}, \citenamefont {Verstraete}, \citenamefont {Hamann}, \citenamefont {Gonze},\ and\ \citenamefont {Rignanese}}]{VanSetten2018}%
  \BibitemOpen
  \bibfield  {author} {\bibinfo {author} {\bibfnamefont {M.}~\bibnamefont {van Setten}}, \bibinfo {author} {\bibfnamefont {M.}~\bibnamefont {Giantomassi}}, \bibinfo {author} {\bibfnamefont {E.}~\bibnamefont {Bousquet}}, \bibinfo {author} {\bibfnamefont {M.}~\bibnamefont {Verstraete}}, \bibinfo {author} {\bibfnamefont {D.}~\bibnamefont {Hamann}}, \bibinfo {author} {\bibfnamefont {X.}~\bibnamefont {Gonze}},\ and\ \bibinfo {author} {\bibfnamefont {G.-M.}\ \bibnamefont {Rignanese}},\ }\bibfield  {title} {\bibinfo {title} {{The PseudoDojo: Training and grading a 85 element optimized norm-conserving pseudopotential table}},\ }\href {https://doi.org/10.1016/j.cpc.2018.01.012} {\bibfield  {journal} {\bibinfo  {journal} {Computer Physics Communications}\ }\textbf {\bibinfo {volume} {226}},\ \bibinfo {pages} {39} (\bibinfo {year} {2018})}\BibitemShut {NoStop}%
\bibitem [{\citenamefont {{Dal Corso}}(2014)}]{DalCorso2014}%
  \BibitemOpen
  \bibfield  {author} {\bibinfo {author} {\bibfnamefont {A.}~\bibnamefont {{Dal Corso}}},\ }\bibfield  {title} {\bibinfo {title} {{Pseudopotentials periodic table: From H to Pu}},\ }\href {https://doi.org/10.1016/j.commatsci.2014.07.043} {\bibfield  {journal} {\bibinfo  {journal} {Computational Materials Science}\ }\textbf {\bibinfo {volume} {95}},\ \bibinfo {pages} {337} (\bibinfo {year} {2014})}\BibitemShut {NoStop}%
\bibitem [{psl()}]{pslib_suggest}%
  \BibitemOpen
  \href@noop {} {\bibinfo {title} {Suggested pseudopotentials}},\ \bibinfo {howpublished} {\url{https://dalcorso.github.io/pslibrary/PP_list.html}},\ \bibinfo {note} {online; accessed 20 January 2025}\BibitemShut {NoStop}%
\bibitem [{\citenamefont {Giannozzi}\ \emph {et~al.}(2009)\citenamefont {Giannozzi}, \citenamefont {Baroni}, \citenamefont {Bonini}, \citenamefont {Calandra}, \citenamefont {Car}, \citenamefont {Cavazzoni}, \citenamefont {Ceresoli}, \citenamefont {Chiarotti}, \citenamefont {Cococcioni}, \citenamefont {Dabo}, \citenamefont {{Dal Corso}}, \citenamefont {{De Gironcoli}}, \citenamefont {Fabris}, \citenamefont {Fratesi}, \citenamefont {Gebauer}, \citenamefont {Gerstmann}, \citenamefont {Gougoussis}, \citenamefont {Kokalj}, \citenamefont {Lazzeri}, \citenamefont {Martin-Samos}, \citenamefont {Marzari}, \citenamefont {Mauri}, \citenamefont {Mazzarello}, \citenamefont {Paolini}, \citenamefont {Pasquarello}, \citenamefont {Paulatto}, \citenamefont {Sbraccia}, \citenamefont {Scandolo}, \citenamefont {Sclauzero}, \citenamefont {Seitsonen}, \citenamefont {Smogunov}, \citenamefont {Umari},\ and\ \citenamefont {Wentzcovitch}}]{Giannozzi2009}%
  \BibitemOpen
  \bibfield  {author} {\bibinfo {author} {\bibfnamefont {P.}~\bibnamefont {Giannozzi}}, \bibinfo {author} {\bibfnamefont {S.}~\bibnamefont {Baroni}}, \bibinfo {author} {\bibfnamefont {N.}~\bibnamefont {Bonini}}, \bibinfo {author} {\bibfnamefont {M.}~\bibnamefont {Calandra}}, \bibinfo {author} {\bibfnamefont {R.}~\bibnamefont {Car}}, \bibinfo {author} {\bibfnamefont {C.}~\bibnamefont {Cavazzoni}}, \bibinfo {author} {\bibfnamefont {D.}~\bibnamefont {Ceresoli}}, \bibinfo {author} {\bibfnamefont {G.~L.}\ \bibnamefont {Chiarotti}}, \bibinfo {author} {\bibfnamefont {M.}~\bibnamefont {Cococcioni}}, \bibinfo {author} {\bibfnamefont {I.}~\bibnamefont {Dabo}}, \bibinfo {author} {\bibfnamefont {A.}~\bibnamefont {{Dal Corso}}}, \bibinfo {author} {\bibfnamefont {S.}~\bibnamefont {{De Gironcoli}}}, \bibinfo {author} {\bibfnamefont {S.}~\bibnamefont {Fabris}}, \bibinfo {author} {\bibfnamefont {G.}~\bibnamefont {Fratesi}}, \bibinfo {author} {\bibfnamefont {R.}~\bibnamefont {Gebauer}}, \bibinfo {author} {\bibfnamefont {U.}~\bibnamefont {Gerstmann}}, \bibinfo {author} {\bibfnamefont {C.}~\bibnamefont {Gougoussis}}, \bibinfo {author} {\bibfnamefont {A.}~\bibnamefont {Kokalj}}, \bibinfo {author} {\bibfnamefont {M.}~\bibnamefont {Lazzeri}}, \bibinfo {author} {\bibfnamefont {L.}~\bibnamefont {Martin-Samos}}, \bibinfo {author} {\bibfnamefont {N.}~\bibnamefont {Marzari}}, \bibinfo {author} {\bibfnamefont {F.}~\bibnamefont {Mauri}}, \bibinfo {author} {\bibfnamefont {R.}~\bibnamefont {Mazzarello}}, \bibinfo {author} {\bibfnamefont {S.}~\bibnamefont {Paolini}}, \bibinfo {author} {\bibfnamefont {A.}~\bibnamefont {Pasquarello}}, \bibinfo {author} {\bibfnamefont {L.}~\bibnamefont {Paulatto}}, \bibinfo {author} {\bibfnamefont {C.}~\bibnamefont {Sbraccia}}, \bibinfo {author} {\bibfnamefont {S.}~\bibnamefont {Scandolo}}, \bibinfo {author} {\bibfnamefont {G.}~\bibnamefont {Sclauzero}}, \bibinfo {author} {\bibfnamefont {A.~P.}\ \bibnamefont {Seitsonen}}, \bibinfo {author} {\bibfnamefont {A.}~\bibnamefont {Smogunov}}, \bibinfo {author} {\bibfnamefont {P.}~\bibnamefont {Umari}},\ and\ \bibinfo {author} {\bibfnamefont {R.~M.}\ \bibnamefont {Wentzcovitch}},\ }\bibfield  {title} {\bibinfo {title} {{Q\textsc{uantum} ESPRESSO: A modular and open-source software project for quantum simulations of materials}},\ }\bibfield  {journal} {\bibinfo  {journal} {Journal of Physics Condensed Matter}\ }\textbf {\bibinfo {volume} {21}},\ \href {https://doi.org/10.1088/0953-8984/21/39/395502} {10.1088/0953-8984/21/39/395502} (\bibinfo {year} {2009})\BibitemShut {NoStop}%
\bibitem [{\citenamefont {Giannozzi}\ \emph {et~al.}(2020)\citenamefont {Giannozzi}, \citenamefont {Baseggio}, \citenamefont {Bonf{\`{a}}}, \citenamefont {Brunato}, \citenamefont {Car}, \citenamefont {Carnimeo}, \citenamefont {Cavazzoni}, \citenamefont {{De Gironcoli}}, \citenamefont {Delugas}, \citenamefont {{Ferrari Ruffino}}, \citenamefont {Ferretti}, \citenamefont {Marzari}, \citenamefont {Timrov}, \citenamefont {Urru},\ and\ \citenamefont {Baroni}}]{Giannozzi2020}%
  \BibitemOpen
  \bibfield  {author} {\bibinfo {author} {\bibfnamefont {P.}~\bibnamefont {Giannozzi}}, \bibinfo {author} {\bibfnamefont {O.}~\bibnamefont {Baseggio}}, \bibinfo {author} {\bibfnamefont {P.}~\bibnamefont {Bonf{\`{a}}}}, \bibinfo {author} {\bibfnamefont {D.}~\bibnamefont {Brunato}}, \bibinfo {author} {\bibfnamefont {R.}~\bibnamefont {Car}}, \bibinfo {author} {\bibfnamefont {I.}~\bibnamefont {Carnimeo}}, \bibinfo {author} {\bibfnamefont {C.}~\bibnamefont {Cavazzoni}}, \bibinfo {author} {\bibfnamefont {S.}~\bibnamefont {{De Gironcoli}}}, \bibinfo {author} {\bibfnamefont {P.}~\bibnamefont {Delugas}}, \bibinfo {author} {\bibfnamefont {F.}~\bibnamefont {{Ferrari Ruffino}}}, \bibinfo {author} {\bibfnamefont {A.}~\bibnamefont {Ferretti}}, \bibinfo {author} {\bibfnamefont {N.}~\bibnamefont {Marzari}}, \bibinfo {author} {\bibfnamefont {I.}~\bibnamefont {Timrov}}, \bibinfo {author} {\bibfnamefont {A.}~\bibnamefont {Urru}},\ and\ \bibinfo {author} {\bibfnamefont {S.}~\bibnamefont {Baroni}},\ }\bibfield  {title} {\bibinfo {title} {{Q\textsc{uantum} ESPRESSO toward the exascale}},\ }\bibfield  {journal} {\bibinfo  {journal} {Journal of Chemical Physics}\ }\textbf {\bibinfo {volume} {152}},\ \href {https://doi.org/10.1063/5.0005082} {10.1063/5.0005082} (\bibinfo {year} {2020})\BibitemShut {NoStop}%
\bibitem [{mc3()}]{mc3d}%
  \BibitemOpen
  \href@noop {} {\bibinfo {title} {Materials cloud three-dimensional structure database ({MC3D})}},\ \bibinfo {howpublished} {\url{https://mc3d.materialscloud.org}},\ \bibinfo {note} {online; accessed 20 January 2025}\BibitemShut {NoStop}%
\bibitem [{\citenamefont {Ozaki}(2003)}]{Ozaki2003}%
  \BibitemOpen
  \bibfield  {author} {\bibinfo {author} {\bibfnamefont {T.}~\bibnamefont {Ozaki}},\ }\bibfield  {title} {\bibinfo {title} {{Variationally optimized atomic orbitals for large-scale electronic structures}},\ }\href {https://doi.org/10.1103/PhysRevB.67.155108} {\bibfield  {journal} {\bibinfo  {journal} {Physical Review B}\ }\textbf {\bibinfo {volume} {67}},\ \bibinfo {pages} {155108} (\bibinfo {year} {2003})}\BibitemShut {NoStop}%
\bibitem [{\citenamefont {Leon}\ \emph {et~al.}(2013)\citenamefont {Leon}, \citenamefont {Bj{\"{o}}rck},\ and\ \citenamefont {Gander}}]{Leon2013}%
  \BibitemOpen
  \bibfield  {author} {\bibinfo {author} {\bibfnamefont {S.~J.}\ \bibnamefont {Leon}}, \bibinfo {author} {\bibfnamefont {{\AA}.}~\bibnamefont {Bj{\"{o}}rck}},\ and\ \bibinfo {author} {\bibfnamefont {W.}~\bibnamefont {Gander}},\ }\bibfield  {title} {\bibinfo {title} {{Gram-Schmidt orthogonalization: 100 years and more}},\ }\href {https://doi.org/10.1002/nla.1839} {\bibfield  {journal} {\bibinfo  {journal} {Numerical Linear Algebra with Applications}\ }\textbf {\bibinfo {volume} {20}},\ \bibinfo {pages} {492} (\bibinfo {year} {2013})}\BibitemShut {NoStop}%
\bibitem [{aww()}]{aww}%
  \BibitemOpen
  \href@noop {} {\bibinfo {title} {{aiida-wannier90-workflows: A collection of advanced au- tomated workflows to compute Wannier functions using AiiDA and the Wannier90 code}}},\ \bibinfo {howpublished} {\url{https://github.com/aiidateam/aiida-wannier90-workflows}},\ \bibinfo {note} {online; accessed 22 January 2025}\BibitemShut {NoStop}%
\bibitem [{\citenamefont {Mostofi}\ \emph {et~al.}(2008)\citenamefont {Mostofi}, \citenamefont {Yates}, \citenamefont {Lee}, \citenamefont {Souza}, \citenamefont {Vanderbilt},\ and\ \citenamefont {Marzari}}]{Mostofi2008}%
  \BibitemOpen
  \bibfield  {author} {\bibinfo {author} {\bibfnamefont {A.~A.}\ \bibnamefont {Mostofi}}, \bibinfo {author} {\bibfnamefont {J.~R.}\ \bibnamefont {Yates}}, \bibinfo {author} {\bibfnamefont {Y.-S.}\ \bibnamefont {Lee}}, \bibinfo {author} {\bibfnamefont {I.}~\bibnamefont {Souza}}, \bibinfo {author} {\bibfnamefont {D.}~\bibnamefont {Vanderbilt}},\ and\ \bibinfo {author} {\bibfnamefont {N.}~\bibnamefont {Marzari}},\ }\bibfield  {title} {\bibinfo {title} {{wannier90: A tool for obtaining maximally-localised Wannier functions}},\ }\href {https://doi.org/https://doi.org/10.1016/j.cpc.2007.11.016} {\bibfield  {journal} {\bibinfo  {journal} {Computer Physics Communications}\ }\textbf {\bibinfo {volume} {178}},\ \bibinfo {pages} {685} (\bibinfo {year} {2008})}\BibitemShut {NoStop}%
\bibitem [{\citenamefont {Pizzi}\ \emph {et~al.}(2020)\citenamefont {Pizzi}, \citenamefont {Vitale}, \citenamefont {Arita}, \citenamefont {Bl{\"{u}}gel}, \citenamefont {Freimuth}, \citenamefont {G{\'{e}}ranton}, \citenamefont {Gibertini}, \citenamefont {Gresch}, \citenamefont {Johnson}, \citenamefont {Koretsune}, \citenamefont {Iba{\~{n}}ez-Azpiroz}, \citenamefont {Lee}, \citenamefont {Lihm}, \citenamefont {Marchand}, \citenamefont {Marrazzo}, \citenamefont {Mokrousov}, \citenamefont {Mustafa}, \citenamefont {Nohara}, \citenamefont {Nomura}, \citenamefont {Paulatto}, \citenamefont {Ponc{\'{e}}}, \citenamefont {Ponweiser}, \citenamefont {Qiao}, \citenamefont {Th{\"{o}}le}, \citenamefont {Tsirkin}, \citenamefont {Wierzbowska}, \citenamefont {Marzari}, \citenamefont {Vanderbilt}, \citenamefont {Souza}, \citenamefont {Mostofi},\ and\ \citenamefont {Yates}}]{Pizzi2020}%
  \BibitemOpen
  \bibfield  {author} {\bibinfo {author} {\bibfnamefont {G.}~\bibnamefont {Pizzi}}, \bibinfo {author} {\bibfnamefont {V.}~\bibnamefont {Vitale}}, \bibinfo {author} {\bibfnamefont {R.}~\bibnamefont {Arita}}, \bibinfo {author} {\bibfnamefont {S.}~\bibnamefont {Bl{\"{u}}gel}}, \bibinfo {author} {\bibfnamefont {F.}~\bibnamefont {Freimuth}}, \bibinfo {author} {\bibfnamefont {G.}~\bibnamefont {G{\'{e}}ranton}}, \bibinfo {author} {\bibfnamefont {M.}~\bibnamefont {Gibertini}}, \bibinfo {author} {\bibfnamefont {D.}~\bibnamefont {Gresch}}, \bibinfo {author} {\bibfnamefont {C.}~\bibnamefont {Johnson}}, \bibinfo {author} {\bibfnamefont {T.}~\bibnamefont {Koretsune}}, \bibinfo {author} {\bibfnamefont {J.}~\bibnamefont {Iba{\~{n}}ez-Azpiroz}}, \bibinfo {author} {\bibfnamefont {H.}~\bibnamefont {Lee}}, \bibinfo {author} {\bibfnamefont {J.-M.}\ \bibnamefont {Lihm}}, \bibinfo {author} {\bibfnamefont {D.}~\bibnamefont {Marchand}}, \bibinfo {author} {\bibfnamefont {A.}~\bibnamefont {Marrazzo}}, \bibinfo {author} {\bibfnamefont {Y.}~\bibnamefont {Mokrousov}}, \bibinfo {author} {\bibfnamefont {J.~I.}\ \bibnamefont {Mustafa}}, \bibinfo {author} {\bibfnamefont {Y.}~\bibnamefont {Nohara}}, \bibinfo {author} {\bibfnamefont {Y.}~\bibnamefont {Nomura}}, \bibinfo {author} {\bibfnamefont {L.}~\bibnamefont {Paulatto}}, \bibinfo {author} {\bibfnamefont {S.}~\bibnamefont {Ponc{\'{e}}}}, \bibinfo {author} {\bibfnamefont {T.}~\bibnamefont {Ponweiser}}, \bibinfo {author} {\bibfnamefont {J.}~\bibnamefont {Qiao}}, \bibinfo {author} {\bibfnamefont {F.}~\bibnamefont {Th{\"{o}}le}}, \bibinfo {author} {\bibfnamefont {S.~S.}\ \bibnamefont {Tsirkin}}, \bibinfo {author} {\bibfnamefont {M.}~\bibnamefont {Wierzbowska}}, \bibinfo {author} {\bibfnamefont {N.}~\bibnamefont {Marzari}}, \bibinfo {author} {\bibfnamefont {D.}~\bibnamefont {Vanderbilt}}, \bibinfo {author} {\bibfnamefont {I.}~\bibnamefont {Souza}}, \bibinfo {author} {\bibfnamefont {A.~A.}\ \bibnamefont {Mostofi}},\ and\ \bibinfo {author} {\bibfnamefont {J.~R.}\ \bibnamefont {Yates}},\ }\bibfield  {title} {\bibinfo {title} {{Wannier90 as a community code: new features and applications}},\ }\href {https://doi.org/10.1088/1361-648X/ab51ff} {\bibfield  {journal} {\bibinfo  {journal} {Journal of Physics: Condensed Matter}\ }\textbf {\bibinfo {volume} {32}},\ \bibinfo {pages} {165902} (\bibinfo {year} {2020})}\BibitemShut {NoStop}%
\bibitem [{\citenamefont {Pizzi}\ \emph {et~al.}(2016)\citenamefont {Pizzi}, \citenamefont {Cepellotti}, \citenamefont {Sabatini}, \citenamefont {Marzari},\ and\ \citenamefont {Kozinsky}}]{Pizzi2016}%
  \BibitemOpen
  \bibfield  {author} {\bibinfo {author} {\bibfnamefont {G.}~\bibnamefont {Pizzi}}, \bibinfo {author} {\bibfnamefont {A.}~\bibnamefont {Cepellotti}}, \bibinfo {author} {\bibfnamefont {R.}~\bibnamefont {Sabatini}}, \bibinfo {author} {\bibfnamefont {N.}~\bibnamefont {Marzari}},\ and\ \bibinfo {author} {\bibfnamefont {B.}~\bibnamefont {Kozinsky}},\ }\bibfield  {title} {\bibinfo {title} {{AiiDA: automated interactive infrastructure and database for computational science}},\ }\href {https://doi.org/10.1016/j.commatsci.2015.09.013} {\bibfield  {journal} {\bibinfo  {journal} {Computational Materials Science}\ }\textbf {\bibinfo {volume} {111}},\ \bibinfo {pages} {218} (\bibinfo {year} {2016})}\BibitemShut {NoStop}%
\bibitem [{\citenamefont {Huber}\ \emph {et~al.}(2020)\citenamefont {Huber}, \citenamefont {Zoupanos}, \citenamefont {Uhrin}, \citenamefont {Talirz}, \citenamefont {Kahle}, \citenamefont {H{\"{a}}uselmann}, \citenamefont {Gresch}, \citenamefont {M{\"{u}}ller}, \citenamefont {Yakutovich}, \citenamefont {Andersen}, \citenamefont {Ramirez}, \citenamefont {Adorf}, \citenamefont {Gargiulo}, \citenamefont {Kumbhar}, \citenamefont {Passaro}, \citenamefont {Johnston}, \citenamefont {Merkys}, \citenamefont {Cepellotti}, \citenamefont {Mounet}, \citenamefont {Marzari}, \citenamefont {Kozinsky},\ and\ \citenamefont {Pizzi}}]{Huber2020}%
  \BibitemOpen
  \bibfield  {author} {\bibinfo {author} {\bibfnamefont {S.~P.}\ \bibnamefont {Huber}}, \bibinfo {author} {\bibfnamefont {S.}~\bibnamefont {Zoupanos}}, \bibinfo {author} {\bibfnamefont {M.}~\bibnamefont {Uhrin}}, \bibinfo {author} {\bibfnamefont {L.}~\bibnamefont {Talirz}}, \bibinfo {author} {\bibfnamefont {L.}~\bibnamefont {Kahle}}, \bibinfo {author} {\bibfnamefont {R.}~\bibnamefont {H{\"{a}}uselmann}}, \bibinfo {author} {\bibfnamefont {D.}~\bibnamefont {Gresch}}, \bibinfo {author} {\bibfnamefont {T.}~\bibnamefont {M{\"{u}}ller}}, \bibinfo {author} {\bibfnamefont {A.~V.}\ \bibnamefont {Yakutovich}}, \bibinfo {author} {\bibfnamefont {C.~W.}\ \bibnamefont {Andersen}}, \bibinfo {author} {\bibfnamefont {F.~F.}\ \bibnamefont {Ramirez}}, \bibinfo {author} {\bibfnamefont {C.~S.}\ \bibnamefont {Adorf}}, \bibinfo {author} {\bibfnamefont {F.}~\bibnamefont {Gargiulo}}, \bibinfo {author} {\bibfnamefont {S.}~\bibnamefont {Kumbhar}}, \bibinfo {author} {\bibfnamefont {E.}~\bibnamefont {Passaro}}, \bibinfo {author} {\bibfnamefont {C.}~\bibnamefont {Johnston}}, \bibinfo {author} {\bibfnamefont {A.}~\bibnamefont {Merkys}}, \bibinfo {author} {\bibfnamefont {A.}~\bibnamefont {Cepellotti}}, \bibinfo {author} {\bibfnamefont {N.}~\bibnamefont {Mounet}}, \bibinfo {author} {\bibfnamefont {N.}~\bibnamefont {Marzari}}, \bibinfo {author} {\bibfnamefont {B.}~\bibnamefont {Kozinsky}},\ and\ \bibinfo {author} {\bibfnamefont {G.}~\bibnamefont {Pizzi}},\ }\bibfield  {title} {\bibinfo {title} {{AiiDA 1.0, a scalable computational infrastructure for automated reproducible workflows and data provenance}},\ }\href {https://doi.org/10.1038/s41597-020-00638-4} {\bibfield  {journal} {\bibinfo  {journal} {Scientific Data}\ }\textbf {\bibinfo {volume} {7}},\ \bibinfo {pages} {300} (\bibinfo {year} {2020})}\BibitemShut {NoStop}%
\bibitem [{\citenamefont {Uhrin}\ \emph {et~al.}(2021)\citenamefont {Uhrin}, \citenamefont {Huber}, \citenamefont {Yu}, \citenamefont {Marzari},\ and\ \citenamefont {Pizzi}}]{Uhrin2021}%
  \BibitemOpen
  \bibfield  {author} {\bibinfo {author} {\bibfnamefont {M.}~\bibnamefont {Uhrin}}, \bibinfo {author} {\bibfnamefont {S.~P.}\ \bibnamefont {Huber}}, \bibinfo {author} {\bibfnamefont {J.}~\bibnamefont {Yu}}, \bibinfo {author} {\bibfnamefont {N.}~\bibnamefont {Marzari}},\ and\ \bibinfo {author} {\bibfnamefont {G.}~\bibnamefont {Pizzi}},\ }\bibfield  {title} {\bibinfo {title} {{Workflows in AiiDA: Engineering a high-throughput, event-based engine for robust and modular computational workflows}},\ }\href {https://doi.org/10.1016/j.commatsci.2020.110086} {\bibfield  {journal} {\bibinfo  {journal} {Computational Materials Science}\ }\textbf {\bibinfo {volume} {187}},\ \bibinfo {pages} {110086} (\bibinfo {year} {2021})}\BibitemShut {NoStop}%
\bibitem [{\citenamefont {Liu}\ \emph {et~al.}(2024)\citenamefont {Liu}, \citenamefont {Zhang}, \citenamefont {Fang}, \citenamefont {Weng},\ and\ \citenamefont {Wu}}]{Liu2024}%
  \BibitemOpen
  \bibfield  {author} {\bibinfo {author} {\bibfnamefont {Z.}~\bibnamefont {Liu}}, \bibinfo {author} {\bibfnamefont {S.}~\bibnamefont {Zhang}}, \bibinfo {author} {\bibfnamefont {Z.}~\bibnamefont {Fang}}, \bibinfo {author} {\bibfnamefont {H.}~\bibnamefont {Weng}},\ and\ \bibinfo {author} {\bibfnamefont {Q.}~\bibnamefont {Wu}},\ }\bibfield  {title} {\bibinfo {title} {Combined first-principles and boltzmann transport theory methodology for studying magnetotransport in magnetic materials},\ }\href {https://doi.org/10.1103/PhysRevResearch.6.043185} {\bibfield  {journal} {\bibinfo  {journal} {Phys. Rev. Res.}\ }\textbf {\bibinfo {volume} {6}},\ \bibinfo {pages} {043185} (\bibinfo {year} {2024})}\BibitemShut {NoStop}%
\bibitem [{\citenamefont {Bercx}\ \emph {et~al.}(2025)\citenamefont {Bercx}, \citenamefont {Poncé}, \citenamefont {Zhang}, \citenamefont {Trezza}, \citenamefont {Ghezeljehmeidan}, \citenamefont {Bastonero}, \citenamefont {Qiao}, \citenamefont {von Rohr}, \citenamefont {Pizzi}, \citenamefont {Chiavazzo},\ and\ \citenamefont {Marzari}}]{Bercx2025}%
  \BibitemOpen
  \bibfield  {author} {\bibinfo {author} {\bibfnamefont {M.}~\bibnamefont {Bercx}}, \bibinfo {author} {\bibfnamefont {S.}~\bibnamefont {Poncé}}, \bibinfo {author} {\bibfnamefont {Y.}~\bibnamefont {Zhang}}, \bibinfo {author} {\bibfnamefont {G.}~\bibnamefont {Trezza}}, \bibinfo {author} {\bibfnamefont {A.~G.}\ \bibnamefont {Ghezeljehmeidan}}, \bibinfo {author} {\bibfnamefont {L.}~\bibnamefont {Bastonero}}, \bibinfo {author} {\bibfnamefont {J.}~\bibnamefont {Qiao}}, \bibinfo {author} {\bibfnamefont {F.~O.}\ \bibnamefont {von Rohr}}, \bibinfo {author} {\bibfnamefont {G.}~\bibnamefont {Pizzi}}, \bibinfo {author} {\bibfnamefont {E.}~\bibnamefont {Chiavazzo}},\ and\ \bibinfo {author} {\bibfnamefont {N.}~\bibnamefont {Marzari}},\ }\href {https://arxiv.org/abs/2503.10943} {\bibinfo {title} {{Charting the landscape of Bardeen-Cooper-Schrieffer superconductors in experimentally known compounds}}} (\bibinfo {year} {2025}),\ \Eprint {https://arxiv.org/abs/2503.10943} {arXiv:2503.10943 [cond-mat.supr-con]} \BibitemShut {NoStop}%
\bibitem [{aat()}]{aatom}%
  \BibitemOpen
  \href@noop {} {\bibinfo {title} {{AiiDA-Atomistic: AiiDA plugin for atomistic materials-science simulations}}},\ \bibinfo {howpublished} {\url{https://github.com/aiidateam/aiida-atomistic}},\ \bibinfo {note} {online; accessed 27 January 2025}\BibitemShut {NoStop}%
\bibitem [{\citenamefont {Talirz}\ \emph {et~al.}(2020)\citenamefont {Talirz}, \citenamefont {Kumbhar}, \citenamefont {Passaro}, \citenamefont {Yakutovich}, \citenamefont {Granata}, \citenamefont {Gargiulo}, \citenamefont {Borelli}, \citenamefont {Uhrin}, \citenamefont {Huber}, \citenamefont {Zoupanos}, \citenamefont {Adorf}, \citenamefont {Andersen}, \citenamefont {Sch{\"u}tt}, \citenamefont {Pignedoli}, \citenamefont {Passerone}, \citenamefont {VandeVondele}, \citenamefont {Schulthess}, \citenamefont {Smit}, \citenamefont {Pizzi},\ and\ \citenamefont {Marzari}}]{Talirz2020}%
  \BibitemOpen
  \bibfield  {author} {\bibinfo {author} {\bibfnamefont {L.}~\bibnamefont {Talirz}}, \bibinfo {author} {\bibfnamefont {S.}~\bibnamefont {Kumbhar}}, \bibinfo {author} {\bibfnamefont {E.}~\bibnamefont {Passaro}}, \bibinfo {author} {\bibfnamefont {A.~V.}\ \bibnamefont {Yakutovich}}, \bibinfo {author} {\bibfnamefont {V.}~\bibnamefont {Granata}}, \bibinfo {author} {\bibfnamefont {F.}~\bibnamefont {Gargiulo}}, \bibinfo {author} {\bibfnamefont {M.}~\bibnamefont {Borelli}}, \bibinfo {author} {\bibfnamefont {M.}~\bibnamefont {Uhrin}}, \bibinfo {author} {\bibfnamefont {S.~P.}\ \bibnamefont {Huber}}, \bibinfo {author} {\bibfnamefont {S.}~\bibnamefont {Zoupanos}}, \bibinfo {author} {\bibfnamefont {C.~S.}\ \bibnamefont {Adorf}}, \bibinfo {author} {\bibfnamefont {C.~W.}\ \bibnamefont {Andersen}}, \bibinfo {author} {\bibfnamefont {O.}~\bibnamefont {Sch{\"u}tt}}, \bibinfo {author} {\bibfnamefont {C.~A.}\ \bibnamefont {Pignedoli}}, \bibinfo {author} {\bibfnamefont {D.}~\bibnamefont {Passerone}}, \bibinfo {author} {\bibfnamefont {J.}~\bibnamefont {VandeVondele}}, \bibinfo {author} {\bibfnamefont {T.~C.}\ \bibnamefont {Schulthess}}, \bibinfo {author} {\bibfnamefont {B.}~\bibnamefont {Smit}}, \bibinfo {author} {\bibfnamefont {G.}~\bibnamefont {Pizzi}},\ and\ \bibinfo {author} {\bibfnamefont {N.}~\bibnamefont {Marzari}},\ }\bibfield  {title} {\bibinfo {title} {Materials cloud, a platform for open computational science},\ }\href@noop {} {\bibfield  {journal} {\bibinfo  {journal} {Scientific Data}\ }\textbf {\bibinfo {volume} {7}},\ \bibinfo {pages} {299} (\bibinfo {year} {2020})}\BibitemShut {NoStop}%
\bibitem [{\citenamefont {Jiang}\ \emph {et~al.}(2025)\citenamefont {Jiang}, \citenamefont {Qiao}, \citenamefont {Paulish}, \citenamefont {Weisheng}, \citenamefont {Marzari},\ and\ \citenamefont {Pizzi}}]{MCA}%
  \BibitemOpen
  \bibfield  {author} {\bibinfo {author} {\bibfnamefont {Y.}~\bibnamefont {Jiang}}, \bibinfo {author} {\bibfnamefont {J.}~\bibnamefont {Qiao}}, \bibinfo {author} {\bibfnamefont {N.}~\bibnamefont {Paulish}}, \bibinfo {author} {\bibfnamefont {Z.}~\bibnamefont {Weisheng}}, \bibinfo {author} {\bibfnamefont {N.}~\bibnamefont {Marzari}},\ and\ \bibinfo {author} {\bibfnamefont {G.}~\bibnamefont {Pizzi}},\ }\bibfield  {title} {\bibinfo {title} {{Robust Wannierization including magnetization and spin-orbit coupling via projectability disentanglement}},\ }\href {https://doi.org/10.24435/materialscloud:9g-ds} {10.24435/materialscloud:9g-ds} (\bibinfo {year} {2025})\BibitemShut {NoStop}%
\end{thebibliography}
\end{document}